\documentclass[aps,pra,twocolumn]{revtex4-1}
\usepackage{amsmath,amssymb}
\usepackage{natbib}
\usepackage{braket}
\usepackage{hyperref}
\usepackage{graphicx}
\usepackage{color}
\usepackage[caption=false]{subfig}
\usepackage{epstopdf}
\DeclareMathOperator{\Tr}{Tr}
\renewcommand{\Im}{\text{Im}}
\renewcommand{\Re}{\text{Re}}
\begin{document}
\title{Non-Markovian dynamics of a 
superconducting qubit in an open multimode resonator}
\author{Moein Malekakhlagh}
\author{Alexandru Petrescu}
\author{Hakan E. T\"ureci}
\affiliation{Department of Electrical Engineering, Princeton University, Princeton, New Jersey, 08544}
\date{\today}
\begin{abstract}
We study the dynamics of a transmon qubit that is capacitively coupled to an open multimode superconducting resonator. Our effective equations are derived by eliminating resonator degrees of freedom while encoding their effect in the Green's function of the electromagnetic background. We account for the dissipation of the resonator exactly by employing a spectral representation for the Green's function in terms of a set of non-Hermitian modes and show that it is possible to derive effective  Heisenberg-Langevin equations without resorting to the rotating wave, two level, Born or Markov approximations. A well-behaved time domain perturbation theory is derived to systematically account for the nonlinearity of the transmon. We apply this method to the problem of spontaneous emission, capturing accurately the non-Markovian features of the qubit dynamics, valid for any qubit-resonator coupling strength.
\end{abstract}
\maketitle
\section{Introduction}
Superconducting circuits are of interest for gate based quantum information processing \cite{Devoret_Implementing_2004, Blais_Quantum-Information_2007, Devoret_Superconducting_2013} and for fundamental studies of collective quantum phenomena away from equilibrium \cite{Houck_On-chip_2012, Schmidt_Circuit_2013, LeHur_Many-body_2016}. In these circuits, Josephson junctions provide the nonlinearity required to define a qubit or a pseudo-spin degree of freedom, and low loss microwave waveguides and resonators provide a convenient linear environment to mediate interactions between Josephson junctions \cite{Majer_Coupling_2007, Sillanpaa_Coherent_2007, Filipp_Preparation_2011, Filipp_Multimode_2011, Loo_Photon-mediated_2013, Shankar_Autonomously_2013, Kimchi-Schwartz_Stabilizing_2016}, act as Purcell filters \cite{Houck_Controlling_2008, Jeffrey_Fast_2014, Bronn_Broadband_2015} or as suitable access ports for efficient state preparation and readout.
Fabrication capabilities have reached a stage where coherent interactions between multiple qubits occur through a waveguide \cite{Loo_Photon-mediated_2013}, active coupling elements \cite{Roushan_Chiral_2016} or cavity arrays \cite{McKay_High-Contrast_2015}, while allowing manipulation and readout of individual qubits in the circuit.  In addition, experiments started deliberately probing regimes featuring very high qubit coupling strengths \cite{Niemczyk_Circuit_2010, Diaz_Observation_2010, Todorov_Ultrastrong_2010} or setups where multimode effects cannot be avoided \cite{Sundaresan_Beyond_2015}. Accurate modeling of these complex circuits has not only become important for designing such circuits, e.g. to avoid cross talk and filter out the electromagnetic environment, but also for the fundamental question of the collective quantum dynamics of qubits \cite{Mlynek_Observation_2014}. In this work, we introduce a first principles Heisenberg-Langevin framework that accounts for such complexity.

\begin{figure}[t!]
\centering
\subfloat[\label{subfig:cQEDopenSybolic}]{%
\includegraphics{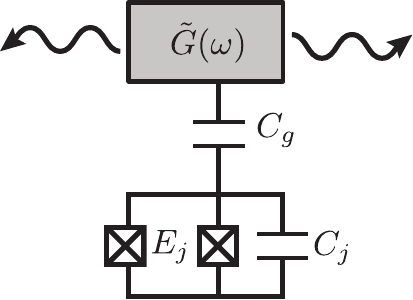}%
} 
\subfloat[\label{subfig:cQEDopenSybolicSpider}]{%
\includegraphics{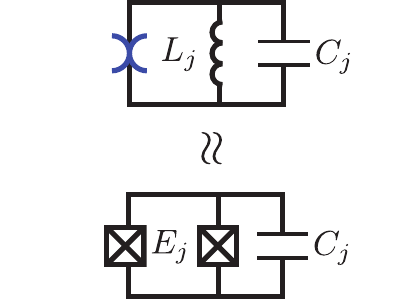}%
} 
\caption{a) Transmon qubit linearly (capacitively) coupled to an open harmonic electromagnetic background, i.e. a multimode superconducting resonator, characterized by Green's function $\tilde{G}(\omega)$. b) Separation of linear and anharmonic parts of the Josephson potential.}
\label{Fig:cQEDopenSybolic}
\end{figure}

The inadequacy of the standard Cavity QED models based on the interaction of a pseudo-spin degree of freedom with a single cavity mode was recognized early on \cite{Houck_Controlling_2008}. In principle, the Rabi model could straightforwardly be extended to include many cavity electromagnetic modes and the remaining qubit transitions (See Sec. III of \cite{Malekakhlagh_Origin_2016}), but this does not provide a computationally viable approach for several reasons. Firstly, we do not know of a systematic approach for the truncation of this multimode multilevel system. Secondly, the truncation itself will depend strongly on the spectral range that is being probed in a given experiment (typically around a transition frequency of the qubit), and the effective model for a given frequency would have to accurately describe the resonator loss in a broad frequency range. It is then unclear whether the Markov approximation would be sufficient to describe such losses. 

Multimode effects come to the fore in the accurate computation of the effective Purcell decay of a qubit \cite{Houck_Controlling_2008} or the photon-mediated effective exchange interaction between qubits in the dispersive regime \cite{Filipp_Multimode_2011}, where the perturbation theory is divergent. A phenomenological semiclassical approach to the accurate modeling of Purcell loss has been suggested~\cite{Houck_Controlling_2008}, based on the availability of the effective impedance seen by the qubit. A full quantum model that incorporates the effective impedance of the linear part of the circuit at its core was later presented~\cite{Nigg_Black-Box_2012}. This approach correctly recognizes that a better behaved perturbation theory in the nonlinearity can be developed if the hybridization of the qubit with the linear multimode environment is taken into account at the outset \cite{Bourassa_Josephson_2012}. Incorporating the dressing of the modes into the basis that is used to expand the nonlinearity gives then rise to self- and cross-Kerr interactions between hybridized modes. This basis however does not account for the open nature of the resonator. Qubit loss is then extracted from the poles of the linear circuit impedance at the qubit port, $Z(\omega)$. This quantity can in principle be measured or obtained from a simulation of the classical Maxwell equations. Finding the poles of $Z(\omega)$ through Foster's theorem introduces potential numerical complications \cite{solgun_blackbox_2014}. Moreover, the interplay of the qubit nonlinearity and dissipation is not addressed within Rayleigh-Schr\"odinger perturbation theory. An exact treatment of dissipation is important for the calculation of multimode Purcell rates of qubits as well as the dynamics of driven dissipative qubit networks \cite{aron_photon-mediated_2016}.

The difficulty with incorporating dissipation on equal footing with energetics in open systems is symptomatic of more general issues in the quantization of radiation in finite inhomogeneous media. One of the earliest thorough treatments of this problem \cite{Glauber_Quantum_1991} proposes to use a complete set of states in the unbounded space including the finite body as a scattering object. This ``modes of the universe'' approach \cite{lang_why_1973, ching_quasinormal-mode_1998} is well-defined but has an impractical aspect: one has to deal with a continuum of modes, and as a consequence simple properties characterizing the scatterer itself (e.g. its resonance frequencies and widths) are not effectively utilized. Several methods have been proposed since then to address this shortcoming, which discussed quantization using quasi-modes (resonances) of the finite-sized open resonator \cite{dalton_quasi_1999, lamprecht_quantized_1999, dutra_quantized_2000, hackenbroich_field_2002}. Usually, these methods treat the atomic degree of freedom as a two-level system and use the rotating wave and the Markov approximations. 

In the present work, rather than using a Hamiltonian description, we derive an effective Heisenberg-Langevin equation to describe the dynamics of a transmon qubit \cite{Koch_Charge_2007} capacitively coupled to an open multimode resonator (See Fig.~\ref{subfig:cQEDopenSybolic}). Our treatment illustrates a general framework that does not rely on the Markov, rotating wave or two level approximations. We show that the electromagnetic degrees of freedom of the entire circuit can be integrated out and appear in the equation of motion through the classical electromagnetic Green's function (GF) corresponding to the Maxwell operator and the associated boundary conditions. A spectral representation of the GF in terms of a complete set of non-Hermitian modes \cite{Tureci_SelfConsistent_2006, martin_claassen_constant_2010} accounts for dissipative effects from first principles. This requires the solution of a boundary-value problem of the Maxwell operator {\it only} in the finite domain of the resonator. Our main result is the effective equation of motion~(\ref{eqn:Eff Dyn before trace}), which is a Heisenberg-Langevin \cite{Scully_Quantum_1997, Gardiner_Quantum_2004, Carmichael_Statistical_1998} integro-differential equation for the phase operator of the transmon. Outgoing fields, which may be desired to calculate the homodyne field at the input of an amplifier chain, can be conveniently related through the GF to the qubit phase operator.   

As an immediate application, we use the effective Heisenberg-Langevin equation of motion to study spontaneous emission. The spontaneous emission of a two level system in a finite polarizable medium was calculated \cite{Dung_Spontaneous_2000} in the Schr\"odinger-picture in the spirit of Wigner-Weisskopf theory \cite{Scully_Quantum_1997}. These calculations are based on a radiation field quantization procedure which incorporates continuity and boundary conditions corresponding to the finite dielectric \cite{matloob_electromagnetic_1995, gruner_green-function_1996}, but only focus on separable geometries where the GF can be calculated semianalytically. A generalization of this methodology to an arbitrary geometry \cite{Krimer_Route_2014} uses an expansion of the GF in terms of a set of non-Hermitian modes for the appropriate boundary value problem \cite{Tureci_SelfConsistent_2006, martin_claassen_constant_2010}. This approach is able to consistently account for multimode effects where the atom-field coupling strength is of the order of the free spectral range of the cavity \cite{Meiser_Superstrong_2006, Krimer_Route_2014, Sundaresan_Beyond_2015} for which the atom is found to emit narrow pulses at the cavity roundtrip period \cite{Krimer_Route_2014}. A drawback of these previous calculations performed in the Schr\"odinger picture is that without the rotating wave approximation, no truncation scheme has been proposed so far to reduce the infinite hierarchy of equations to a tractable Hilbert space dimension. The employment of the rotating wave approximation breaks this infinite hierarchy through the existence of a conserved excitation number. The Heisenberg-Langevin method introduced here is valid for arbitrary light-matter coupling, and therefore can access the dynamics accurately where the rotating-wave approximation is not valid.

In summary, our microscopic treatment of the openness is one essential difference between our study and previous works on the collective excitations of circuit-QED systems with a localized Josephson nonlinearity \cite{Wallquist_Selective_2006, Bourassa_Josephson_2012, Nigg_Black-Box_2012, Leib_Networks_2012}. In our work, the lifetime of the collective excitations arises from a proper treatment of the resonator boundary conditions \cite{Clerk_Introduction_2010}. The harmonic theory of the coupled transmon-resonator system is exactly solvable via Laplace transform. Transmon qubits typically operate in a weakly nonlinear regime, where charge dispersion is negligible \cite{Koch_Charge_2007}. We treat the Josephson anharmonicity on top of the non-Hermitian linear theory (See Fig~\ref{subfig:cQEDopenSybolicSpider}) using multi-scale perturbation theory (MSPT) \cite{Bender_Advanced_1999, Nayfeh_Nonlinear_2008, Strogatz_Nonlinear_2014}. First, it resolves the anomaly of secular contributions in conventional time-domain perturbation theories via a resummation \cite{Bender_Advanced_1999, Nayfeh_Nonlinear_2008, Strogatz_Nonlinear_2014}. While this perturbation theory is equivalent to the Rayleigh-Schr\"odinger perturbation theory when the electromagnetic environment is closed, it allows a systematic expansion even when the environment is open and the dynamics is non-unitary. Second, we account for the self-Kerr and cross-Kerr interactions \cite{Drummond_Quantum_1980} between the collective non-Hermitian excitations extending \cite{Bourassa_Josephson_2012, Nigg_Black-Box_2012}. Third, treating the transmon qubit as a weakly nonlinear bosonic degree of freedom allows us to include the linear coupling to the environment non-perturbatively. This is unlike the dispersive limit treatment of the light-matter coupling as a perturbation \cite{Boissonneault_Dispersive_2009}. Therefore, the effective equation of motion is valid for all experimentally accessible coupling strengths \cite{Thompson_Observation_1992, Strong_Wallraff_2004, Reithmaier_Strong_2004, Anappara_Signatures_2009, Niemczyk_Circuit_2010, Diaz_Observation_2010, Todorov_Ultrastrong_2010,  Sundaresan_Beyond_2015}. 

We finally present a perturbative procedure to reduce the computational complexity of the solution of Eq.~(\ref{eqn:Eff Dyn before trace}), originating from the enormous Hilbert space size, when the qubit is weakly anharmonic. Electromagnetic degrees of freedom can then be perturbatively traced out resulting in an effective equation of motion~(\ref{eqn:NumSim-QuDuffingOscMemReduced}) in the qubit Hilbert space \textit{only}, which makes its numerical simulation tractable.

The paper is organized as follows: In Sec.~\ref{Sec:Toy Model}, we introduce a toy model to familiarize the reader with the main ideas and notation. In Sec.~\ref{Sec:Eff Dyn Of Transmon}, we present an ab initio effective Heisenberg picture dynamics for the transmon qubit. The derivation for this effective model has been discussed in detail in Apps.~\ref{App:Quantum EOM} and \ref{App:Eff Dyn of transmon}. In Sec.~\ref{Sec:Lin SE Theory}, we study linear theory of spontaneous emission. In Sec.~\ref{Sec:PertCor}, we employ quantum multi-scale perturbation theory to investigate the effective dynamics beyond linear approximation. The details of multi-scale calculations are presented in App.~\ref{App:MSPT}. In Sec.~\ref{Sec:NumSimul} we compare these results with the pure numerical simulation. We summarize the main results of this paper in Sec.~\ref{Sec:Conclusion}.
\section{Toy model}
\label{Sec:Toy Model}
In this section, we discuss a toy model that captures the basic elements of the effective equations (Eq.~(\ref{eqn:Eff Dyn before trace})), which we derive in full microscopic detail in Sec.~\ref{Sec:Eff Dyn Of Transmon}. This will also allow us to introduce the notation and concepts relevant to the rest of this paper, in the context of a tractable and well-known model. We consider the single-mode Cavity QED model, consisting of a nonlinear quantum oscillator (qubit) that couples linearly to a single bosonic degree of freedom representing the cavity mode (Fig.~\ref{Fig:cQEDopenSybolic}). This mode itself is coupled to a continuum set of bosons playing the role of the waveguide modes. When the nonlinear oscillator is truncated to the lowest two levels, this reduces to the standard open Rabi Model, which is generally studied using Master equation \cite{Ridolfo_Photon_2012} or stochastic Schr\"odinger equation \cite{Loic_Quantum_2014} approaches. Here we will discuss a Heisenberg-picture approach to arrive at an equation of motion for qubit quadratures. The Hamiltonian for the toy model is ($\hbar=1$)
\begin{align}
\begin{split}
\hat{\mathcal{H}} &\equiv \frac{\omega_j}{4}\left(\hat{\mathcal{X}}_j^2+\hat{\mathcal{Y}}_j^2\right)+\frac{\omega_j}{2}U(\hat{\mathcal{X}}_j)\\
&+\frac{\omega_c}{4}\left(\hat{\mathcal{X}}_c^2+\hat{\mathcal{Y}}_c^2\right)+g\hat{\mathcal{Y}}_j\hat{\mathcal{Y}}_c\\
&+\sum\limits_{b}\left[\frac{\omega_b}{4}\left(\hat{\mathcal{X}}_b^2+\hat{\mathcal{Y}}_b^2\right)+g_b\hat{\mathcal{Y}}_c\hat{\mathcal{Y}}_b\right],
\end{split}
\label{eqn:ToyModel-H}
\end{align}
where $\omega_j$, $\omega_c$ and $\omega_b$ are bare oscillation frequencies of qubit, the cavity and the bath modes, respectively. We have defined the canonically conjugate variables
\begin{align}
\hat{\mathcal{X}}_l\equiv(\hat{a}_l+\hat{a}_l^{\dag}), \ \hat{\mathcal{Y}}_l\equiv -i(\hat{a}_l-\hat{a}_l^{\dag}),
\label{Eq:ToyModel-Def of X&Y}
\end{align}
where $\hat{a}_{l}$ represent the boson annihilation operator of sector $l\equiv j,c,b$. Furthermore, $g$ and $g_b$ are qubit-cavity and cavity-bath couplings. $U(\hat{\mathcal{X}}_j)$ represents the nonlinear part of the potential shown in Fig.~\ref{subfig:cQEDopenSybolicSpider} with a blue spider symbol.

The remainder of this section is structured as follows. In Sec.~\ref{Sec:ToyModel-Eff Dyn}, we eliminate the cavity and bath degrees of freedom to obtain an effective Heisenberg-Langevin equation of motion for the qubit. We dedicate Sec.~\ref{Sec:ToyModel-LinearTheory} to the resulting characteristic function describing the hybridized modes of the linear theory. 
\subsection{Effective dynamics of the qubit}
\label{Sec:ToyModel-Eff Dyn}
In this subsection, we derive the equations of motion for the Hamiltonian~(\ref{eqn:ToyModel-H}). We first integrate out the bath degrees of freedom via Markov approximation to obtain an effective dissipation for the cavity. Then, we eliminate the degrees of freedom of the leaky cavity mode to arrive at an effective equation of motion for the qubit, expressed in terms of the GF of the cavity. The Heisenberg equations of motion are found as
\begin{subequations}
\begin{align}
&\hat{\dot{\mathcal{X}}}_j(t)=\omega_j\hat{\mathcal{Y}}_j(t)+2g\hat{\mathcal{Y}}_c(t),
\label{eqn:ToyModel-dotX_j}\\
&\hat{\dot{\mathcal{Y}}}_j(t)=-\omega_j\left\{\hat{\mathcal{X}}_j(t)+U'[\hat{\mathcal{X}}_j(t)]\right\},
\label{eqn:ToyModel-dotY_j}\\
&\hat{\dot{\mathcal{X}}}_c(t)=\omega_c\hat{\mathcal{Y}}_c(t)+2g\hat{\mathcal{Y}}_j(t)+\sum
\limits_{b} 2g_b\hat{\mathcal{Y}}_b(t),
\label{eqn:ToyModel-dotX_c}\\
&\hat{\dot{\mathcal{Y}}}_c(t)=-\omega_c\hat{\mathcal{X}}_c(t),
\label{eqn:ToyModel-dotY_c}\\
&\hat{\dot{\mathcal{X}}}_b(t)=\omega_b\hat{\mathcal{Y}}_b(t)+2g_b\hat{\mathcal{Y}}_c(t)
\label{eqn:ToyModel-dotX_b}\\
&\hat{\dot{\mathcal{Y}}}_b(t)=-\omega_b\hat{\mathcal{X}}_b(t),
\label{eqn:ToyModel-dotY_b}
\end{align}
\end{subequations}
where $U'[\hat{\mathcal{X}}_j]\equiv dU/d\hat{\mathcal{X}}_j$. Eliminating $\hat{\mathcal{Y}}_{j,c,b}(t)$ using Eqs.~(\ref{eqn:ToyModel-dotY_j}), (\ref{eqn:ToyModel-dotY_c}) and (\ref{eqn:ToyModel-dotY_b}) first, and integrating out the bath degree of freedom via Markov approximation \cite{Scully_Quantum_1997, Walls_Quantum_2008} we obtain effective equations for the qubit and cavity as 
\begin{subequations}
\begin{align}
\hat{\ddot{\mathcal{X}}}_j(t)+\omega_j^2\left\{\hat{\mathcal{X}}_j(t)+U'[\hat{\mathcal{X}}_j(t)]\right\}=-2g\omega_c\hat{\mathcal{X}}_c(t),
\label{eqn:ToyModel-ddotX_j}
\end{align}
\begin{align}
\begin{split}
&\hat{\ddot{\mathcal{X}}}_c(t)+2\kappa_c\hat{\dot{\mathcal{X}}}_c(t)+\omega_c^2\hat{\mathcal{X}}_c(t)\\
&=-2g\omega_j\left\{\hat{\mathcal{X}}_j(t)+U'[\hat{\mathcal{X}}_j(t)]\right\}-\hat{f}_{B}(t),
\end{split}
\label{eqn:ToyModel-ddotX_c}
\end{align}
\end{subequations}
where $2\kappa_c$ is the effective dissipation \cite{Senitzky_Dissipation_1960, Caldeira_Influence_1981, Clerk_Introduction_2010} and $\hat{f}_B(t)$ is the noise operator of the bath seen by the cavity
\begin{align}
\hat{f}_B(t)=\sum\limits_{b}2g_b\left[\omega_b\hat{\mathcal{X}}_b(0)\cos(\omega_b t)+\hat{\dot{\mathcal{X}}}_b(0)\sin(\omega_b t)\right].
\label{eqn:ToyModel-f_B(t)}
\end{align}
 
Note that Eq.~(\ref{eqn:ToyModel-ddotX_c}) is a linear non-homogoneous ODE in terms of $\hat{\mathcal{X}}_c(t)$. Therefore, it is possible to find its general solution in terms of its impulse response, i.e. the GF of the associated classical cavity oscillator:
\begin{align}
\ddot{G}_c(t,t')+2\kappa_c\dot{G}_c(t,t')+\omega_c^2G_c(t,t')=-\delta(t-t').
\label{eqn:ToyModel-Def of G(t,t')}
\end{align}
Following the Fourier transform conventions
\begin{subequations}
\begin{align}
&\tilde{G}_c(\omega)\equiv \int_{-\infty}^{\infty}dtG_c(t,t')e^{i\omega(t-t')},\\
&G_c(t,t')\equiv \int_{-\infty}^{\infty}\frac{d\omega}{2\pi}\tilde{G}_c(\omega)e^{-i\omega(t-t')},
\end{align}
\end{subequations}
we obtain an algebraic solution for $\tilde{G}_c(\omega)$ as
\begin{align}
\tilde{G}_c(\omega)=\frac{1}{(\omega-\omega_C)(\omega+\omega_C^*)},
\label{eqn:ToyModel-Sol of G(Om)}
\end{align}
with $\omega_C\equiv \nu_c-i\kappa_c$ and $\nu_c\equiv\sqrt{\omega_c^2-\kappa_c^2}$. Taking the inverse Fourier transform of Eq.~(\ref{eqn:ToyModel-Sol of G(Om)}) we find the single mode GF of the cavity oscillator
\begin{align}
G_c(t,t')=-\frac{1}{\nu_c}\sin\left[\nu_c(t-t')\right]e^{-\kappa_c(t-t')}\Theta(t-t'),
\label{eqn:ToyModel-Sol of G(t,t')}
\end{align}   
where since the poles of $\tilde{G}_c(\omega)$ reside in the lower-half of the complex $\omega$-plane, $G_c(t,t')$ is retarded (causal) and $\Theta(t)$ stands for the Heaviside step function \cite{Abramowitz_Handbook_1964}.

Then, the general solution to Eq.~(\ref{eqn:ToyModel-ddotX_c}) can be expressed in terms of $G_c(t,t')$ as \cite{Morse_Methods_1953}
\begin{align}
\begin{split}
&\hat{\mathcal{X}}_c(t)=2g\omega_j\int_{0}^{t}dt'G_c(t,t')\left\{\hat{\mathcal{X}}_j(t')+U'[\hat{\mathcal{X}}_j(t')]\right\}\\
&+\left.\left(\partial_{t'}+2\kappa_c\right)G_c(t,t')\right|_{t'=0}\hat{\mathcal{X}}_c(0)-G_c(t,0)\hat{\dot{\mathcal{X}}}_c(0)\\
&+\int_{0}^{t}dt'G_c(t,t')\hat{f}_B(t').
\end{split}
\label{eqn:ToyModel-Gen Sol of Xc}
\end{align}
Substituting Eq.~(\ref{eqn:ToyModel-Gen Sol of Xc}) into the RHS of Eq.~(\ref{eqn:ToyModel-ddotX_j}) and defining 
\begin{subequations}
\begin{align}
&\mathcal{K}(t)\equiv 4g^2\frac{\omega_c}{\omega_j}G_c(t,0),
\label{eqn:ToyModel-Def of K(t)}\\
&\mathcal{D}(t)\equiv -2g\omega_c G_c(t,0),
\label{eqn:ToyModel-Def of D(t)}\\
&\mathcal{I}(\omega)\equiv-2g\omega_c\tilde{G}_c(\omega),
\label{eqn:ToyModel-Def of I(om)}
\end{align}
\end{subequations}
we find the effective dynamics of the nonlinear oscillator in terms of $\hat{\mathcal{X}}_j(t)$ as
\begin{align}
\begin{split}
&\hat{\ddot{\mathcal{X}}}_j(t)+\omega_j^2\left\{\hat{\mathcal{X}}_j(t)+U'[\hat{\mathcal{X}}_j(t)]\right\}=\\
&-\int_{0}^{t}dt'\mathcal{K}(t-t')\omega_j^2\left\{\hat{\mathcal{X}}_j(t')+U'[\hat{\mathcal{X}}_j(t')]\right\}\\
&+\int_{0}^{t}dt'\mathcal{D}(t-t')\hat{f}_B(t')\\
&+\int_{-\infty}^{\infty}\frac{d\omega}{2\pi}
\mathcal{I}(\omega)\left[(i\omega+2\kappa_c)\hat{\mathcal{X}}_c(0)-\hat{\dot{\mathcal{X}}}_c(0)\right]e^{-i\omega t}.
\end{split}
\label{eqn:ToyModel-Eff Dyn}
\end{align}
The LHS of Eq.~(\ref{eqn:ToyModel-Eff Dyn}) is the free dynamics of the qubit. The first term on the RHS includes the memory of all past events encoded in the memory kernel $\mathcal{K}(t)$. The second term incorporates the influence of bath noise on qubit dynamics and plays the role of a drive term. Finally, the last term captures the effect of the initial operator conditions of the cavity. Note that even though Eq.~(\ref{eqn:ToyModel-Eff Dyn}) is an effective equation for the qubit, all operators act on the full Hilbert space of the qubit and the cavity.
\subsection{Linear theory}
\label{Sec:ToyModel-LinearTheory}
In the absence of the nonlinearity, i.e. $U[\hat{\mathcal{X}}_j]=0$, Eq.~(\ref{eqn:ToyModel-Eff Dyn}) is a linear integro-differential equation that can be solved exactly via unilateral Laplace transform
\begin{align}
\tilde{f}(s)\equiv \int_{0}^{\infty}dt e^{-st}f(t), 
\label{eqn:Def of Laplace}
\end{align} 
since the memory integral on the RHS appears as a convolution between the kernel $\mathcal{K}(t)$ and earlier values of $\hat{\mathcal{X}}_j(t')$ for $0<t'<t$. Employing the convolution identity
\begin{align}
\mathfrak{L}\left\{\int_{0}^{t}dt' \mathcal{K}(t-t')\hat{\mathcal{X}}_j(t')\right\}=\tilde{\mathcal{K}}(s)\hat{\tilde{\mathcal{X}}}_j(s),
\end{align}
we find that the Laplace solution to Eq.~(\ref{eqn:ToyModel-Eff Dyn}) takes the general form 
\begin{align}
\hat{\tilde{\mathcal{X}}}_j(s)=\frac{\hat{\mathcal{N}}_j(s)}{D_j(s)},
\label{eqn:ToyModel-FormSol of mathcal(X)_j}
\end{align}
where the numerator
\begin{align}
\begin{split}
\hat{\mathcal{N}}_j(s)&=s\hat{\mathcal{X}}_j(0)+\hat{\dot{\mathcal{X}}}_j(0)\\
&-\frac{2g\omega_c\left[(s+2\kappa_c)\hat{\mathcal{X}}_c(0)+\hat{\dot{\mathcal{X}}}_c(0)-\hat{\tilde{f}}_B(s)\right]}{s^2+2\kappa_c s+\omega_c^2},
\end{split}
\label{eqn:ToyModel-N_j(s)}
\end{align}
contains the information regarding the initial conditions and the noise operator. The characteristic function $D_j(s)$ is defined as
\begin{align}
\begin{split}
D_j(s)\equiv s^2+\omega_j^2\left[1+\tilde{\mathcal{K}}(s)\right]&=s^2+\omega_j^2\\
&-\frac{4g^2\omega_j\omega_c}{s^2+2\kappa_cs+\omega_c^2},
\end{split}
\label{eqn:ToyModel-D_j(s)}
\end{align}
which is the denominator of the algebraic Laplace solution~(\ref{eqn:ToyModel-FormSol of mathcal(X)_j}). Therefore, its roots determine the complex resonances of the coupled system. The poles of $D_j(s)$ are, on the other hand, the bare complex frequencies of the dissipative cavity oscillator found before, $z_c\equiv -i\omega_C$. Therefore, $D_j(s)$ can always be represented formally as
\begin{align}
D_j(s)=(s-p_j)(s-p_j^*)\frac{(s-p_c)(s-p_c^*)}{(s-z_c)(s-z_c^*)},
\label{eqn:ToyModel-Formal Rep of D(s)}
\end{align}
where $p_j$ and $p_c$ are the qubit-like and cavity-like poles such that for $g\to 0$  we get $p_j\to -i\omega_j$ and $p_c\to -i\omega_C\equiv z_c$. In writing Eq.~(\ref{eqn:ToyModel-Formal Rep of D(s)}), we have used the fact that the roots of a polynomial with real coefficients come in complex conjugate pairs.

\begin{figure}
\subfloat[\label{subfig:ToyModelRWA}]{%
\includegraphics[scale=0.27]{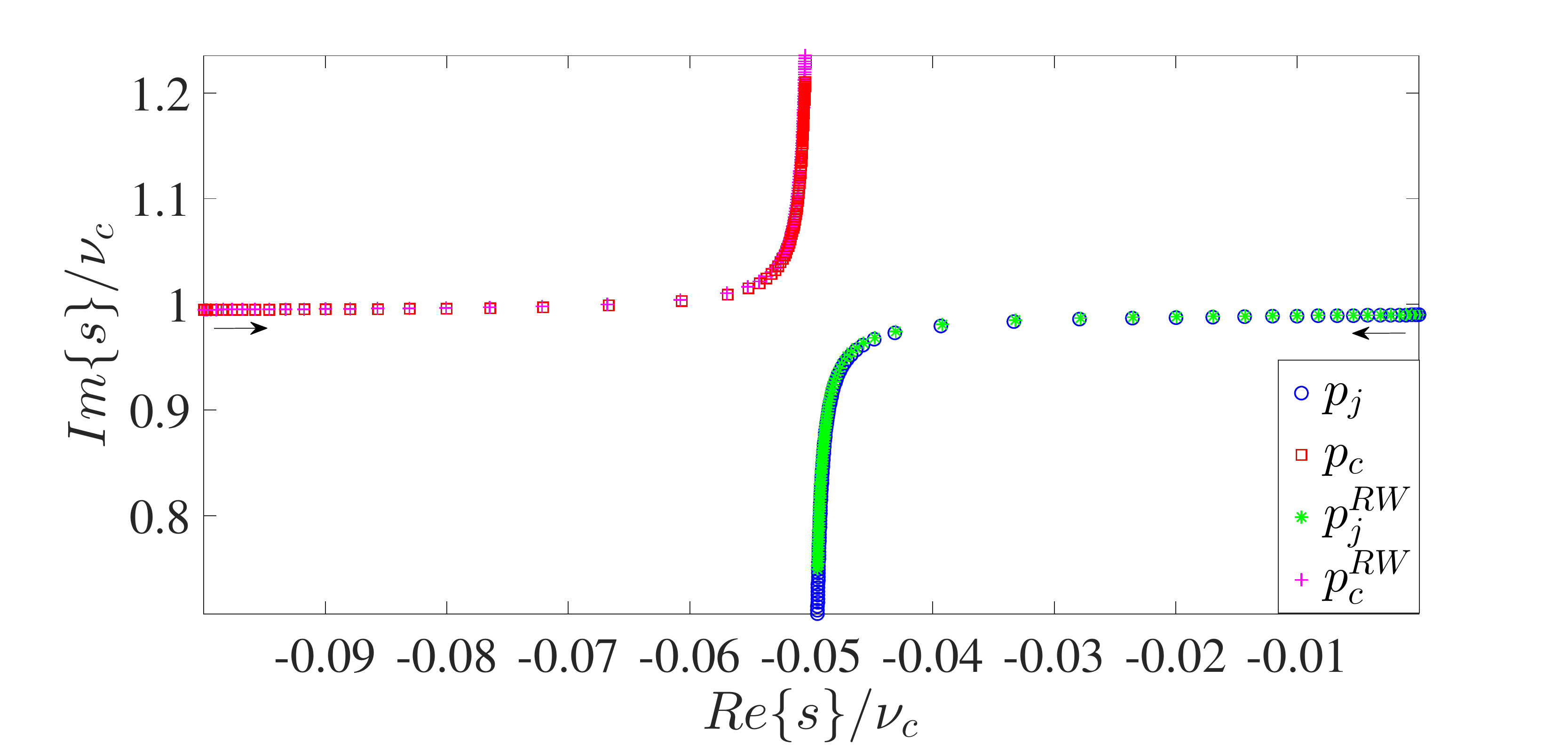}%
}\hfill 
\subfloat[\label{subfig:ToyModelRWADeltaPc}]{%
\includegraphics[scale=0.335]{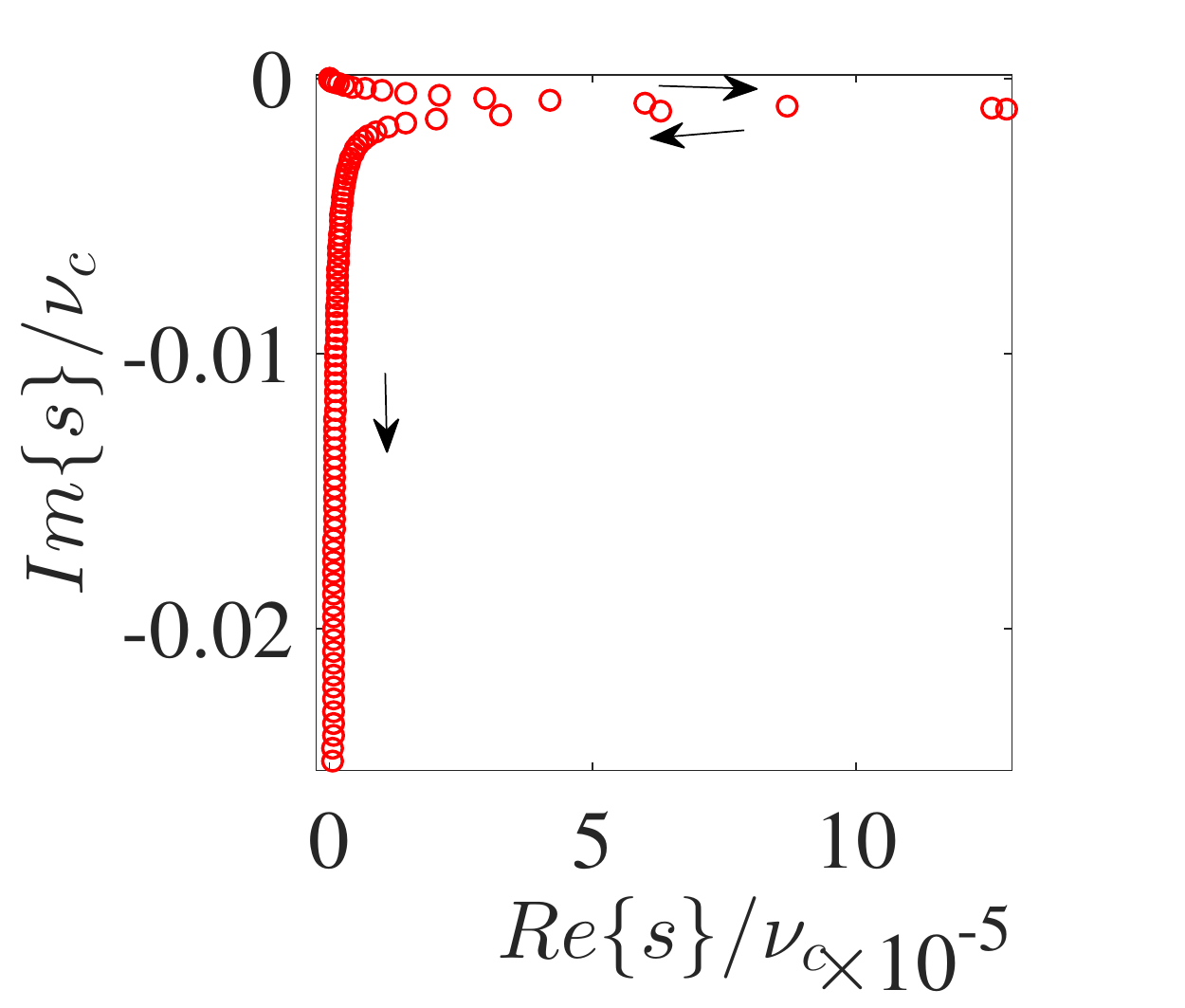}%
}\hfill 
\subfloat[\label{subfig:ToyModelRWADeltaPj}]{%
\includegraphics[scale=0.335]{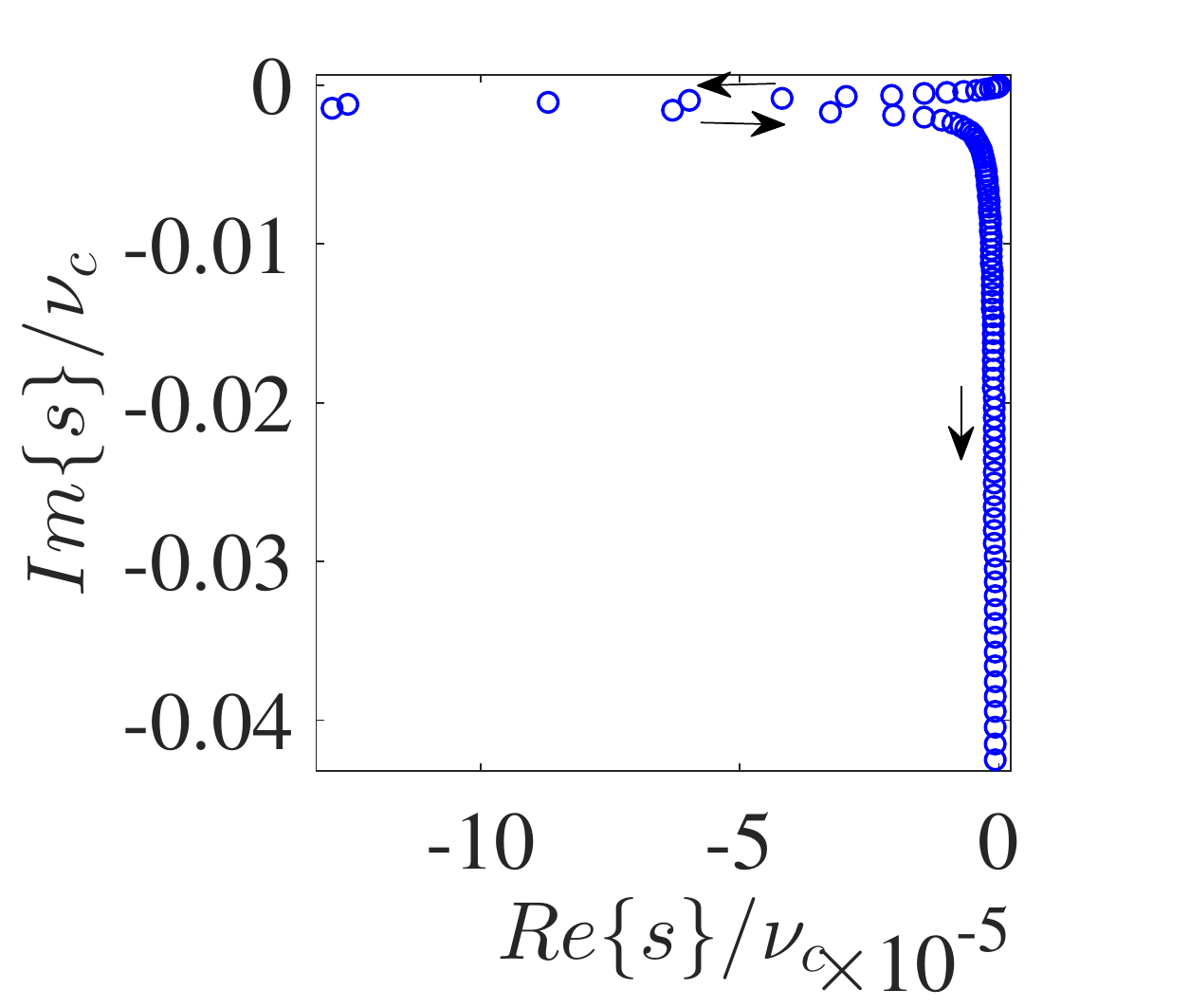}%
} 
\caption{(Color online) a) Hybridized poles of the linear theory, $p_j$ and $p_c$, obtained from Eqs.~(\ref{eqn:ToyModel-D_j(s)}) and (\ref{eqn:ToyModel-RW Dq(s)}) for the resonant case  $\omega_j=\nu_c^-$, $\kappa_c=0.1\nu_c$ as a function of $g\in [0,0.5\omega_j]$ with increment $\Delta g=0.005\omega_j$. The blue circles and green stars show the qubit-like pole $p_j$ with and without RW, respectively. Similarly, the red squares and purple crosses show the cavity-like pole $p_c$. b) and c) represent the difference $\Delta p_{j,c}\equiv p_{j,c}-p_{j,c}^{RW}$ between the two solutions. The black arrows show the direction of increase in $g$.}
\label{Fig:ToyModelRWA}
\end{figure}

It is worth emphasizing that our toy model avoids the rotating wave (RW) approximation. This approximation is known to break down in the ultrastrong coupling regime \cite{Bourassa_Ultrastrong_2009, Anappara_Signatures_2009, Niemczyk_Circuit_2010, Diaz_Observation_2010, Todorov_Ultrastrong_2010}. In order to understand its consequence and make a quantitative comparison, we have to find how the RW approximation modifies $D_j(s)$. Note that by applying the RW approximation, only the coupling Hamiltonian in Eq.~(\ref{eqn:ToyModel-H}) transforms as
\begin{align}
\hat{\mathcal{Y}}_j\hat{\mathcal{Y}}_c\underset{\text{RW}}{\longrightarrow}\frac{1}{2}\left(\hat{\mathcal{X}}_j\hat{\mathcal{X}}_c+\hat{\mathcal{Y}}_j\hat{\mathcal{Y}}_c\right).
\label{eqn:ToyModel-RW Dq(s)}
\end{align}
Then, the modified equations of motion for $\hat{\mathcal{X}}_j(t)$ and $\hat{\mathcal{X}}_c(t)$ read
\begin{subequations}
\begin{align}
\hat{\ddot{\mathcal{X}}}_j(t)+\left(\omega_j^2+g^2\right)\hat{\mathcal{X}}_j(t)=-g(\omega_j+\omega_c)\hat{\mathcal{X}}_c(t),
\label{eqn:ToyModelRWA-ddotX_j}
\end{align}
\begin{align}
\begin{split}
\hat{\ddot{\mathcal{X}}}_c(t)+2\kappa_c\hat{\dot{\mathcal{X}}}_c(t)&+\left(\omega_c^2+g^2\right)\hat{\mathcal{X}}_c(t)\\
&=-g(\omega_j+\omega_c)\hat{\mathcal{X}}_j(t)-\hat{f}_B(t).
\end{split}
\label{eqn:ToyModelRWA-ddotX_c}
\end{align}
\end{subequations}
Note that the form of Eqs.~(\ref{eqn:ToyModelRWA-ddotX_j}-\ref{eqn:ToyModelRWA-ddotX_c}) is the same as Eqs.~(\ref{eqn:ToyModel-ddotX_j}-\ref{eqn:ToyModel-ddotX_c}) except for the modified parameters. Therefore, following the same calculation as in Sec.~\ref{Sec:ToyModel-Eff Dyn} we find a new characteristic function $D_j^{RW}(s)$ which reads
\begin{align}
\begin{split}
D_j^{RW}(s)&=s^2+\left(\omega_j^2+g^2\right)\\
&-\frac{g^2(\omega_j+\omega_c)^2}{s^2+2\kappa_cs+\left(\omega_c^2+g^2\right)}.
\end{split}
\label{eqn:ToyModel-D^RW(s)}
\end{align}

We compare the complex roots of $D_j(s)$ and $D_j^{RW}(s)$ in Fig.~\ref{Fig:ToyModelRWA} as a function of $g$. For $g=0$, the poles start from their bare values $i\omega_j$ and $i\nu_c-\kappa_c$ and the results with and without RW match exactly. As $g$ increases both theories predict that the dissipative cavity oscillator passes some of its decay rate to the qubit oscillator. This is seen in Fig.~\ref{subfig:ToyModelRWA} where the poles move towards each other in the $s$-plane while the oscillation frequency is almost unchanged. As $g$ is increased more, there is an avoided crossing and the poles resolve into two distinct frequencies. After this point, the predictions from $D_j(s)$ and $D_j^{RW}(s)$ for $p_j$ and $p_c$ deviate more significantly. This is more visible in Figs.~\ref{subfig:ToyModelRWADeltaPc} and \ref{subfig:ToyModelRWADeltaPj} that show the difference between the two solutions in the complex $s$-plane. In addition, there is a saturation of the decay rates to half of the bare decay rate of the dissipative cavity oscillator.

In summary, we have obtained the effective equation of motion~(\ref{eqn:ToyModel-Eff Dyn}) for the quadrature $\hat{\mathcal{X}}_j(t)$ of the nonlinear oscillator. This equation incorporates the effects of memory, initial conditions of the cavity and drive. It admits an exact solution via Laplace transform in the absence of nonlinearity. To lowest order, the Josephson nonlinearity is a time-domain perturbation $\propto \hat{\mathcal{X}}_j^3(t)$ in Eq.~(\ref{eqn:ToyModel-Eff Dyn}). This amounts to a quantum Duffing oscillator \cite{Bowen_Quantum_2015} coupled to a linear environment. Time-domain perturbation theory consists of an order by order solution of Eq.~(\ref{eqn:ToyModel-Eff Dyn}). A naive application leads to the appearance of resonant coupling between the solutions at successive orders. The resulting solution contains secular contributions, i.e. terms that grow unbounded in time. We present the resolution of this problem using multi-scale perturbation theory (MSPT) \cite{Bender_Advanced_1999, Nayfeh_Nonlinear_2008, Strogatz_Nonlinear_2014} in Sec.~\ref{Sec:PertCor}.

\section{Effective dynamics of a transmon qubit}
\label{Sec:Eff Dyn Of Transmon}
\begin{figure}
\centering
\includegraphics[scale=0.58]{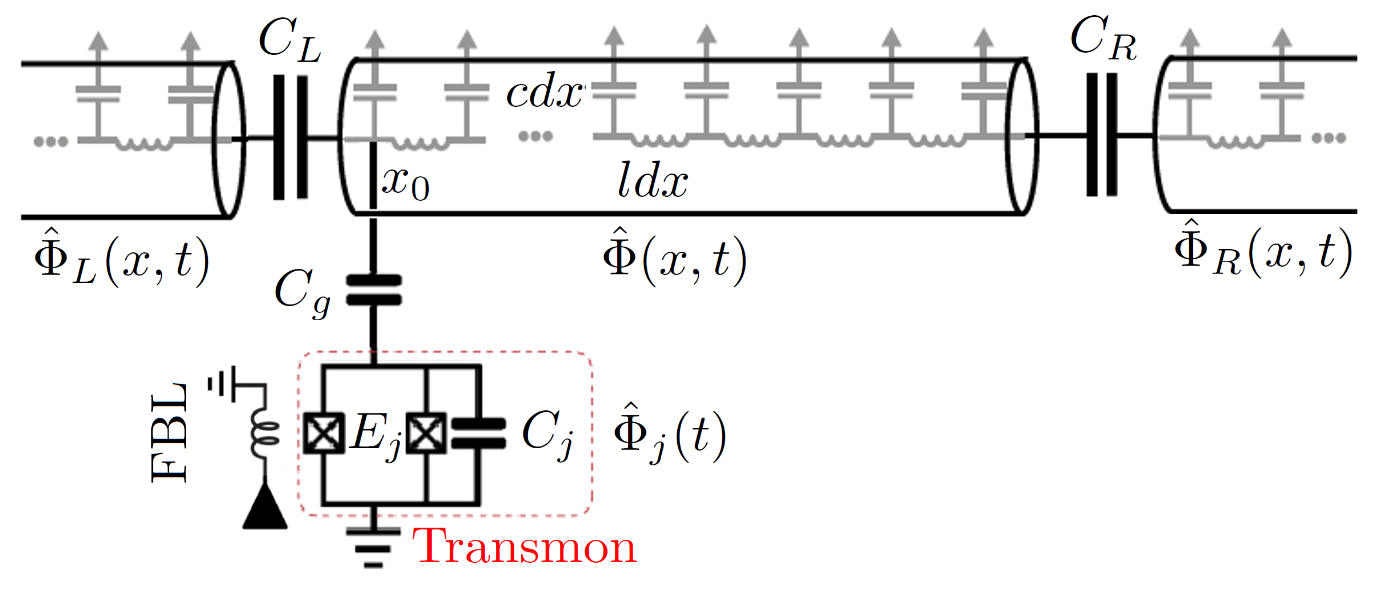}
\caption{A transmon qubit coupled to an open superconducting resonator.}
\label{Fig:cQED-open}
\end{figure}
In this section, we present a first principles calculation for the problem of a transmon qubit that couples capacitively to an open multimode resonator (see Fig.~\ref{Fig:cQED-open}). Like the toy model in Sec.~\ref{Sec:Toy Model}, this calculation relies on an effective equation of motion for the transmon qubit quadratures, in which the photonic degrees of freedom are integrated out. In contrast to the toy model where the decay rate was obtained via Markov approximation, we use a microscopic model for dissipation \cite{Senitzky_Dissipation_1960, Caldeira_Influence_1981}. We model our bath as a pair of semi-infinite waveguides capacitively coupled to each end of a resonator.

As shown in Fig.~\ref{Fig:cQED-open}, the transmon qubit is coupled to a superconducting resonator of finite length $L$ by a capacitance $C_g$. The resonator itself is coupled to the two waveguides at its ends by capacitances $C_R$ and $C_L$, respectively. For all these elements, the capacitance and inductance per length are equal and given as $c$ and $l$, correspondingly. The transmon qubit is characterized by its Josephson energy $E_j$, which is tunable by an external flux bias line (FBL) \cite{Johnson_Quantum_2010}, and its charging energy $E_c$, which is related to the capacitor $C_j$ as $E_c=e^2/(2C_j)$. The explicit circuit quantization is explained in App.~\ref{App:Quantum EOM} following a standard approach \cite{ Devoret_Quantum_1995, Clerk_Introduction_2010, Bishop_Circuit_2010, Devoret_Quantum_2014}. We describe the system in terms of flux operator $\hat{\Phi}_j(t)$ for transmon and flux fields $\hat{\Phi}(x,t)$ and $\hat{\Phi}_{R,L}(x,t)$ for the resonator and waveguides.

The dynamics for the quantum flux operators of the transmon and each resonator shown in Fig.~\ref{Fig:cQED-open} is derived in App.~\ref{App:Quantum EOM}. In what follows, we work with unitless variables 
\begin{align}
\begin{split}
\frac{x}{L}\rightarrow x,\quad
\frac{t}{\sqrt{lc}L}\rightarrow t,\quad
\sqrt{lc}L\omega\rightarrow \omega,\ 2\pi \frac{\hat{\Phi}}{\Phi_0} \rightarrow\hat{\varphi}, 
\end{split}
\label{eqn:unitless vars}
\end{align}
where $\Phi_0\equiv h/(2e)$ is the flux quantum and $1/\sqrt{lc}$ is the phase velocity. We also define unitless parameters
\begin{align}
&\chi_i\equiv\frac{C_i}{cL}, \quad i=R,L,j,g,s\\
&\mathcal{E}_{j,c}\equiv \sqrt{lc} L\frac{E_{j,c}}{\hbar}. 
\end{align}
\begin{table}[t!]
\centering
\begin{tabular}{c|c|c}
\textbf{Notation} & \textbf{Definition} & \textbf{Physical Meaning} \\ \hline
$\chi$ & $C/cL$ & unitless capacitance \\ \hline
$\chi_s$ & $\chi_g \chi_j/(\chi_g+\chi_j)$ & series capacitance \\ \hline
$\gamma$ & $\chi_g/(\chi_g+\chi_j)$ & capacitive ratio \\ \hline
$\chi(x,x_0)$ & $1+\chi_s\delta(x-x_0)$ & capacitance per length \\ \hline
$\mathcal{E}_{j,c}$ & $\sqrt{lc}LE_{j,c}/\hbar$ & unitless energy \\ \hline
$\omega_j$ & $\sqrt{8\mathcal{E}_c\mathcal{E}_j}$ & bare transmon frequency \\ \hline
$\epsilon$ & $\left(\mathcal{E}_c/\mathcal{E}_j\right)^{1/2}$ & nonlinearity measure \\ \hline
$\varepsilon$ & $\frac{\sqrt{2}}{6}\left(\mathcal{E}_c/\mathcal{E}_j\right)^{1/2}$ & small expansion parameter \\ \hline 
$\Phi_0$ & $h/(2e)$ & flux quantum \\ \hline
$\phi_{\text{zpf}}$ & $(2\mathcal{E}_c/\mathcal{E}_j)^{1/4}$ & zero-point fluctuation phase \\ \hline
$\hat{\Phi}(t)$ & $\int_{0}^{t}dt'\hat{V}(t)$ & flux\\ \hline
$\hat{\varphi}(t)$ & $2\pi\hat{\Phi}/\Phi_0$ & phase \\ \hline
$\hat{\phi}_j(t)$ & $\Tr_{ph}\{\hat{\rho}_{ph}(0)\hat{\varphi}_j(t)\}$ & reduced phase\\ \hline
$\hat{\mathcal{X}}(t)$ & $\hat{\varphi}(t)/\phi_{\text{zpf}}$ & unitless quadrature \\ \hline
$\hat{X}_j(t)$ & $\hat{\phi}_j(t)/\phi_{\text{zpf}}$ &  reduced unitless quadrature
\end{tabular}
\caption{Summary of definitions for some parameters and variables. Operators are denoted by a hat notation.}
\label{Tab:Def of Pars&Vars}
\end{table}
The Heisenberg equation of motion for the transmon reads
\begin{align}
\hat{\ddot{\varphi}}_j(t)+(1-\gamma)\omega_j^2\sin{[\hat{\varphi}_j(t)]}=\gamma \partial_{t}^2\hat{\varphi}(x_0,t),
\label{eqn:Transmon Dyn}
\end{align}
where $\gamma\equiv\chi_g/(\chi_g+\chi_j)$ is a capacitive  ratio, $\omega_j\equiv\sqrt{8\mathcal{E}_c\mathcal{E}_j}$ is the unitless bare transmon frequency and $x_0$ is the location of transmon. The phase field $\hat{\varphi}(x,t)$ of the resonator satisfies an inhomogeneous wave equation 
\begin{align}
\left[\partial_{x}^2-\chi(x,x_0)\partial_{t}^2\right]\hat{\varphi}(x,t)=\chi_s\omega_j^2 \sin{[\hat{\varphi}_j(t)]}\delta(x-x_0),
\label{eqn:Res Dyn}
\end{align}
where $\chi(x,x_0)=1+\chi_s\delta(x-x_0)$ is the unitless capacitance per unit length modified due to coupling to the transmon qubit, and $\chi_s\equiv\chi_g\chi_j/(\chi_g+\chi_j)$ is the unitless series capacitance of $C_j$ and $C_g$. The effect of a nonzero $\chi_s$ reflects the modification of the cavity modes due to the action of the transmon as a classical scatterer \cite{Malekakhlagh_Origin_2016}. We note that this modification is distinct from, and in addition to, the modification of the cavity modes due to the linear part of the transmon potential discussed in \cite{Nigg_Black-Box_2012}. Table~\ref{Tab:Def of Pars&Vars} lists the unitless variables and parameters used in the remainder of this paper.

The flux field in each waveguide obeys a homogeneous wave equation    
\begin{align}
\left(\partial_{x}^2-\partial_{t}^2\right)\hat{\varphi}_{R,L}(x,t)=0.
\label{eqn:Side Res Dyn}
\end{align}
The boundary conditions (BC) are derived from conservation of current at each end of the resonator as
\begin{subequations}
\begin{align}
\begin{split}
-\left.\partial_{x}\hat{\varphi}\right|_{x=1^-}&=-\left.\partial_{x}\hat{\varphi}_R\right|_{x=1^+}\\
&=\chi_R\partial_{t}^2\left[\hat{\varphi}(1^-,t)-\hat{\varphi}_R(1^+,t)\right],
\end{split}
\label{eqn:BC-Cons of current at 1}\\
\begin{split}
-\left.\partial_{x}\hat{\varphi}\right|_{x=0^+}&=-\left.\partial_{x}\hat{\varphi}_L\right|_{x=0^-}\\
&=\chi_L\partial_{t}^2\left[\hat{\varphi}_L(0^-,t)-\hat{\varphi}(0^+,t)\right].
\label{eqn:BC-Cons of current at 0}
\end{split}
\end{align}
\end{subequations}

Equations~(\ref{eqn:Transmon Dyn}-\ref{eqn:BC-Cons of current at 0}) completely describe the dynamics of a transmon qubit coupled to an open resonator. Note that according to Eq.~(\ref{eqn:Transmon Dyn}) the bare dynamics of the transmon is modified due to the force term $\gamma\partial_t^2\hat{\varphi}(x_0,t)$. Therefore, in order to find the effective dynamics for the transmon, we need to solve for $\hat{\varphi}(x,t)$ first and evaluate it at the point of connection $x=x_0$. This can be done using the {\it classical} electromagnetic GF by virtue of the homogeneous part of Eqs.~(\ref{eqn:Res Dyn},\ref{eqn:Side Res Dyn}) being linear in the quantum fields (see App.~\ref{SubApp:Def of G}). Substituting it into the LHS of Eq.~(\ref{eqn:Transmon Dyn}) and further simplifying leads to the effective dynamics for the transmon phase operator
\begin{align}
\begin{split}
&\hat{\ddot{\varphi}}_j(t)+(1-\gamma)\omega_j^2\sin{\left[\hat{\varphi}_j(t)\right]}=\\
+&\frac{d^2}{dt^2}\int_{0}^{t}dt'\mathcal{K}_0(t-t')\omega_j^2\sin{\left[\hat{\varphi}_j(t')\right]}\\
+&\int_{-\infty}^{+\infty}\frac{d\omega}{2\pi}\mathcal{D}_R(\omega)\hat{\tilde{\varphi}}_R^{inc}(1^+,\omega)e^{-i\omega t}\\
+&\int_{-\infty}^{+\infty}\frac{d\omega}{2\pi}\mathcal{D}_L(\omega)\hat{\tilde{\varphi}}_L^{inc}(0^-,\omega)e^{-i\omega t}\\
+&\int_{0^-}^{1^+}dx'\int_{-\infty}^{+\infty}\frac{d\omega}{2\pi}\mathcal{I}(x',\omega)\left[i\omega\hat{\varphi}(x',0)-\hat{\dot{\varphi}}(x',0)\right]e^{-i\omega t}.
\end{split}
\label{eqn:Eff Dyn before trace}
\end{align}
The electromagnetic GF is the basic object that appears in the various kernels constituting the above integro-differential equation:
\begin{subequations}
\begin{align}
&\mathcal{K}_n(\tau)\equiv\gamma\chi_s\int_{-\infty}^{+\infty} \frac{d\omega}{2\pi} \omega^n\tilde{G}(x_0,x_0,\omega)e^{-i\omega\tau},
\label{eqn:Def of K_n(tau)}\\
&\mathcal{D}_R(\omega)\equiv -2i\gamma\omega^3\tilde{G}(x_0,1^+,\omega),
\label{eqn:Def of D_R(om)}\\
&\mathcal{D}_L(\omega)\equiv -2i\gamma\omega^3\tilde{G}(x_0,0^-,\omega),
\label{eqn:Def of D_L(om)}\\
&\mathcal{I}(x',\omega)\equiv \gamma\omega^2\chi(x',x_0)\tilde{G}(x_0,x',\omega).
\label{eqn:Def of I(x',om)}
\end{align}
\end{subequations}
Equation~(\ref{eqn:Eff Dyn before trace}) fully describes the effective dynamics of the transmon phase operator. The various terms appearing in this equation have transparent physical interpretation. The first integral on the RHS of Eq.~(\ref{eqn:Eff Dyn before trace}) represents the retarded self-interaction of the qubit. It contains the GF in the form $\tilde{G}(x_0,x_0,\omega)$ and describes all processes in which the electromagnetic radiation is emitted from the transmon at $x=x_0$ and is scattered back again. We will see later on that this term is chiefly responsible for the spontaneous emission of the qubit. The boundary terms include only the incoming part of the waveguide phase fields. They describe the action of the electromagnetic fluctuations in the waveguides on the qubit, as described by the propagators from cavity interfaces to the qubit, $\tilde{G}(x_0,0^-,\omega)$ and $\tilde{G}(x_0,1^+,\omega)$. The phase fields $\hat{\varphi}_{L}(0^-,t)$ and $\hat{\varphi}_{R}(1^+,t)$ may contain a classical (coherent) part as well. Finally, the last integral adds up all contributions of a nonzero initial value for the electromagnetic field inside the resonator that propagates from the point $0<x'<1$ to the position of transmon $x_0$. 

The solution to the effective dynamics (\ref{eqn:Eff Dyn before trace}) requires knowledge of $\tilde{G}(x,x',\omega)$. To this end, we employ the spectral representation of the GF in terms of a set of constant flux (CF) modes \cite{Tureci_SelfConsistent_2006, Tureci_Strong_2008} 
\begin{align}
\tilde{G}(x,x',\omega)=\sum\limits_{n}\frac{\tilde{\Phi}_n(x,\omega)\bar{\tilde{\Phi}}_n^*(x',\omega)}{\omega^2-\omega_n^2(\omega)},
\label{eqn:Spec rep of G-Open}
\end{align}
where $\tilde{\Phi}_n(x,\omega)$ and $\bar{\tilde{\Phi}}_n(x,\omega)$ are the right and left eigenfunctions of the Helmholtz eigenvalue problem with outgoing BC and hence carry a constant flux when $x\to\pm \infty$. Note that in this representation, both the CF frequencies $\omega_n(\omega)$ and the CF modes $\tilde{\Phi}_n(x,\omega)$ parametrically depend on the source frequency $\omega$. The expressions for $\omega_n(\omega)$ and $\tilde{\Phi}_n(x,\omega)$ are given in App.~\ref{SubApp:Spec Rep of G-open}.  

\begin{figure}
\subfloat[\label{subfig:NHEigFreqsXrXl1Em5}]{%
\includegraphics[scale=0.355]{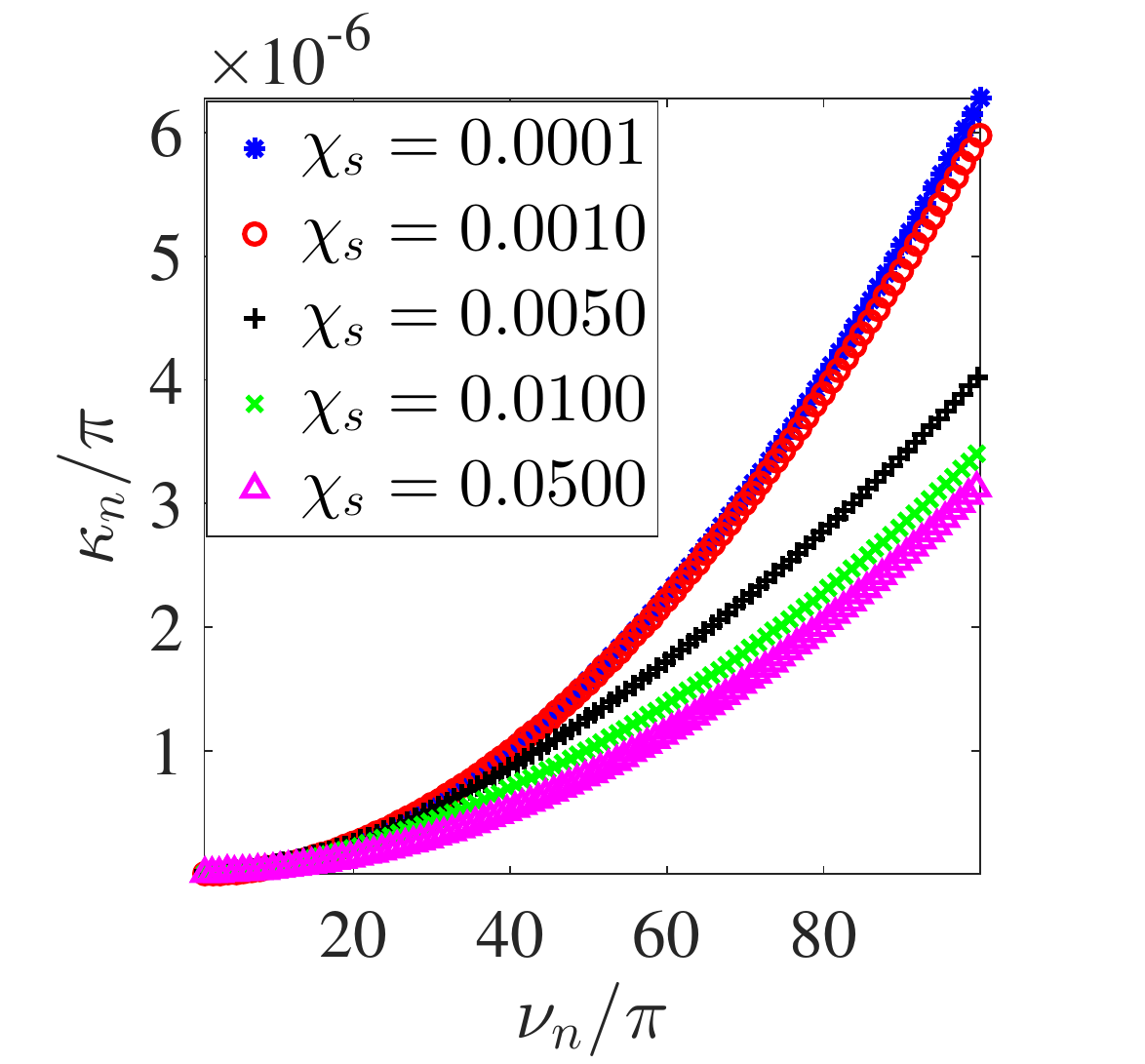}%
}\hfill 
\subfloat[\label{subfig:NHEigFreqsXrXl1Em3}]{%
\includegraphics[scale=0.355]{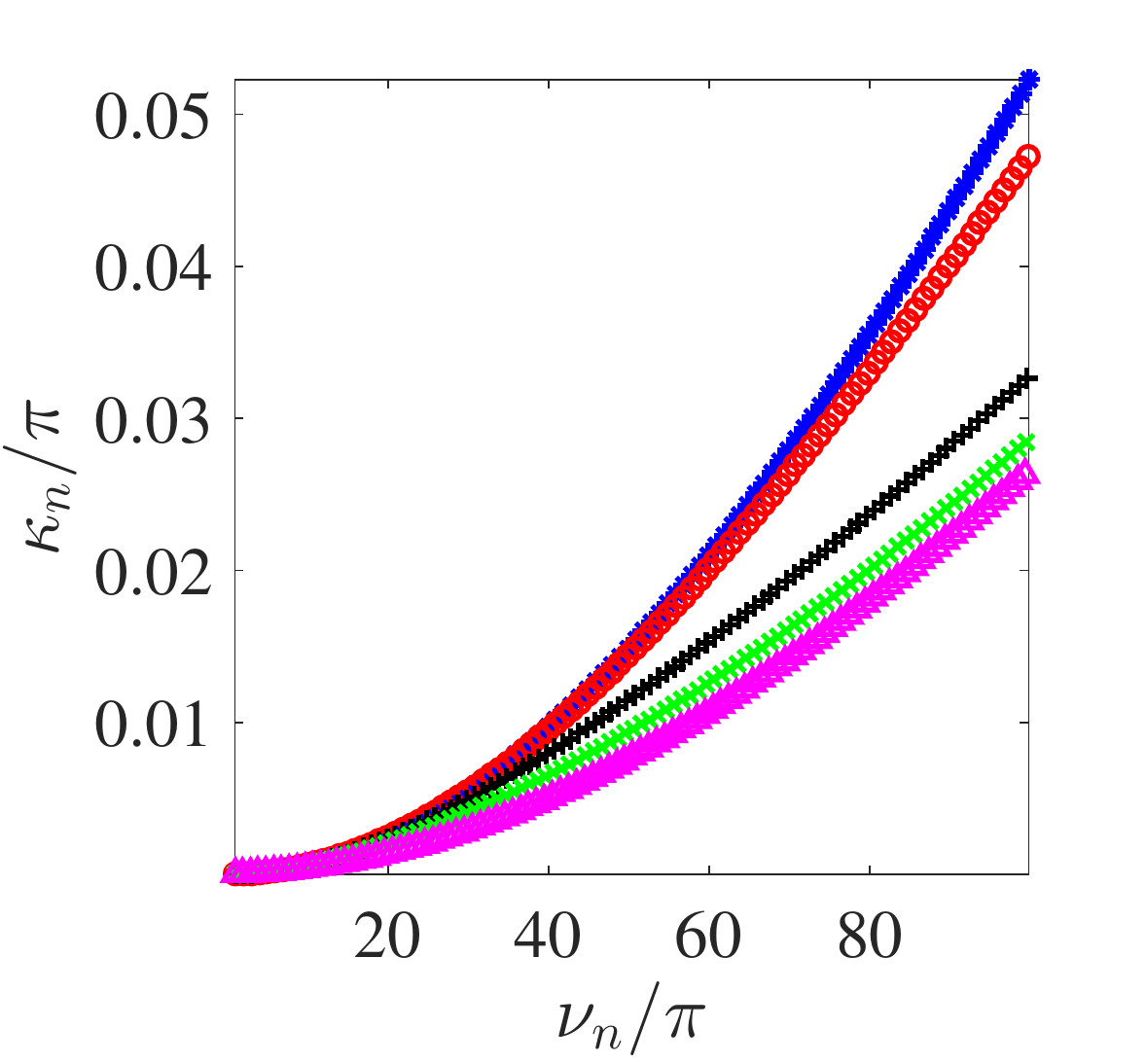}%
}
\hfill
\subfloat[\label{subfig:NHEigFreqsXrXl1Em2}]{%
\includegraphics[scale=0.355]{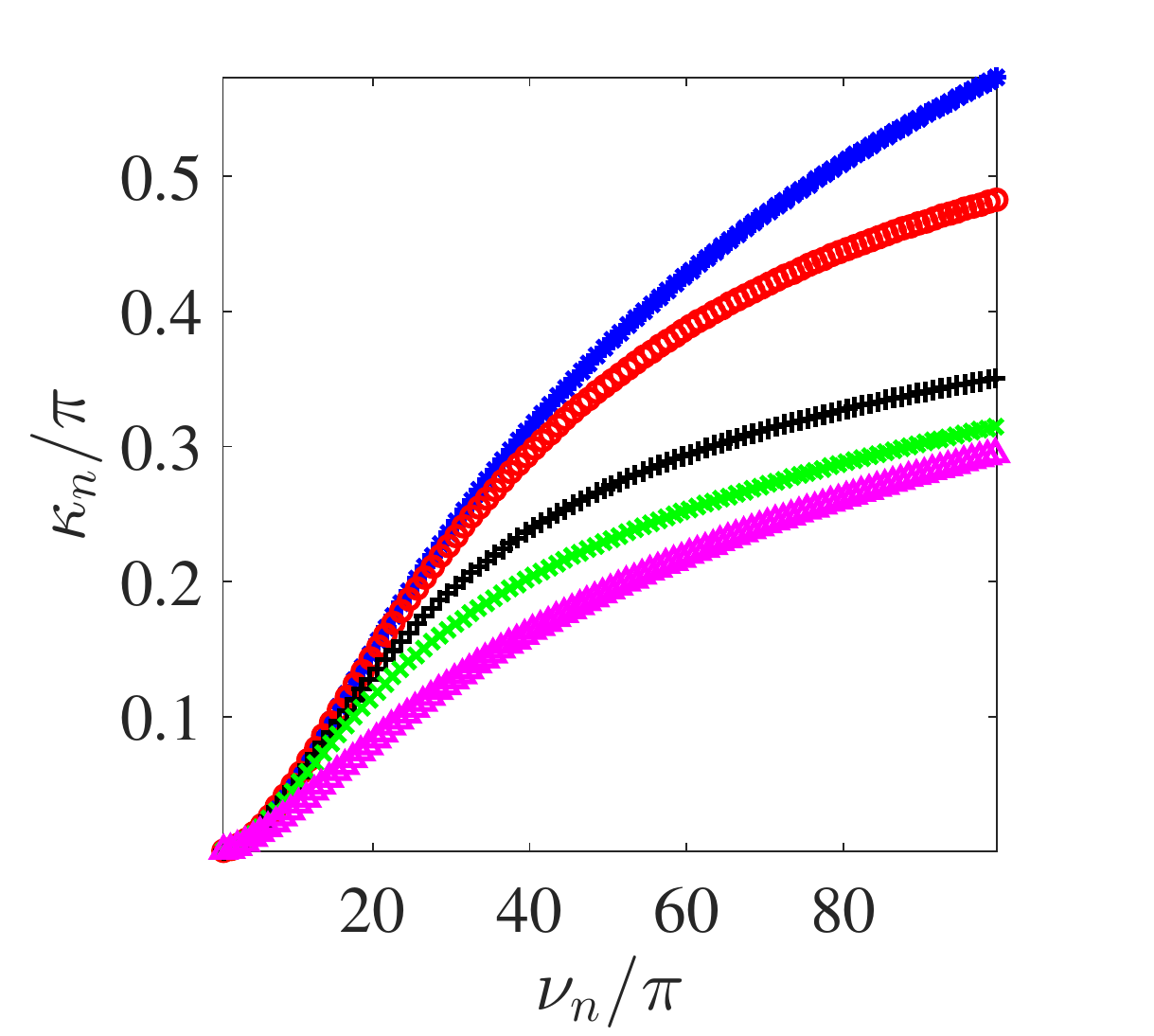}%
}
\hfill
\subfloat[\label{subfig:NHEigFreqsXrXl1Em1}]{%
\includegraphics[scale=0.355]{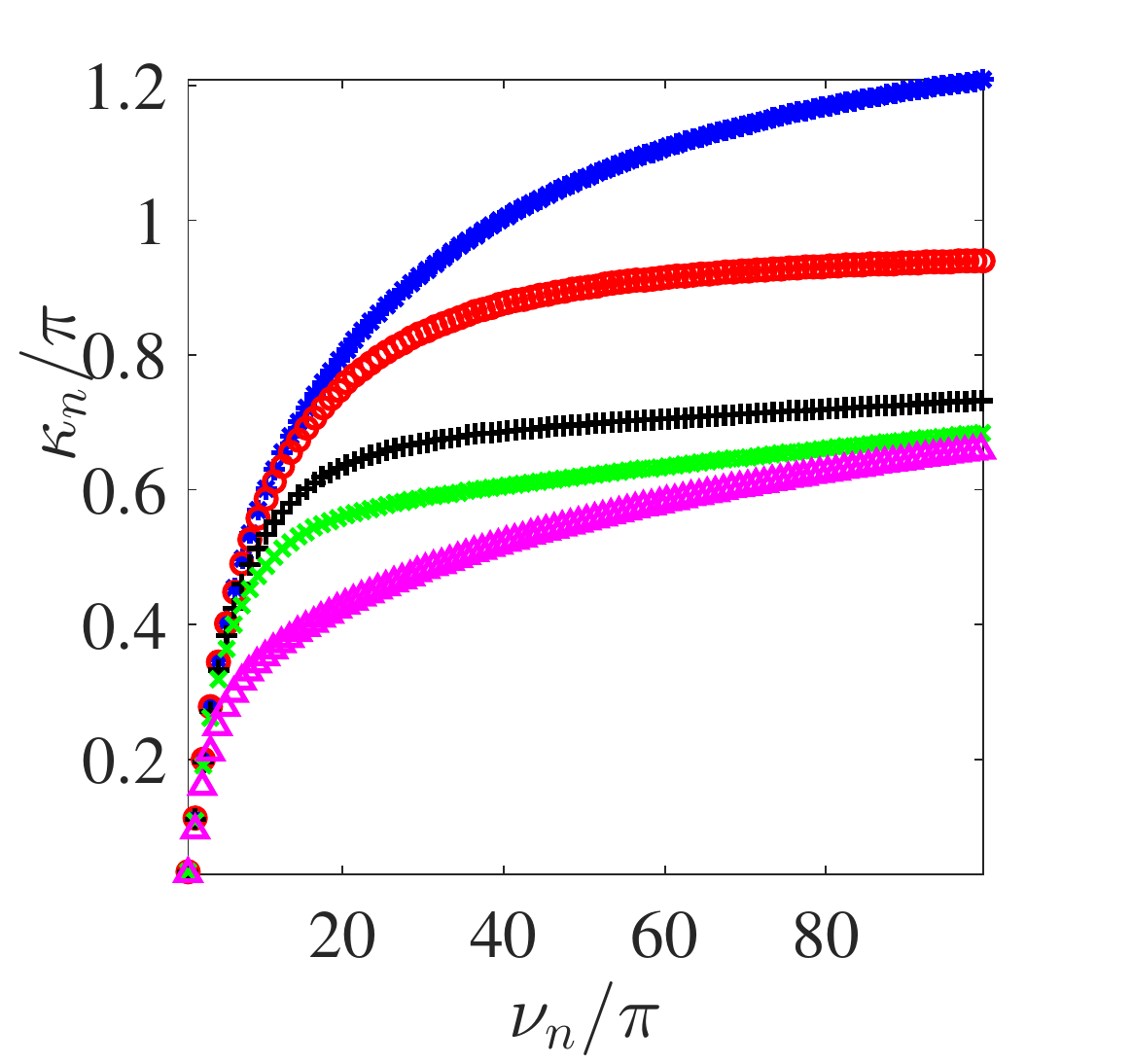}%
}
\caption{(Color online) Decay rate $\kappa_n$ versus oscillation frequency $\nu_n$ for the first 100 non-Hermitian modes for $x_0=0$ and different values of $\chi_s$. a) $\chi_R=\chi_L=10^{-5}$, b) $\chi_R=\chi_L=10^{-3}$, c) $\chi_R=\chi_L=10^{-2}$ and d) $\chi_R=\chi_L=10^{-1}$.} 
\label{Fig:NHEigFreqs}
\end{figure}

The poles of the GF are the solutions to $\omega=\omega_n(\omega)$ that satisfy the transcendental equation
\begin{align}
\begin{split}
&\left[e^{2i\omega_n}-(1-2i\chi_L\omega_n)(1-2i\chi_R\omega_n)\right]\\
&+\frac{i}{2}\chi_s\omega_n[e^{2i\omega_n x_0}+(1-2i\chi_L\omega_n)]\\
&\times[e^{2i\omega_n (1-x_0)}+(1-2i\chi_R\omega_n)]=0.
\end{split}
\label{eqn:Generic NHEigfreq}
\end{align}
The solutions to Eq.~(\ref{eqn:Generic NHEigfreq}) all reside in the lower half of $\omega$-plane resulting in a finite lifetime for each mode that is characterized by the imaginary part of $\omega_n\equiv \nu_n-i\kappa_n$.  In Fig.~\ref{Fig:NHEigFreqs} we plotted the decay rate $\kappa_n$ versus the oscillation frequency $\nu_n$ of the first 100 modes for $x_0=0$ and different values of $\chi_R=\chi_L$ and $\chi_s$. There is a transition from a super-linear \cite{Houck_Controlling_2008} dependence on mode number for smaller opening to a sub-linear dependence for larger openings. Furthermore, increasing $\chi_s$ always decreases the decay rate $\kappa_n$. Intuitively, $\chi_s$ is the strength of a $\delta$-function step in the susceptibility at the position of the transmon. An increase in the average refractive index inside the resonator generally tends to redshift the cavity resonances, while decreasing their decay rate.

In summary, we have derived an effective equation of motion, Eq.~(\ref{eqn:Eff Dyn before trace}), for the transmon qubit flux operator $\hat{\varphi}_j$, in which the resonator degrees of freedom enter via the   electromagnetic GF $\tilde{G}(x,x',\omega)$ given in Eq.~(\ref{eqn:Spec rep of G-Open}). 
\section{Spontaneous emission into a leaky resonator}
In this section, we revisit the problem of spontaneous emission \cite{Purcell_Resonance_1946, Kleppner_Inhibited_1981, Goy_Observation_1983, Hulet_Inhibited_1985, Jhe_Suppression_1987, Dung_Spontaneous_2000, Houck_Controlling_2008, Krimer_Route_2014}, where the system starts from the initial density matrix
\begin{align}
\hat{\rho}(0)=\hat{\rho}_j(0)\otimes\ket{0}_{ph}\bra{0}_{ph},
\label{eqn:SE-IC}
\end{align} 
such that the initial excitation exists in the transmon sector of Hilbert space with zero photons in the resonator and waveguides. $\hat{\rho}_j(0)$ is a general density matrix in the qubit subspace. For our numerical simulation of the spontaneous emission dynamics in terms of quadratures, we will consider $\hat{\rho}_j(0)=\ket{\Psi_j(0)}\bra{\Psi_j(0)}$ with $\ket{\Psi_j(0)}=(\ket{0}_j+\ket{1}_j)/\sqrt{2}$. The spontaneous emission was conventionally studied through the Markov approximation of the memory term which results only in a modification of the qubit-like pole. This is the Purcell modified spontaneous decay where, depending on the density of the states of the environment, the emission rate can be suppressed or enhanced \cite{Purcell_Resonance_1946, Kleppner_Inhibited_1981, Goy_Observation_1983, Hulet_Inhibited_1985, Jhe_Suppression_1987}. We extract the spontaneous decay as the real part of transmon-like pole in a full {\it multimode} calculation that is accurate for any qubit-resonator coupling strength.

A product initial density matrix like Eq.~(\ref{eqn:SE-IC}) allows us to reduce the generic dynamics significantly, since the expectation value of any operator $\hat{\mathcal{O}}(t)$ can be expressed as
\begin{align}
\Tr_{j}\Tr_{ph}\left\{\hat{\rho}_j(0)\otimes\hat{\rho}_{ph}(0)\hat{\mathcal{O}}(t)\right\}=\Tr_j\left\{\hat{\rho}_j(0)\hat{O}(t)\right\}
\end{align}
where $\hat{O}\equiv\Tr_{ph}\{\hat{\mathcal{O}}\}$ is the reduced operator in the Hilbert space of the transmon. Therefore, we define a reduced phase operator
\begin{align}
\hat{\phi}_j(t)\equiv \Tr_{ph}\{\hat{\rho}_{ph}(0)\hat{\varphi}_j(t)\}.
\label{eqn:Def of phi_j(t)}
\end{align}
In the absence of an external drive, the generic effective dynamics in Eq.~(\ref{eqn:Eff Dyn before trace}) reduces to
\begin{align}
\begin{split}
&\hat{\ddot{\phi}}_j(t)+\omega_j^2\left[1-\gamma+i\mathcal{K}_1(0)\right]\Tr_{ph}\left\{\hat{\rho}_{ph}(0)\sin{\left[\hat{\varphi}_j(t)\right]}\right\}\\
&=-\int_0^{t}dt'\mathcal{K}_2(t-t')\omega_j^2\Tr_{ph}\left\{\hat{\rho}_{ph}(0)\sin{\left[\hat{\varphi}_j(t')\right]}\right\}.
\label{eqn:NL SE Problem}
\end{split}
\end{align}
The derivation of Eq.~(\ref{eqn:NL SE Problem}) can be found in Apps.~\ref{SubApp:SE Eff Dyn} and \ref{SubApp:Spec Rep of K}.

Note that, due to the sine nonlinearity, Eq.~(\ref{eqn:NL SE Problem}) is not closed in terms of $\hat{\phi}_j(t)$. However, in the transmon regime \cite{Koch_Charge_2007}, where $\mathcal{E}_j \gg \mathcal{E}_c$, the nonlinearity in the spectrum of transmon is weak. This becomes apparent when we work with the unitless quadratures 
\begin{align}
&\hat{X}_j(t)\equiv\frac{\hat{\phi}_j(t)}{\phi_{\text{zpf}}},
\quad \hat{\mathcal{X}}_j(t)\equiv \frac{\hat{\varphi}_j(t)}{\phi_{\text{zpf}}},
\label{eqn:Def of mathcalX_j(t)}
\end{align}  
where $\phi_{\text{zpf}}\equiv (2\mathcal{E}_c/\mathcal{E}_j)^{1/4}$ is the zero-point fluctuation (zpf) phase amplitude. Then, we can expand the nonlinearity in both sides of Eq.~(\ref{eqn:NL SE Problem}) as
\begin{align}
\begin{split}
\frac{\sin{\left[\hat{\varphi}_j(t)\right]}}{\phi_{\text{zpf}}}&=\frac{\hat{\varphi}_j(t)}{\phi_{\text{zpf}}}
-\frac{\hat{\varphi}_j^3(t)}{3!\phi_{\text{zpf}}}+\mathcal{O}\left[\frac{\hat{\varphi}_j^5(t)}{\phi_{\text{zpf}}}\right]\\
&=\hat{\mathcal{X}}_j(t)-\frac{\sqrt{2}\epsilon}{6}\hat{\mathcal{X}}_j^3(t)+\mathcal{O}\left(\epsilon^2\right),
\end{split}
\label{eqn:Expansion of Sine}
\end{align}
where $\epsilon\equiv(\mathcal{E}_c/\mathcal{E}_j)^{1/2}$ appears as a measure for the strength of the nonlinearity. In experiment, the Josephson energy $\mathcal{E}_j$ can be tuned through the FBL while the charging energy $\mathcal{E}_c$ is fixed. Therefore, a higher transmon frequency $\omega_j=\sqrt{8\mathcal{E}_c\mathcal{E}_j}$ is generally associated with a smaller $\epsilon$ and hence weaker nonlinearity. 

The remainder of this section is organized as follows. In Sec.~\ref{Sec:Lin SE Theory} we study the linear theory. In Sec.~\ref{Sec:PertCor} we develop a perturbation expansion up to leading order in $\epsilon$. In Sec.~\ref{Sec:NumSimul}, we compare our analytical results with numerical simulation. Finally, in Sec.~\ref{Sec:SysOutput} we discuss the output response of the cQED system that can be probed in experiment. 
\subsection{Linear theory}
\label{Sec:Lin SE Theory}
In this subsection, we solve the linear effective dynamics and discuss hybridization of the transmon and the resonator resonances. We emphasize the importance of off-resonant modes as the coupling $\chi_g$ is increased. We next investigate the spontaneous decay rate as a function of transmon frequency $\omega_j$ and coupling $\chi_g$ and find an asymmetric dependence on $\omega_j$ in agreement with a previous experiment \cite{Houck_Controlling_2008}.

Neglecting the cubic term in Eq.~(\ref{eqn:Expansion of Sine}), the partial trace with respect to the resonator modes can be taken directly and we obtain the effective dynamics 
\begin{align}
\begin{split}
\hat{\ddot{X}}_j(t)&+\omega_j^2\left[1-\gamma+i\mathcal{K}_1(0)\right]\hat{X}_j(t)\\
&=-\int_0^{t}dt'\mathcal{K}_2(t-t')\omega_j^2\hat{X}_j(t').
\label{eqn:Lin SE Problem}
\end{split}
\end{align}
Then, using Laplace transform we can solve Eq.~(\ref{eqn:Lin SE Problem}) as
\begin{align}
\hat{\tilde{X}}_j(s)=\frac{s\hat{X}_j(0)+\omega_j\hat{Y}_j(0)}{D_j(s)},
\label{eqn:Sol of X_j(s)}
\end{align}
with $D_j(s)$ defined as
\begin{align}
D_j(s)\equiv s^2+\omega_j^2\left[1-\gamma+i\mathcal{K}_1(0)+\tilde{\mathcal{K}}_2(s)\right].
\label{eqn:Def of D(s)}
\end{align}
Equations~(\ref{eqn:Sol of X_j(s)}) and (\ref{eqn:Def of D(s)}) contain the solution for the reduced quadrature operator of the transmon qubit in the Laplace domain. 

In order to find the time domain solution, it is necessary to study the poles of Eq.~(\ref{eqn:Sol of X_j(s)}) and consequently the roots of $D_j(s)$. The characteristic function $D_j(s)$ can be expressed as (see App.~\ref{App:Char func D(s)}) 
\begin{align}
\begin{split}
&D_j(s)=s^2+\omega_j^2+\\
&\omega_j^2\left\{-\gamma+\sum\limits_{n}M_n\frac{s\{\cos{[2\delta_{n}(x_0)]}s+\sin{[2\delta_{n}(x_0)]}\nu_n\}}{(s+\kappa_n)^2+\nu_n^2}\right\},
\label{eqn:simplified D(s)}
\end{split}
\end{align}
where $\delta_n(x)$ is the phase of the non-Hermitian eigenfunction such that $\tilde{\Phi}_n(x)=|\tilde{\Phi}_n(x)|e^{i\delta_n(x)}$. We identify the term
\begin{align}
M_n\equiv\gamma\chi_s|\tilde{\Phi}_n(x_0)|^2
\label{eqn:Def of Mn}
\end{align}
as the measure of hybridization with individual resonator modes. The form of $M_n$ in Eq.~(\ref{eqn:Def of Mn}) illustrates that the hybridization between the transmon and the resonator is bounded. This strength of hybridization is parameterized by $\gamma\chi_s$ rather than $\chi_g$. This implies that as $\chi_g$, the coupling capacitance, is increased, the qubit-resonator hybridization is limited by the internal capacitance of the qubit, $\chi_j$:
\begin{align}
\lim\limits_{\frac{\chi_g}{\chi_j}\to\infty}\gamma\chi_s=\lim\limits_{\frac{\chi_g}{\chi_j}\to\infty}\left(\frac{\chi_g}{\chi_g+\chi_j}\right)^2\chi_j=\chi_j.
\label{eqn:LinTheory-Asymptote}
\end{align}
For this reason, our numerical results below feature a saturation in hybridization as $\chi_g$ is increased.  
\begin{figure}
\subfloat[\label{subfig:Poles5ModeXrXl1Em2Weak}]{%
\includegraphics[scale=0.36]{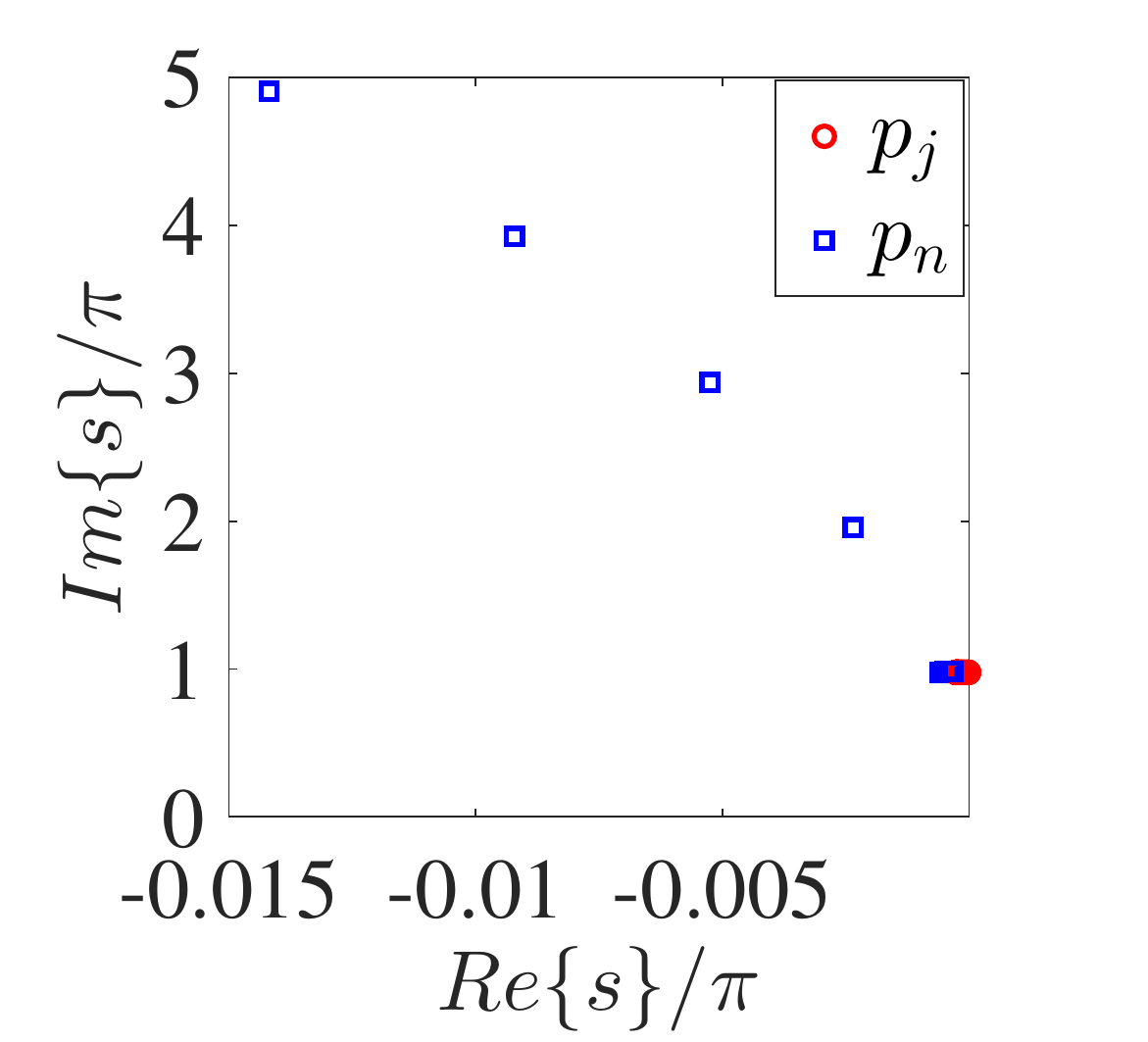}%
}\hfill 
\subfloat[\label{subfig:Poles5ModeXrXl1Em2WeakZoomed}]{%
\includegraphics[scale=0.36]{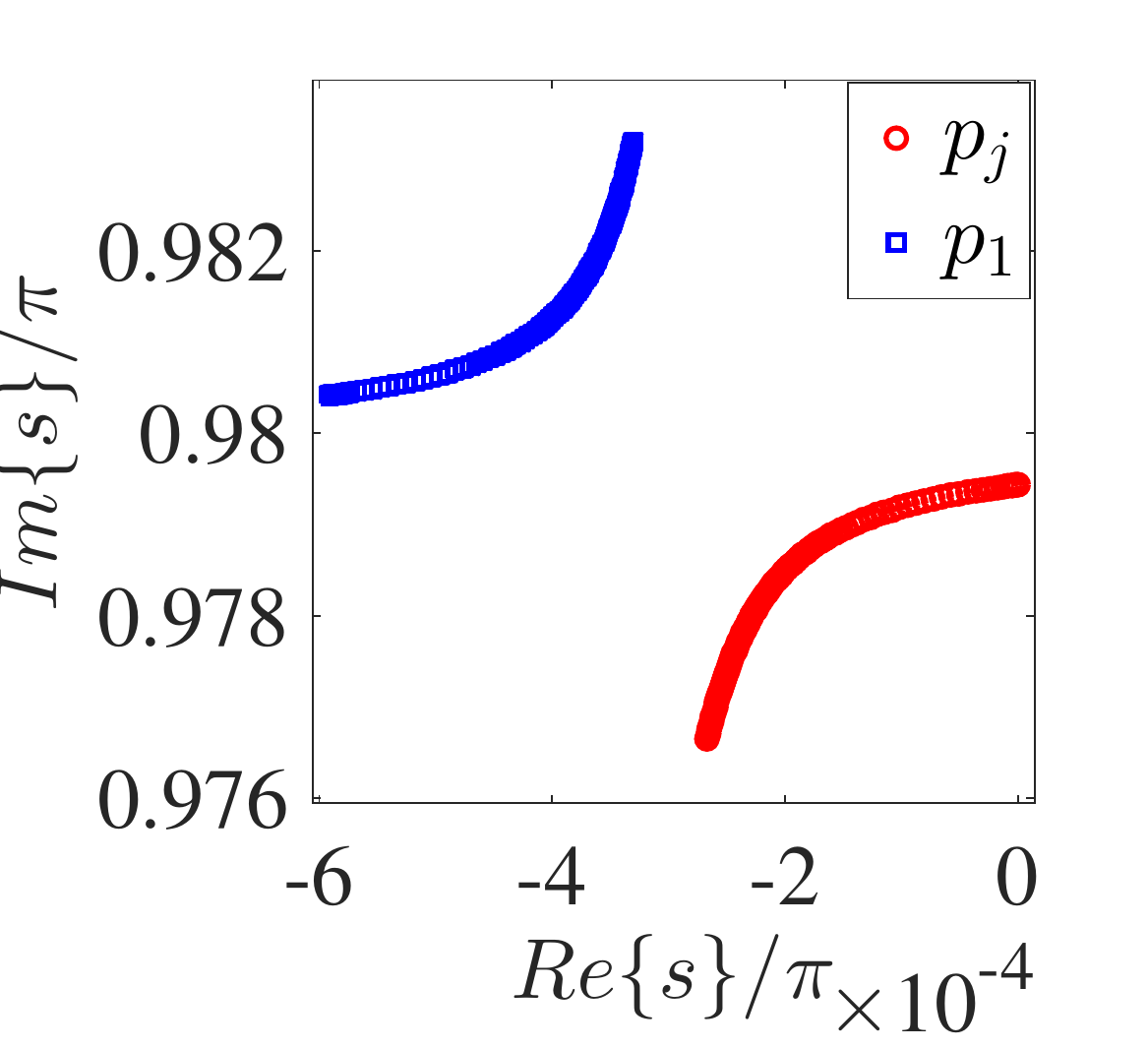}%
} 
\caption{(Color online) a) The first five hybridized poles of the resonator-qubit system, for the case where the transmon is slightly detuned below the fundamental mode, i.e. $\omega_j=\nu_1^-$. The other parameters are set as $\chi_R=\chi_L=0.01$, $\chi_j=0.05$ and $\chi_g\in [0,10^{-3}]$ with increments $\Delta\chi_g=10^{-5}$. b) Zoom-in plot of the hybridization of the most resonant modes. Hybridization of $p_1$ and $p_j$ is much stronger than that of the off-resonant poles $p_n$, $n > 1$.}
\label{Fig:Hybrid Poles Weak Coup}
\end{figure}

The roots of $D_j(s)$ are the hybridized poles of the entire system. If there is no coupling, i.e. $\chi_g=0$, then $D_j(s)=s^2+\omega_j^2=(s+i\omega_j)(s-i\omega_j)$ is the characteristic polynomial that gives the bare transmon resonance. However, for a nonzero $\chi_g$, $D_j(s)$ becomes a meromorphic function whose zeros are the hybridized resonances of the entire system, and whose poles are the bare cavity resonances. Therefore, $D_j(s)$ can be expressed as
\begin{align}
D_j(s)=(s-p_j)(s-p_j^*)\prod\limits_{m}\frac{(s-p_m)(s-p_m^*)}{(s-z_m)(s-z_m^*)}.
\label{eqn:Formal Rep of D(s)}
\end{align}
In Eq.~(\ref{eqn:Formal Rep of D(s)}), $p_j\equiv -\alpha_j-i\beta_j$ and $p_n\equiv -\alpha_n-i\beta_n$ are the zeros of $D_j(s)$ that represent the transmon-like and the $n$th resonator-like poles, accordingly. Furthermore, $z_n \equiv -i\omega_n=-\kappa_n-i\nu_n$ stands for the $n$th bare non-Hermitian resonator resonance. The notation chosen here ($p$ for poles and $z$ for zeroes) reflects the meromorphic structure of $1/D_j(s)$ which enters the solution Eq.~(\ref{eqn:Sol of X_j(s)}).

An important question concerns the convergence of $D_j(s)$ as a function of the number of the resonator modes included in the calculation. The form of $D_j(s)$ given in Eq.~(\ref{eqn:Formal Rep of D(s)}) is suitable for this discussion. Consider the factor corresponding to the $m$th resonator mode in $1/D_j(s)$. We reexpress it as
\begin{align}
\begin{split}
\frac{(s-z_m)(s-z_m^*)}{(s-p_m)(s-p_m^*)}&=\left(1-\frac{z_m-p_m}{s-p_m}\right)\left(1-\frac{z_m^*-p_m^*}{s-p_m^*}\right)\\
&=1+\mathcal{O}\left(\left|\frac{z_m-p_m}{s-p_m}\right|\right).
\end{split}
\end{align}   
The consequence of a small shift $|p_m-z_m|$ as compared to the strongly hybridized resonant mode $|p_1-z_1|$ is that it can be neglected in the expansion for $1/D_j(s)$. The relative size of these contributions is controlled by the coupling $\chi_g$. As rule  of thumb, the less hybridized a resonator pole is, the less it contributes to qubit dynamics. Ultimately, the truncation in this work is established by imposing the convergence of the numerics.

A numerical solution for the roots of Eq.~(\ref{eqn:simplified D(s)}) at weak coupling $\chi_g$ reveals that the mode resonant with the transmon is significantly shifted, with comparatively small shifts $|p_m-z_m|$ in the other resonator modes (See Fig.~\ref{Fig:Hybrid Poles Weak Coup}). At weak coupling, the hybridization of $p_j$ and $p_1$ is captured by a single resonator mode.
\begin{figure}[t!]
\subfloat[\label{subfig:Poles1ModeXrXl1Em2Zoomed}]{%
\includegraphics[scale=0.35]{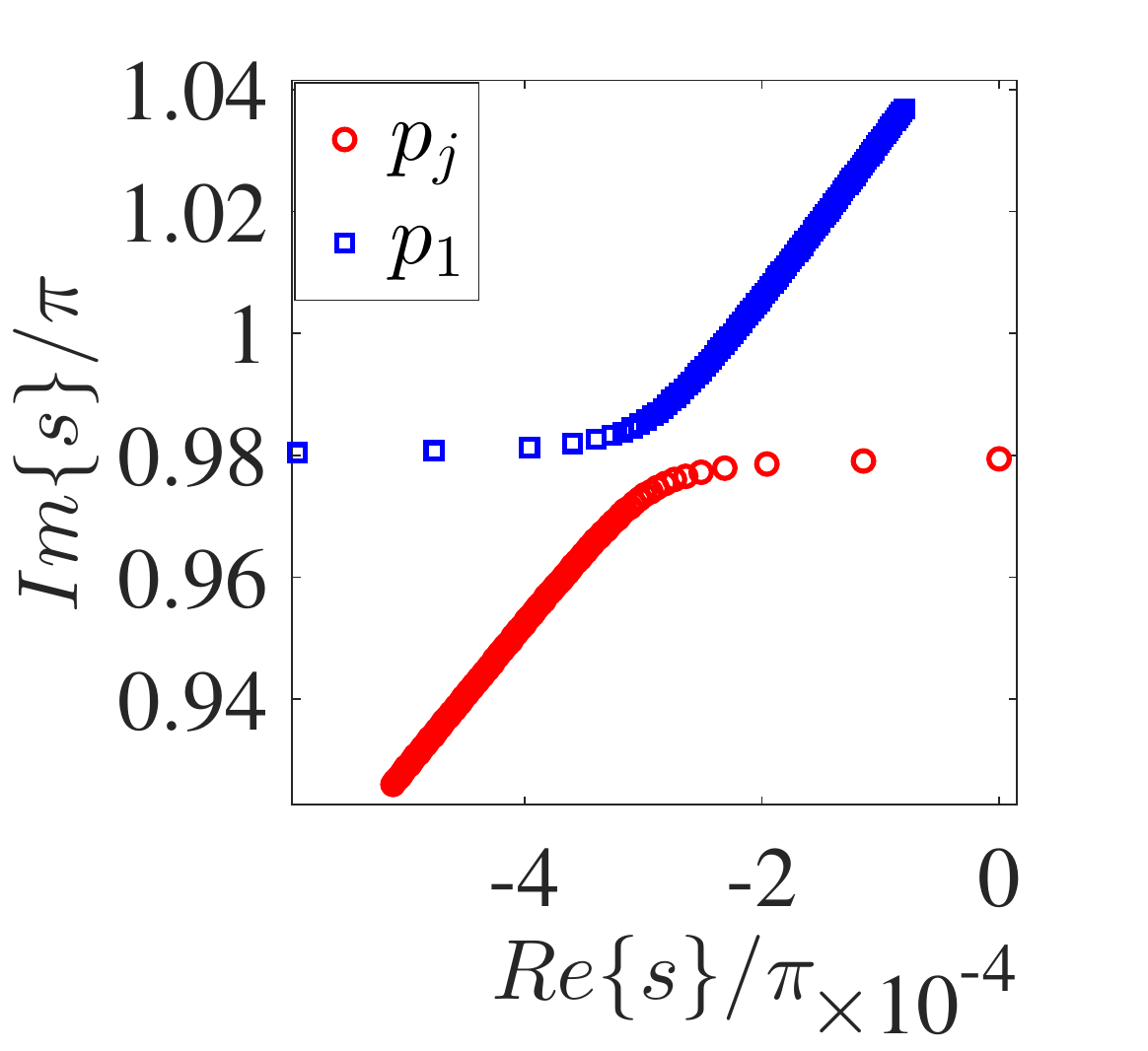}%
}\hfill 
\subfloat[\label{subfig:Poles5ModeXrXl1Em2Zoomed}]{%
\includegraphics[scale=0.35]{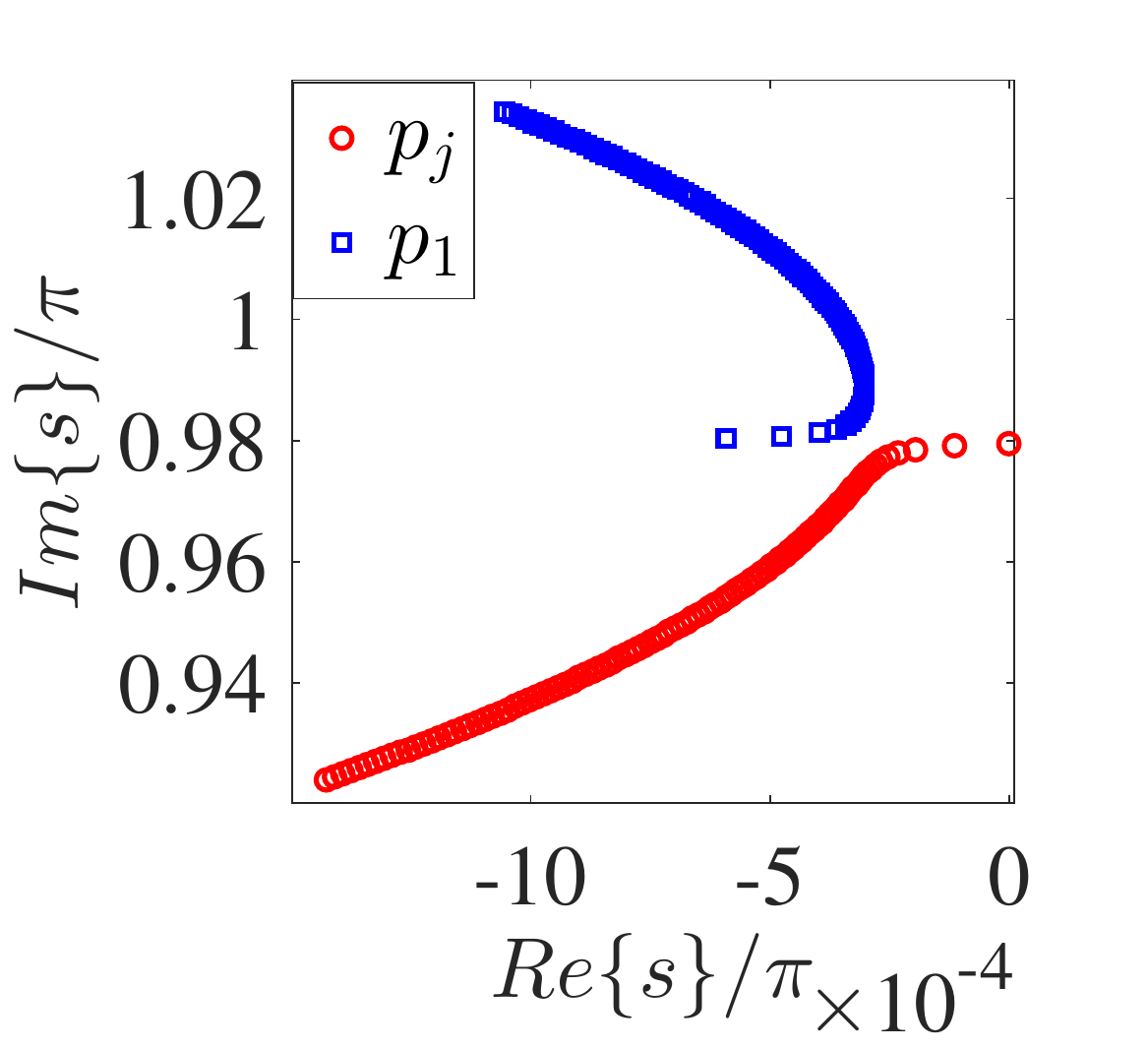}%
}\hfill
\subfloat[\label{subfig:Poles10ModeXrXl1Em2Zoomed}]{%
\includegraphics[scale=0.35]{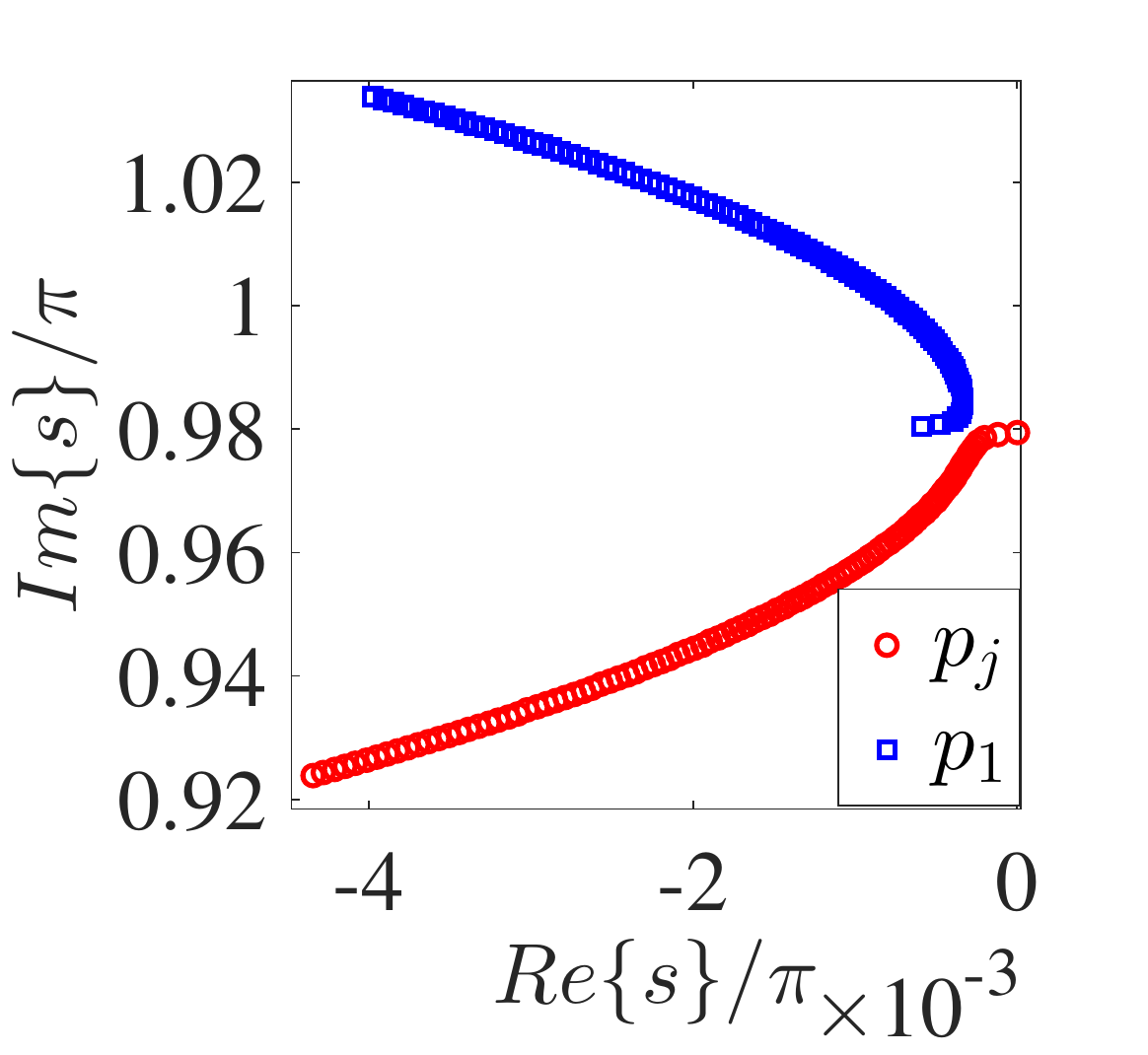}%
}\hfill 
\subfloat[\label{subfig:Poles20ModeXrXl1Em2Zoomed}]{%
\includegraphics[scale=0.35]{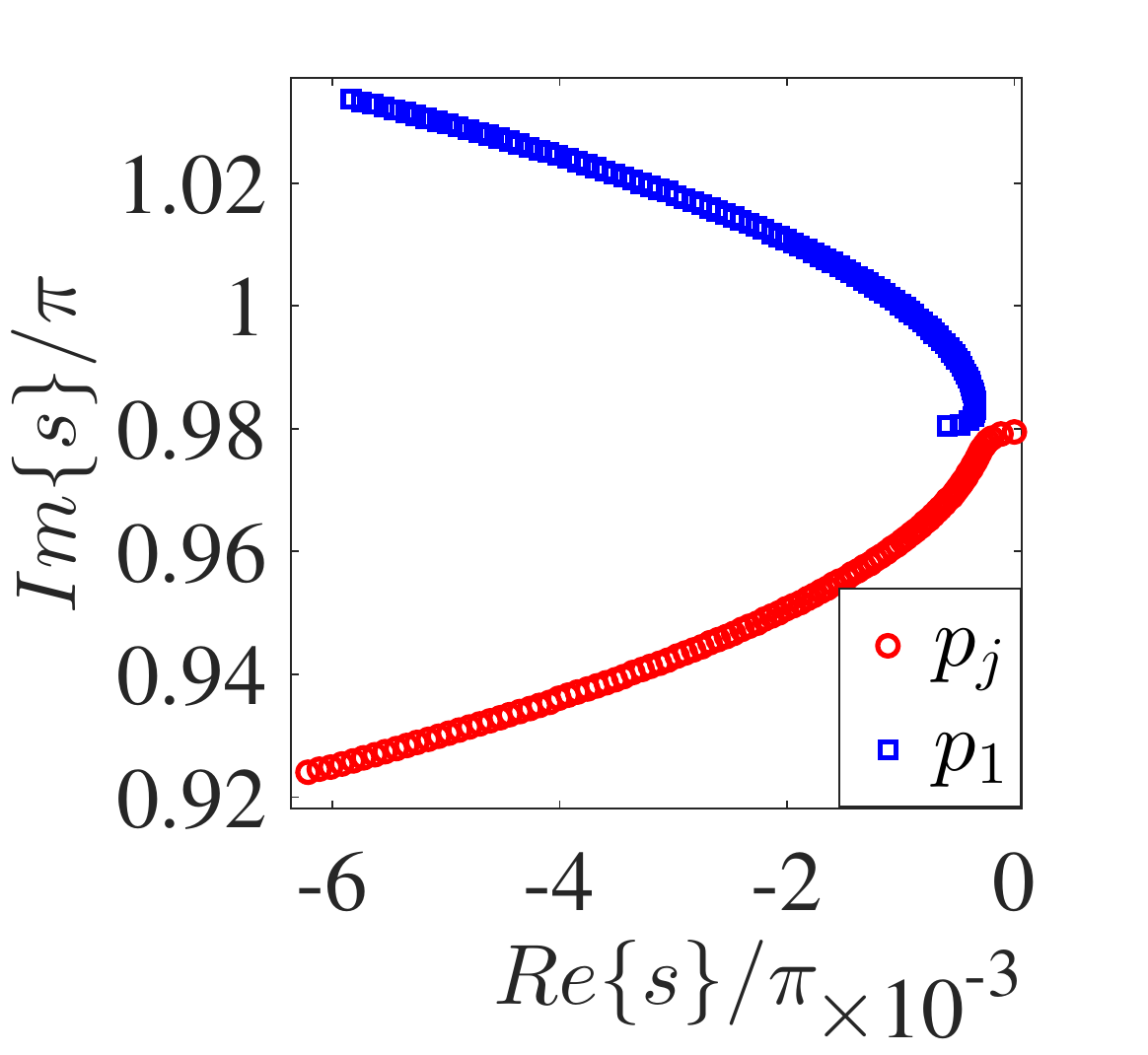}%
}
\caption{(Color online) Convergence of $p_j$ and $p_1$ for the same parameters as Fig.~\ref{Fig:Hybrid Poles Weak Coup}, but for $\chi_g\in[0,0.02]$ and keeping a) 1, b) 5, c) 10 and d) 20 resonator modes in $D_j(s)$.} 
\label{Fig:Hybrid Poles Strong Coup}
\end{figure}
Next, we plot in Fig.~\ref{Fig:Hybrid Poles Strong Coup} the effect of truncation on the response of the multimode system in a band around $s= p_j$. As the coupling $\chi_g$ is increased beyond the avoided crossing, which is also captured by the single mode truncation, the effect of off-resonant modes on $p_j$ and $p_1$ becomes significant. It is important to note that the hybridization occurs in the complex $s$-plane. On the frequency axis $\Im\{s\}$ an increase in $\chi_g$ is associated with a splitting of transmon-like and resonator-like poles. Along the decay rate axis $\Re\{s\}$ we notice that the qubit decay rate is controlled by the resonant mode at weak coupling, with noticeable enhancement of off-resonant mode contribution at strong coupling. If the truncation is not done properly in the strong coupling regime, it may result in spurious unstable roots of $D_j(s)$, i.e. $\Re\{s\}>0$, as seen in Fig.~\ref{subfig:Poles1ModeXrXl1Em2Zoomed}.

The modification of the decay rate of the transmon-like pole, henceforth identified as $\alpha_j\equiv-\Re\{p_j\}$, has an important physical significance. It describes the Purcell modification of the qubit decay (if sources for qubit decay other than the direct coupling to electromagnetic modes can be neglected). The present scheme is able to capture the  full multimode modification, that is out of the reach of conventional single-mode theories of spontaneous emission \cite{Purcell_Resonance_1946, Kleppner_Inhibited_1981, Goy_Observation_1983, Hulet_Inhibited_1985, Jhe_Suppression_1987}.

\begin{figure}
\subfloat[\label{subfig:SpEmRateXrXl1Em2Xg1Em3}]{%
\includegraphics[scale=0.38]{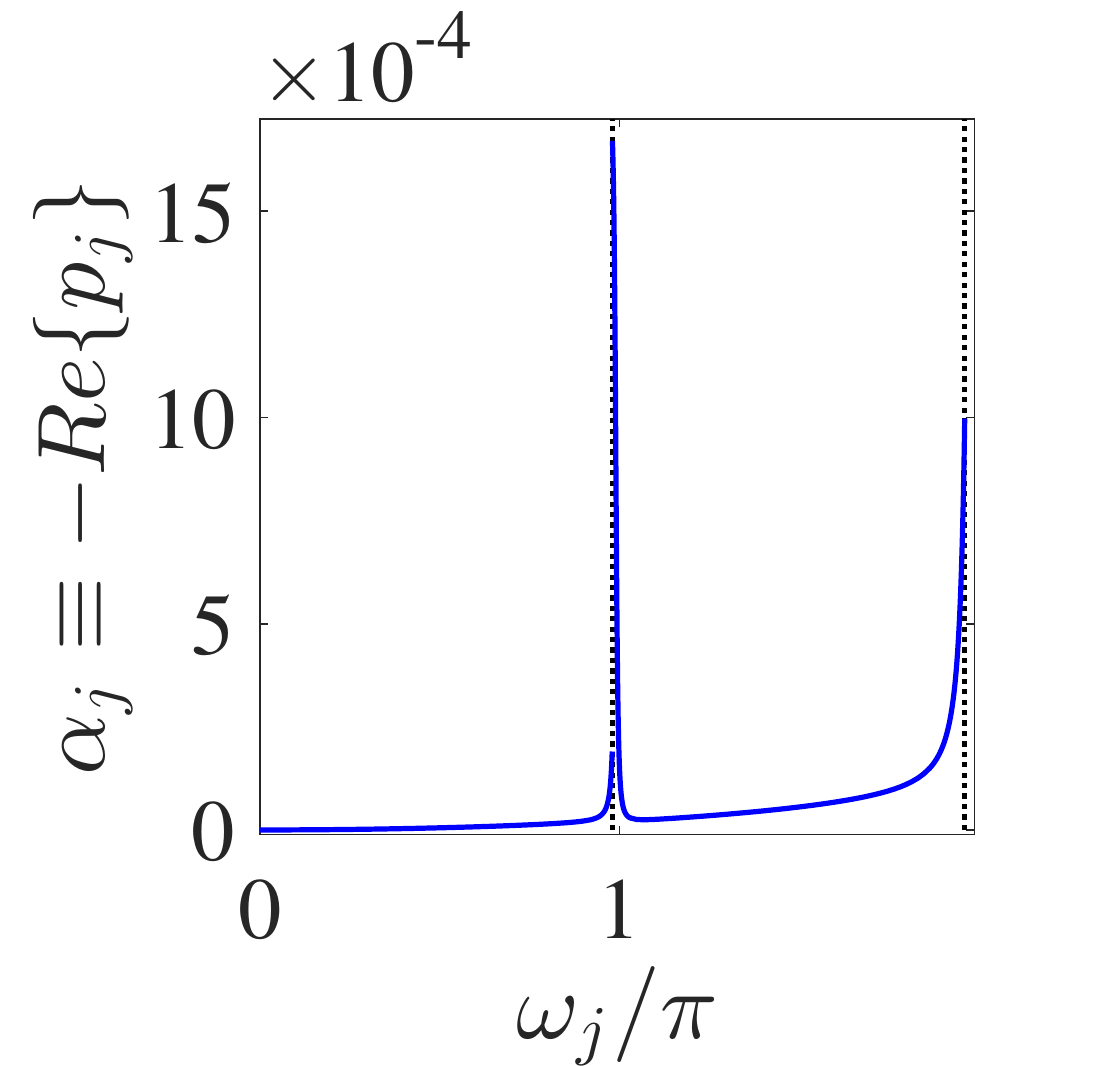}%
}\hfill 
\subfloat[\label{subfig:SpEmRateXrXl1Em2Xg5Em3}]{%
\includegraphics[scale=0.38]{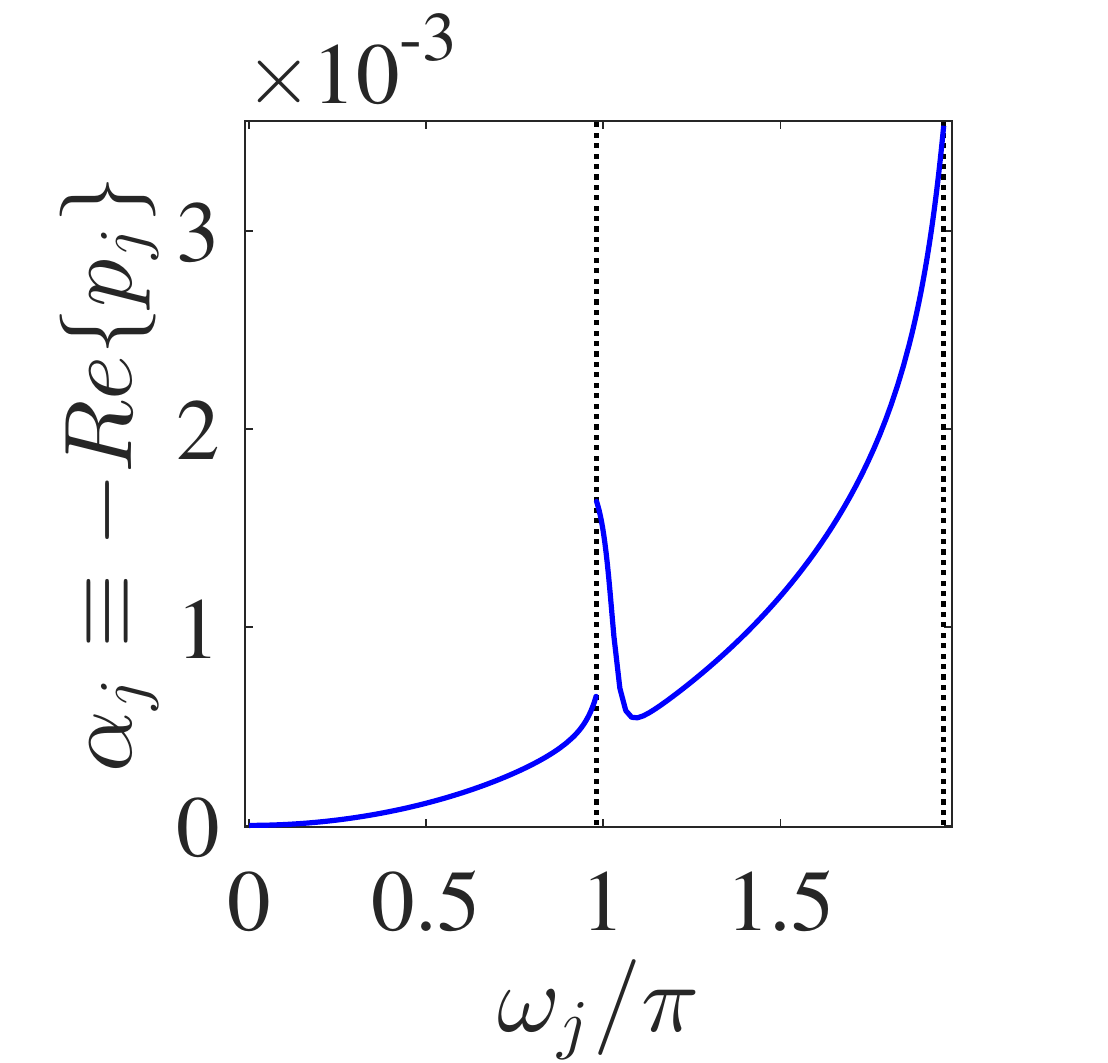}%
}
\caption{(Color online) Spontaneous emission rate defined as $\alpha_j\equiv-\Re\{p_j\}$ as a function of transmon frequency $\omega_j$ for $\chi_R=\chi_L=10^{-2}$, $\chi_j=0.05$, a) $\chi_g=10^{-3}$ and b) $\chi_g=5\times 10^{-3}$. We observe that the asymmetry grows as $\chi_g$ is increased. The black vertical dotted lines show the location of resonator frequencies $\nu_n$.} 
\label{Fig:SpEmRate}
\end{figure}

At fixed $\chi_g$, we observe an asymmetry of $\alpha_j$ when the bare transmon frequency is tuned across the fundamental mode of the resonator, in agreement with a previous experiment \cite{Houck_Controlling_2008}, where a semiclassical model was employed for an accurate fit. Figure~\ref{Fig:SpEmRate} shows that near the resonator-like resonance the spontaneous decay rate is enhanced, as expected. For positive detunings spontaneous decay is significantly larger than for negative detunings, which can be traced back to an asymmetry in the resonator density of states \cite{Houck_Controlling_2008}. We find that this asymmetry grows as $\chi_g$ is increased. Note that besides a systematic inclusion of multimode effects, the presented theory of spontaneous emission goes beyond the rotating wave, Markov and two-level approximations as well.

\begin{figure}
\subfloat[\label{subfig:AxXg0to05Xj005XrXl001}]{%
\includegraphics[scale=0.375]{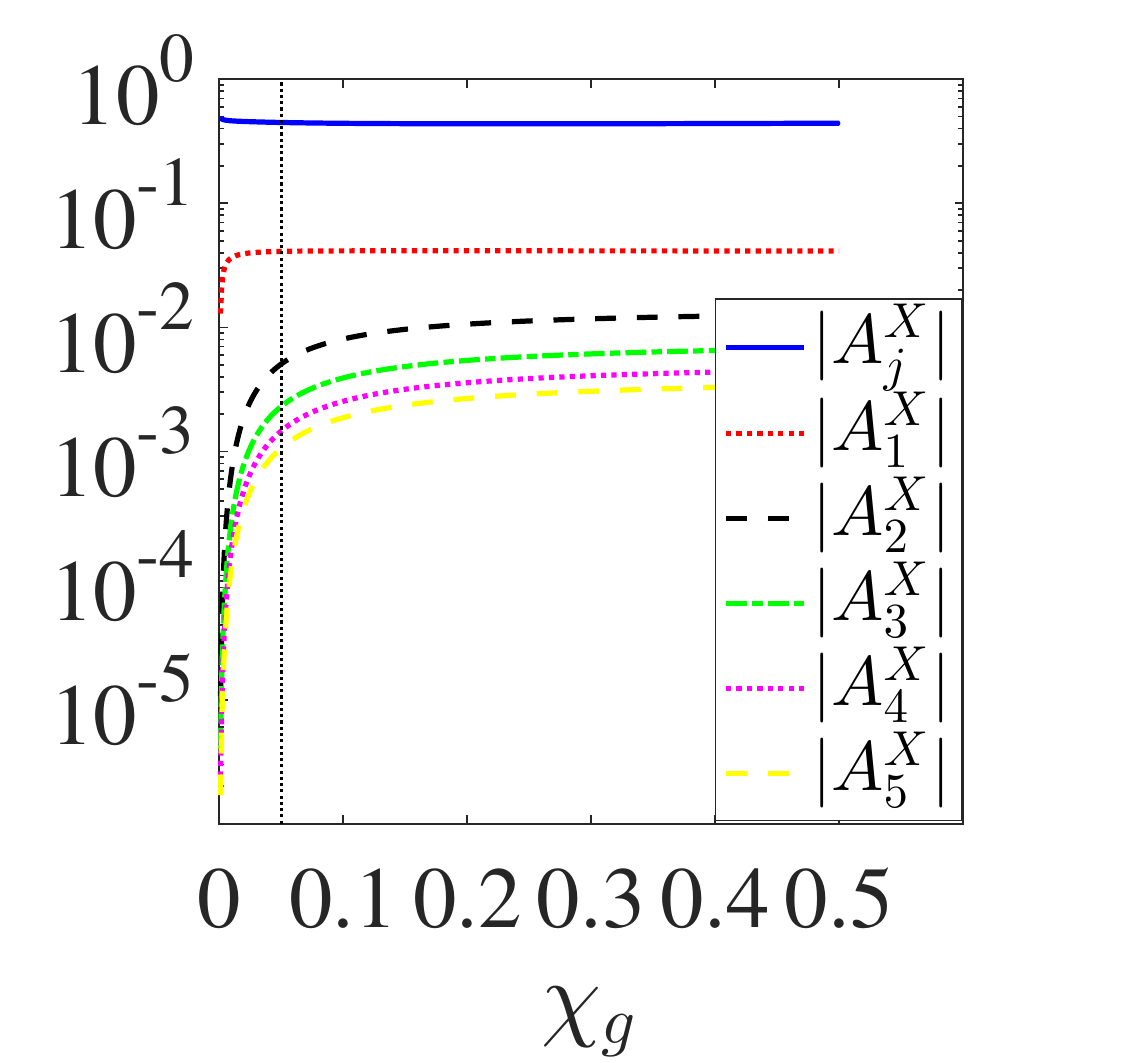}%
}\hfill 
\subfloat[\label{subfig:AyXg0to05Xj005XrXl001}]{%
\includegraphics[scale=0.375]{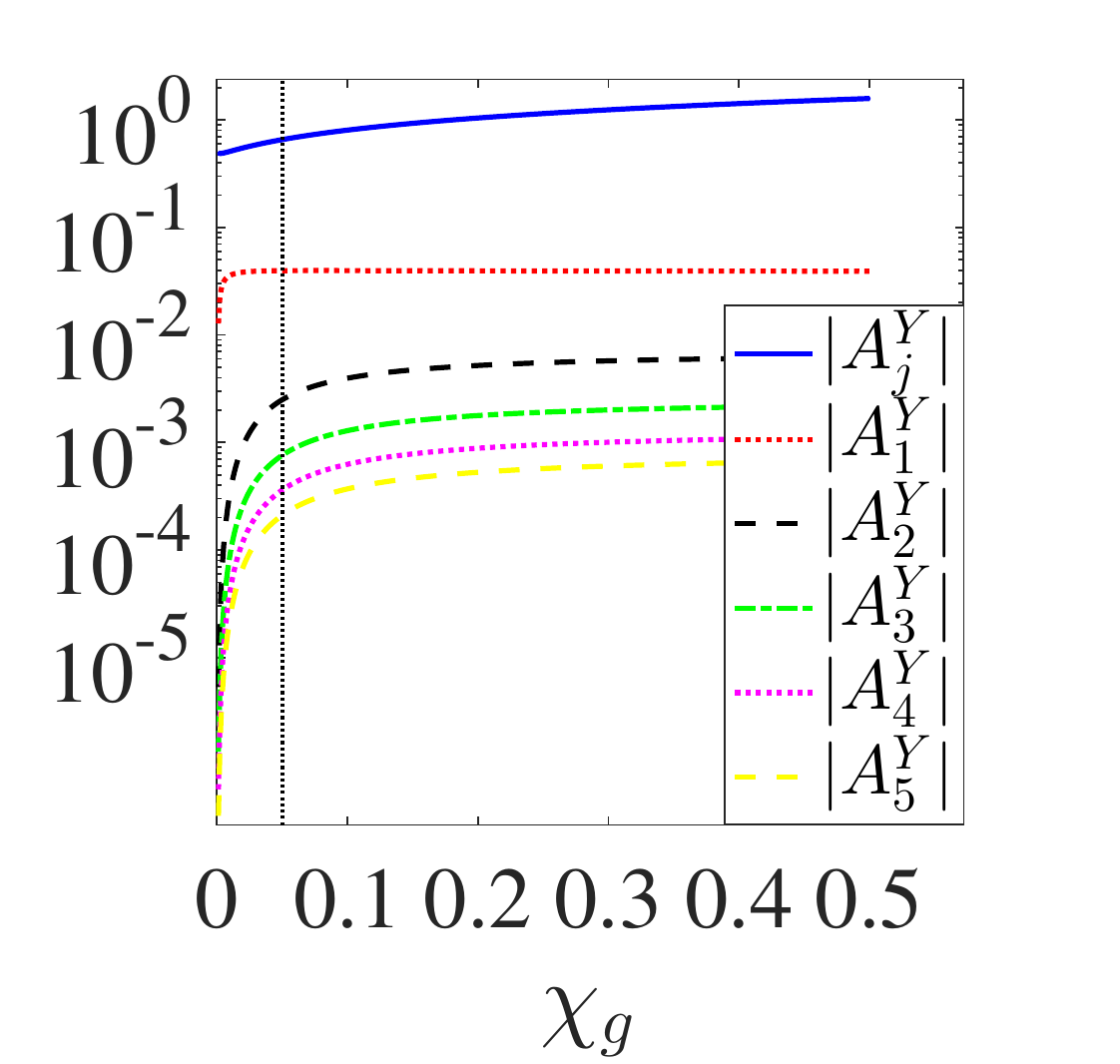}%
}
\caption{(Color online) Dependence of residues defined in Eqs.~(\ref{eqn:Def of A_l^X}-\ref{eqn:Def of A_l^Y}) on $\chi_g$ for $\omega_j=\nu_1^-$, $\chi_R=\chi_L=0.01$ and $\chi_j=0.05$. The black vertical dotted line shows the value of $\chi_j$.} 
\label{Fig:AnxAnyAjxAjy}
\end{figure}

Having studied the hybridized resonances of the entire system, we are now able to provide the time-dependent solution to Eq.~(\ref{eqn:Lin SE Problem}). By substituting Eq.~(\ref{eqn:Formal Rep of D(s)}) into Eq.~(\ref{eqn:Sol of X_j(s)}) we obtain
\begin{align}
\hat{\tilde{X}}_j(s)=\left(\frac{\hat{A}_j}{s-p_j}+\sum\limits_{n}\frac{\hat{A}_n}{s-p_n}\right)+H.c.,
\label{eqn:PFE of X_j(s)}
\end{align}
from which the inverse Laplace transform is immediate
\begin{align}
\hat{X}_j(t)=\left[\left(\hat{A}_je^{p_jt}+\sum\limits_{n}\hat{A}_n e^{p_n t}\right)+H.c.\right]\Theta(t).
\label{eqn:Lin Sol X_j(t)}
\end{align}
The frequency components have operator-valued amplitudes 
\begin{subequations}
\begin{align}
&\hat{A}_j\equiv A_j^{X}\hat{X}_j(0)+A_j^{Y}\hat{Y}_j(0),
\label{eqn:Def of hat(A)_j}\\
&\hat{A}_n\equiv A_n^{X}\hat{X}_j(0)+A_n^{Y}\hat{Y}_j(0),
\label{eqn:Def of hat(A)_n}
\end{align}
\end{subequations}
with the residues given in terms of $D_j(s)$ as
\begin{subequations}
\begin{align}
&A_{j,n}^{X}\equiv \left. \left[(s-p_{j,n})\frac{s}{D_j(s)}\right]\right|_{s=p_{j,n}},
\label{eqn:Def of A_l^X}\\
&A_{j,n}^{Y}\equiv \left. \left[(s-p_{j,n})\frac{\omega_j}{D_j(s)}\right]\right|_{s=p_{j,n}}.
\label{eqn:Def of A_l^Y}
\end{align}
\end{subequations}
The dependence of $A_{j,n}^{X}$ and $A_{j,n}^{Y}$ on coupling $\chi_g$ has been studied in Fig.~\ref{Fig:AnxAnyAjxAjy}. The transmon-like amplitude (blue solid) is always dominant, and further off-resonant modes have smaller amplitudes. By increasing $\chi_g$, the resonator-like amplitude grow significantly first and reach an asymptote as predicted by Eq.~(\ref{eqn:LinTheory-Asymptote}).
\subsection{Perturbative corrections}
\label{Sec:PertCor}
In this section, we develop a well-behaved time-domain perturbative expansion in the transmon qubit nonlinearity as illustrated in Eq.~(\ref{eqn:Expansion of Sine}). Conventional time-domain perturbation theory is inapplicable due to the appearance of resonant coupling between the successive orders which leads to secular contributions, i.e. terms that grow unbounded in time (For a simple example see App.~\ref{SubApp:ClDuffingDiss}). A solution to this is multi-scale perturbation theory (MSPT) \cite{Bender_Advanced_1999, Nayfeh_Nonlinear_2008, Strogatz_Nonlinear_2014}, which considers multiple independent time scales and eliminates secular contributions by a resummation of the conventional perturbation series. 

The effect of the nonlinearity is to mix the hybridized modes discussed in the previous section, leading to transmon mediated self-Kerr and cross-Kerr interactions. Below, we extend MSPT to treat this problem while consistently accounting for the dissipative effects. This goes beyond the extent of Rayleigh-Schr\"odinger perturbation theory, as it will allow us to treat the energetic and dissipative scales on equal footing. 

The outcome of conventional MSPT analysis in a conservative system is frequency renormalization \cite{Bender_Advanced_1999, Bender_Multiple_1996}. We illustrate this point for a classical Duffing oscillator, which amounts to the classical theory of an isolated transmon qubit up to leading order in the nonlinearity. We outline the main steps here leaving the details to App.~\ref{SubApp:ClDuffingDiss}. Consider a classical Duffing oscillator
\begin{align}
\ddot{X}(t)+\omega^2\left[X(t)-\varepsilon X^3(t)\right]=0,
\label{eqn:PertCor-ClDuffing Osc}
\end{align}
with initial conditions $X(0)=X_0$ and $\dot{X}(0)=\omega Y_0$. Equation~(\ref{eqn:PertCor-ClDuffing Osc}) is solved order by order with the Ansatz
\begin{subequations}
\begin{align}
X(t)=x^{(0)}(t,\tau)+\varepsilon x^{(1)}(t,\tau)+\mathcal{O}(\varepsilon^2),
\label{Eq:ClDuffing Expansion of X}
\end{align}
where $\tau\equiv\varepsilon t$ is assumed to be an independent time scale such that
\begin{align}
d_t\equiv \partial_t+\varepsilon\partial_\tau+\mathcal{O}(\varepsilon^2).
\end{align} 
\end{subequations}
This additional time-scale then allows us to remove the secular term that appears in the $\mathcal{O}(\varepsilon)$ equation. This leads to a renormalization in the oscillation frequency of the $\mathcal{O}(1)$ solution as
\begin{subequations}
\begin{align}
&X^{(0)}(t)=x^{(0)}(t,\varepsilon t)=\left[a(0)e^{-i\bar{\omega}t}+c.c.\right],
\label{eqn:PertCor-ClDuffing X^(0)(t)}\\
&\bar{\omega}\equiv \left[1-\frac{3\varepsilon}{2}|a(0)|^2\right]\omega,
\label{eqn:PertCor-ClDuffing FreqRenorm}
\end{align}
\end{subequations}
where $a(0)=(X_0+iY_0)/2$.
\begin{figure}
\centering
\includegraphics[scale=0.375]{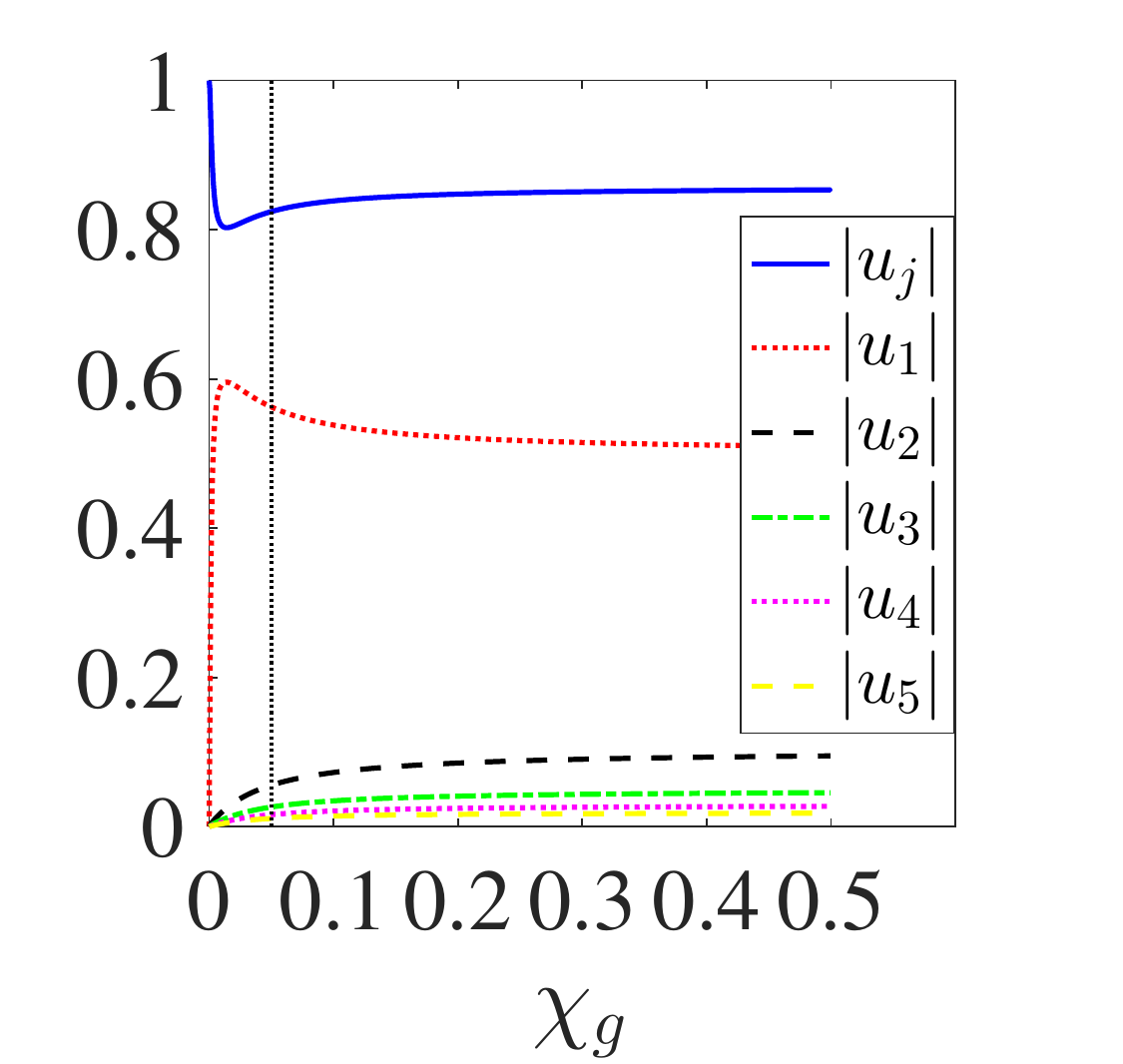} 
\caption{(Color online) Hybridization coefficients $u_j$ and $u_n$ of the first five modes for the case where the transmon is infinitesimally detuned below the fundamental mode, i.e. $\omega_j=\nu_1^-$ as a function of $\chi_g\in[0,0.5]$. Other parameters are set as $\chi_R=\chi_L=0$ and $\chi_j=0.05$. The black vertical dotted line shows the value of $\chi_j$.}
\label{Fig:QuDuffQuHarm-u}
\end{figure}
One may wonder how this leading-order correction is modified in the presence of dissipation. Adding a small damping term $\kappa\dot{X}(t)$ to Eq.~(\ref{eqn:PertCor-ClDuffing Osc}) such that $\kappa\ll\omega$ requires a new time scale $\eta \equiv \frac{\kappa}{\omega}t$ leading to
\begin{subequations}
\begin{align}
&X^{(0)}(t)=e^{-\frac{\kappa}{2}t}\left[a(0)e^{-i\bar{\omega}t}+c.c.\right],
\label{eqn:PertCor-ClDuffingDis X^(0)(t)}\\
&\bar{\omega}\equiv\left[1-\frac{3\varepsilon}{2}|a(0)|^2e^{-\kappa t}\right]\omega.
\label{eqn:PertCor-ClDuffingDiss FreqRenorm}
\end{align}
\end{subequations}
Equations~(\ref{eqn:PertCor-ClDuffingDis X^(0)(t)}-\ref{eqn:PertCor-ClDuffingDiss FreqRenorm}) illustrate a more general fact that the dissipation modifies the frequency renormalization by a decaying envelope. This approach can be extended by introducing higher order (slower) time scales $\varepsilon^2 t, \eta^2 t$, $\eta\varepsilon t$ etc. The lowest order calculation above is valid for times short enough such that $\omega t\ll \varepsilon^{-2} , \eta^{-2} , \eta^{-1}\varepsilon ^{-1}$.

Besides the extra complexity due to non-commuting algebra of quantum mechanics, the principles of MSPT remain the same in the case of a free quantum Duffing oscillator \cite{Bender_Multiple_1996}. The Heisenberg equation of motion is identical to Eq.~(\ref{eqn:PertCor-ClDuffing Osc}) where we promote $X(t)\to \hat{X}(t)$. We obtain the $\mathcal{O}(1)$ solution (see App.~\ref{SubApp:QuDuffingNoMem}) as
\begin{subequations}
\begin{align}
\begin{split}
\hat{X}^{(0)}(t)=e^{-\frac{\kappa}{2}t}\left[\frac{\hat{a}(0)e^{-i\hat{\bar{\omega}} t}+e^{-i\hat{\bar{\omega}} t}\hat{a}(0)}{2\cos\left(\frac{3\omega}{4}\varepsilon te^{-\kappa t}\right)}+H.c.\right]
\label{eqn:QuDuffingNoMem-X^(0)(t) sol}
\end{split}
\end{align}
with an operator-valued renormalization of the frequency 
\begin{align}
&\hat{\bar{\omega}}=\left[1-\frac{3\varepsilon}{2}\hat{\mathcal{H}}(0)e^{-\kappa t}\right]\omega,
\label{eqn:QuDuffingNoMem-bar(W)}\\
&\hat{\mathcal{H}}(0)\equiv\frac{1}{2}\left[\hat{a}^{\dag}(0)\hat{a}(0)+\hat{a}(0)\hat{a}^{\dag}(0)\right].
\label{eqn:QuDuffingNoMem-H(0)}
\end{align}
\end{subequations}
The cosine that appears in the denominator of operator solution~(\ref{eqn:QuDuffingNoMem-X^(0)(t) sol}) cancels when taking the expectation values with respect to the number basis $\{\ket{n}\}$ of $\hat{\mathcal{H}}(0)$:   
\begin{align}
\bra{n-1}\hat{X}^{(0)}(t)\ket{n}=\sqrt{n}e^{-\frac{\kappa}{2}t}e^{-i\left(1-\frac{3n\varepsilon}{2}e^{-\kappa t}\right)\omega t}.
\end{align}

Having learned from these toy problems, we return to the problem of spontaneous emission which can be mapped into a quantum Duffing oscillator with $\varepsilon=\frac{\sqrt{2}}{6}\left(\mathcal{E}_c/\mathcal{E}_j\right)^{1/2}$, up to leading order in perturbation, coupled to multiple leaky quantum harmonic oscillators (see Eq.~(\ref{eqn:Expansion of Sine})). We are interested in finding an analytic expression for the shift of the hybridized poles, $p_j$ and $p_n$, that appear in the reduced dynamics of the transmon. 

The hybridized poles $p_j$ and $p_n$ are the roots of $D_j(s)$ and they are associated with the modal decomposition of the linear theory in Sec.~\ref{Sec:Lin SE Theory}. The modal decomposition can be found from the linear solution $\mathcal{X}_j(t)$ that belongs to the full Hilbert space as   
\begin{align}
\begin{split}
\hat{\mathcal{X}}_j(t)&=\left(\hat{\mathcal{A}}_je^{p_jt}+\sum\limits_{n}\hat{\mathcal{A}}_n e^{p_n t}\right)+H.c.\\
&\equiv\left(u_j\hat{\bar{a}}_je^{p_jt}+\sum\limits_{n}u_n\hat{\bar{a}}_n e^{p_n t}\right)+H.c.
\end{split}
\label{eqn:Lin Sol Mathcal(X)_j(t)}
\end{align}
This is the full-Hilbert space version of Eq.~(\ref{eqn:Lin Sol X_j(t)}). It represents the unperturbed solution upon which we are building our perturbation theory. We have used bar-notation to distinguish the creation and annihilation operators in the hybridized mode basis. Furthermore, $u_j$ and $u_n$ represent the hybridization coefficients, where they determine how much the original transmon operator $\hat{\mathcal{X}}_j(t)$, is transmon-like and resonator-like. They can be obtained from a diagonalization of the linear Heisenberg-Langevin equations of motion (see App.~\ref{SubApp:QuDuffQuHarm}). The dependence of $u_j$ and $u_n$ on coupling $\chi_g$ is shown in Fig.~(\ref{Fig:QuDuffQuHarm-u}) for the case where the transmon is infinitesimally detuned below the fundamental mode of the resonator. For $\chi_g=0$, $u_j=1$ and $u_n=0$ as expected. As $\chi_g$ reaches $\chi_j$, $u_1$ is substantially increased and becomes comparable to $u_j$. By increasing $\chi_g$ further, $u_n$ for the off-resonant modes start to grow as well.

The nonlinearity acting on the transmon mixes all the unperturbed resonances through self- and cross-Kerr contributions \cite{Drummond_Quantum_1980, Nigg_Black-Box_2012, Bourassa_Josephson_2012}. Kerr shifts can be measured in a multimode cQED system \cite{Rehak_Parametric_2014, Weissl_Kerr_2015}. We therefore solve for the equations of motion of each mode. These are (see App. \ref{SubApp:QuDuffQuHarm})
\begin{align}
\begin{split}
&\hat{\ddot{\bar{\mathcal{X}}}}_{l}(t)+2\alpha_{l}\hat{\dot{\bar{\mathcal{X}}}}_{l}(t)\\
&+\beta_{l}^2\left\{\hat{\bar{\mathcal{X}}}_{l}(t)-\varepsilon_{l} \left[u_j\hat{\bar{\mathcal{X}}}_j(t)+\sum\limits_n u_n\hat{\bar{\mathcal{X}}}_n(t)\right]^3\right\}=0,
\end{split}
\label{eqn:QuDuffQuHarm Osc}
\end{align}
where $\hat{\bar{X}}_l\equiv \hat{\bar{a}}_l+\hat{\bar{a}}_l^{\dag}$ is the quadrature of the $l$th mode, and $\alpha_l$ and $\beta_l$ are the decay rate and the oscillation frequency, respectively. Equation~(\ref{eqn:QuDuffQuHarm Osc}) is the leading order approximation in the inverse Q-factor of the $l$th mode, $1/Q_l\equiv\alpha_l/\beta_l$. Each hybridized mode has a distinct strength of the nonlinearity $\varepsilon_{l}\equiv\frac{\omega_j}{\beta_{l}}u_{l}\varepsilon$ for $l\equiv j,n$. In order to do MSPT, we need to introduce as many new time-scales as the number of hybridized modes, i.e. $\tau_j\equiv\varepsilon_j t$ and $\tau_n\equiv \varepsilon_n t$, and do a perturbative expansion in all of these time scales. The details of this calculation can be found in App.~\ref{SubApp:QuDuffQuHarm}. Up to lowest order in $\varepsilon$, we find operator-valued correction of $p_{j}=-\alpha_j-i\beta_j$ as
\begin{figure}
\centering
\subfloat[\label{subfig:FXtXj005XrXl1Em3IC11View1}]{%
\includegraphics[scale=0.26]{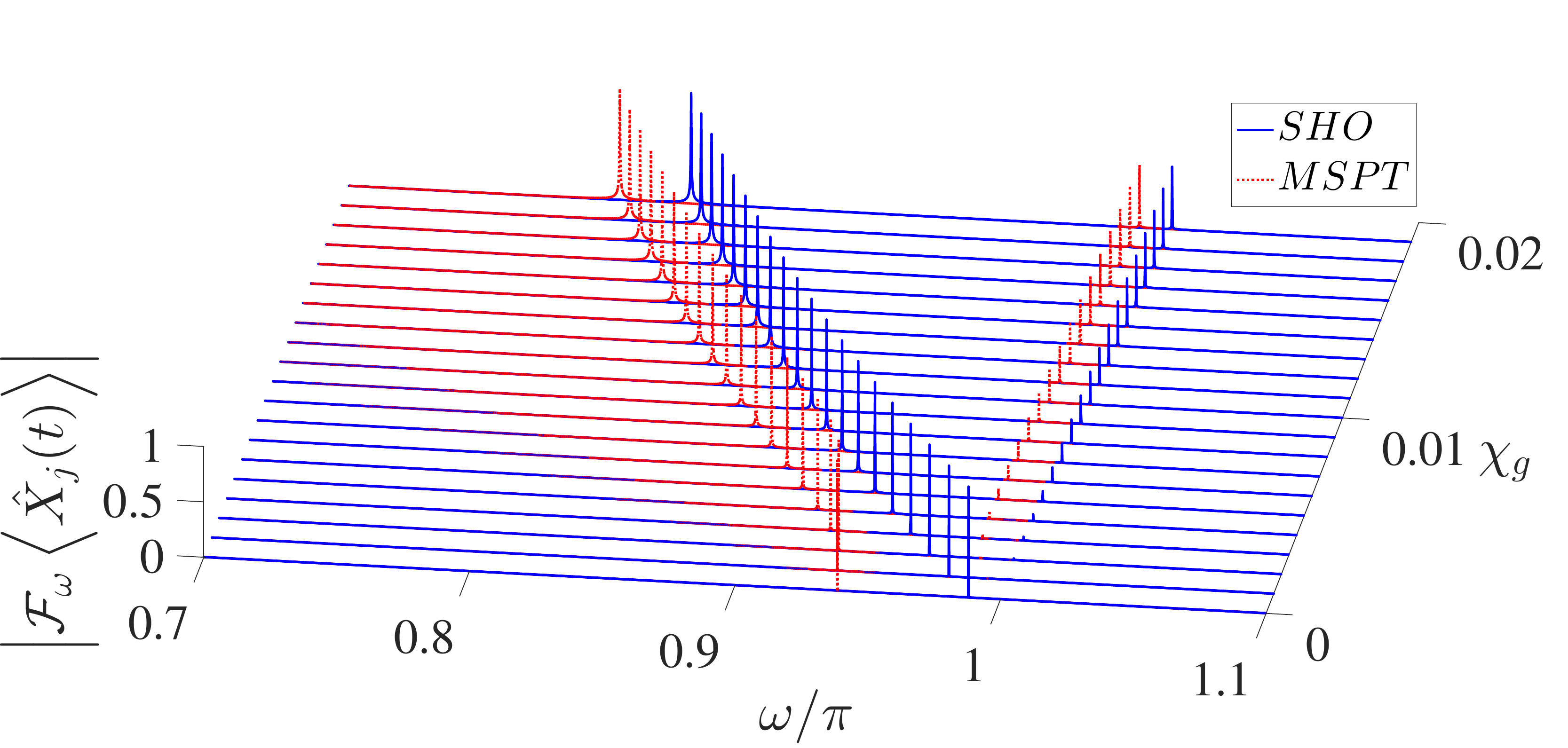}%
}\hfill 
\subfloat[\label{subfig:FXtXj005XrXl1Em3IC11View2}]{%
\includegraphics[scale=0.26]{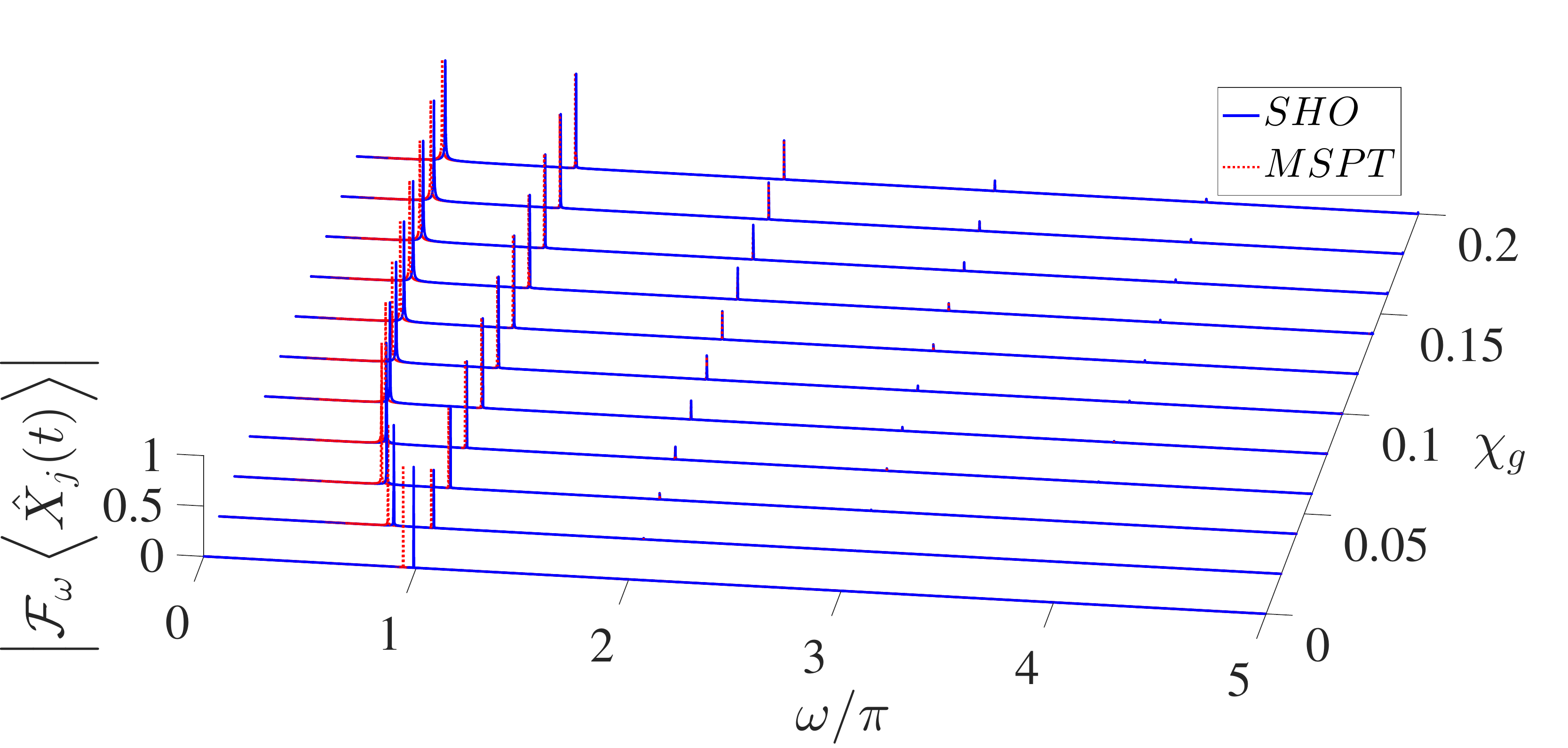}%
}  
\caption{(Color online) Fourier transform of $\braket{\hat{X}_j(t)}$ from linear solution (red dashed) and MSPT (blue solid) for $\chi_j=0.05$, $\chi_R=\chi_L=0.001$, $\mathcal{E}_j=50\mathcal{E}_c$ and initial state $\ket{\Psi_j(0)}=\frac{\ket{0}_j+\ket{1}_j}{\sqrt{2}}$ as a function of $\chi_g$. The maximum value of $\left|\mathcal{F}_{\omega}\braket{\hat{X}_j(t)}\right|$ at each $\chi_g$ is set to $1$. a) $\chi_g\in[0,0.02]$, $\Delta \chi_g=0.001$. b) $\chi_g\in[0,0.2]$, $\Delta \chi_g=0.02$.}
\label{Fig:FXtXj005XrXl1Em3IC11}
\end{figure}
\begin{subequations}
\begin{align}
\hat{\bar{p}}_{j}=p_{j}+i\frac{3\varepsilon}{2}\omega_j\left[u_j^4\hat{\bar{\mathcal{H}}}_j(0)e^{-2\alpha_j t}+\sum\limits_{n}2u_j^2u_n^2\hat{\bar{\mathcal{H}}}_n(0)e^{-2\alpha_n t}\right],
\label{eqn:PertCorr-bar(p)_j}
\end{align}
while $p_{n}=-\alpha_n-i\beta_n$ is corrected as
\begin{align}
\begin{split}
\hat{\bar{p}}_{n}=p_{n}+i\frac{3\varepsilon}{2}\omega_j&\left[u_n^4\hat{\bar{\mathcal{H}}}_n(0)e^{-2\alpha_n t}+2u_n^2u_j^2\hat{\bar{\mathcal{H}}}_j(0)e^{-2\alpha_j t}\right.\\
&+\left.\sum\limits_{m\neq n}2u_n^2u_m^2\hat{\bar{\mathcal{H}}}_m(0)e^{-2\alpha_m t}\right],
\end{split}
\label{eqn:PertCorr-bar(p)_n}
\end{align}
where $\hat{\bar{\mathcal{H}}}_{j}(0)$ and $\hat{\bar{\mathcal{H}}}_{n}(0)$ represent the Hamiltonians of each hybridized mode
\begin{align}
\hat{\bar{\mathcal{H}}}_{l}(0)\equiv \frac{1}{2}\left[\hat{\bar{a}}_{l}^{\dag}(0)\hat{\bar{a}}_{l}(0)+\hat{\bar{a}}_{l}(0)\hat{\bar{a}}_{l}^{\dag}(0)\right], \ l=j,n.
\label{eqn:PertCorr-Def of bar(H)_l(0)}
\end{align}
These are the generalizations of the single quantum Duffing results~(\ref{eqn:QuDuffingNoMem-bar(W)}) and (\ref{eqn:QuDuffingNoMem-H(0)}) and reduce to them as $\chi_g\to 0$ where $u_j=1$ and $u_n=0$. Each hybdridized mode is corrected due to a self-Kerr term proportional to $u_l^4$, and cross-Kerr terms proportional to $u_l^2u_{l'}^2$. Contributions of the form $u_l^2u_{l'}u_{l''}$ \cite{Bourassa_Josephson_2012} do not appear up to the lowest order in MSPT. 

In terms of Eqs.~(\ref{eqn:PertCorr-bar(p)_j}-\ref{eqn:PertCorr-bar(p)_n}), the MSPT solution reads
\end{subequations}
\begin{align}
\begin{split}
\hat{\mathcal{X}}_j^{(0)}(t)&=\frac{\hat{\mathcal{A}}_j(0)e^{\hat{\bar{p}}_j t}+e^{\hat{\bar{p}}_j t}\hat{\mathcal{A}}_j(0)}{2\cos\left(\frac{3\omega_j}{4}u_j^4\varepsilon t e^{-2\alpha_j t}\right)}+H.c.\\
&+\sum\limits_n\left[\frac{\hat{\mathcal{A}}_n(0)e^{\hat{\bar{p}}_n t}+e^{\hat{\bar{p}}_n t}\hat{\mathcal{A}}_n(0)}{2\cos\left(\frac{3\omega_j}{4}u_n^4\varepsilon t  e^{-2\alpha_n t}\right)}+H.c.\right],
\label{eqn:PertCorr-X^(0)(t) MSPT Sol}
\end{split}
\end{align}
where $\hat{\mathcal{A}}_{j,n}$ is defined in Eq.~(\ref{eqn:Lin Sol Mathcal(X)_j(t)}). In Fig.~\ref{Fig:FXtXj005XrXl1Em3IC11}, we have compared the Fourier transform of $\braket{\hat{\mathcal{X}}_j(t)}$ calculated both for the MSPT solution~(\ref{eqn:PertCorr-X^(0)(t) MSPT Sol}) and the linear solution~(\ref{eqn:Lin Sol X_j(t)}) for initial condition $\ket{\Psi(0)}=\frac{\ket{0}_j+\ket{1}_j}{\sqrt{2}}\otimes \ket{0}_{ph}$ as a function of $\chi_g$. At $\chi_g=0$, we notice the bare $\mathcal{O}(\varepsilon)$ nonlinear shift of a free Duffing oscillator as predicted by Eq.~(\ref{eqn:PertCor-ClDuffing FreqRenorm}). As $\chi_g$ is increased, the predominantly self-Kerr nonlinearity on the qubit is gradually passed as cross-Kerr contributions to the resonator modes, as observed from the frequency renormalizations~(\ref{eqn:PertCorr-bar(p)_j}) and (\ref{eqn:PertCorr-bar(p)_n}). As a result of this, interestingly, the effective nonlinear shift in the transmon resonance becomes smaller and saturates at stronger couplings. In other words, the transmon mode becomes {\it more linear} at {\it stronger} coupling $\chi_g$. This counterintuitive result can be understood from Eq.~(\ref{eqn:PertCorr-bar(p)_j}). For initial condition considered here, the last term in Eq.~(\ref{eqn:PertCorr-bar(p)_j}) vanishes, while one can see from Fig.~\ref{Fig:QuDuffQuHarm-u} that $u_j<1$ for $\chi_g>0$.

\subsection{Numerical simulation of reduced equation}
\label{Sec:NumSimul}
The purpose of this section is to compare the results from MSPT and linear theory to a pure numerical solution valid up to $\mathcal{O}(\varepsilon^2)$. A full numerical solution of the Heisenberg equation of motion~(\ref{eqn:Eff Dyn before trace}) requires matrix representation of the qubit operator $\hat{\mathcal{X}}_j(t)$ over the entire Hilbert space, which is impractical due to the exponentially growing dimension. We are therefore led to work with the reduced Eq.~(\ref{eqn:NL SE Problem}). While the nonlinear contribution in Eq.~(\ref{eqn:NL SE Problem}) cannot be traced exactly, it is possible to make progress perturbatively. We substitute the perturbative expansion Eq.~(\ref{eqn:Expansion of Sine}) into Eq.~(\ref{eqn:NL SE Problem}):
\begin{align}
\begin{split}
&\hat{\ddot{X}}_j(t)+\omega_j^2\left[1-\gamma+i\mathcal{K}_1(0)\right]\left[\hat{X}_j(t)-\varepsilon \Tr_{ph}{\{\hat{\rho}_{ph}(0)\hat{\mathcal{X}}_j^3(t)\}}\right]\\
&=-\int_{0}^{t}dt'\omega_j^2 \mathcal{K}_2(t-t')[\hat{X}_j(t')-\varepsilon \Tr_{ph}{\{\hat{\rho}_{ph}(0)\hat{\mathcal{X}}_j^3(t')\}}],
\label{eqn:NumSim-QuDuffingOscMem}
\end{split}
\end{align}
with $\varepsilon\equiv\frac{\sqrt{2}}{6}\epsilon$. If we are interested in the numerical results only up to $\mathcal{O}(\varepsilon^2)$ then the cubic term can be replaced as
\begin{align}
\begin{split}
\varepsilon\hat{\mathcal{X}}_j^3(t)=\varepsilon\left[\left.\hat{\mathcal{X}}_j(t)\right|_{\varepsilon=0}\right]^3+\mathcal{O}\left(\varepsilon^2\right).
\end{split}
\end{align}
Since we know the linear solution~(\ref{eqn:Lin Sol Mathcal(X)_j(t)}) for $ \hat{\mathcal{X}}_j(t)$ analytically, the trace can be performed directly (see App.~\ref{App:RedNumEq}). We obtain the reduced equation in the Hilbert of transmon as
\begin{align}
\begin{split}
&\hat{\ddot{X}}_j(t)+\omega_j^2\left[1-\gamma+i\mathcal{K}_1(0)\right]\left[\hat{X}_j(t)-\varepsilon\hat{X}_j^3(t)\right]\\
&=-\int_{0}^{t}dt'\omega_j^2 \mathcal{K}_2(t-t')\left[\hat{X}_j(t')-\varepsilon\hat{X}_j^3(t')\right]+\mathcal{O}(\varepsilon^2).
\label{eqn:NumSim-QuDuffingOscMemReduced}
\end{split}
\end{align}  
Solving the integro-differential Eq.~(\ref{eqn:NumSim-QuDuffingOscMemReduced}) numerically is a challenging task, since the memory integral on the RHS requires the knowledge of all results for $t'<t$. Therefore, simulation time for Eq.~(\ref{eqn:NumSim-QuDuffingOscMemReduced}) grows polynomially with $t$. The beauty of the Laplace transform in the linear case is that it turns a memory contribution into an algebraic form. However, it is inapplicable to Eq.~(\ref{eqn:NumSim-QuDuffingOscMemReduced}). 
\begin{figure}
\subfloat[\label{subfig:XtQuDuffingEps003Xg0XrXl0005}]{%
\includegraphics[scale=0.28]{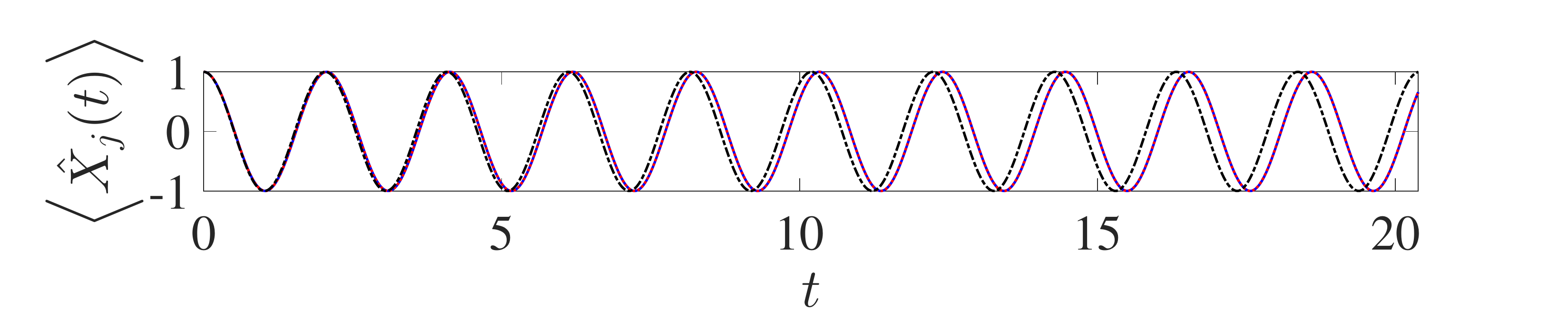}%
}\hfill 
\subfloat[\label{subfig:XtQuDuffingEps003Xg001XrXl0005}]{%
\includegraphics[scale=0.28]{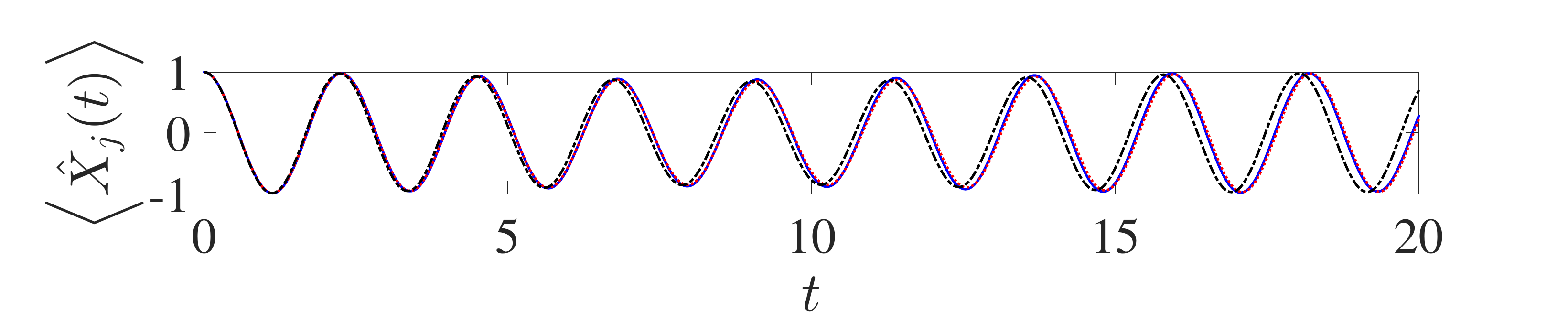}%
}
\hfill
\subfloat[\label{subfig:XtQuDuffingEps003Xg01XrXl0005}]{%
\includegraphics[scale=0.28]{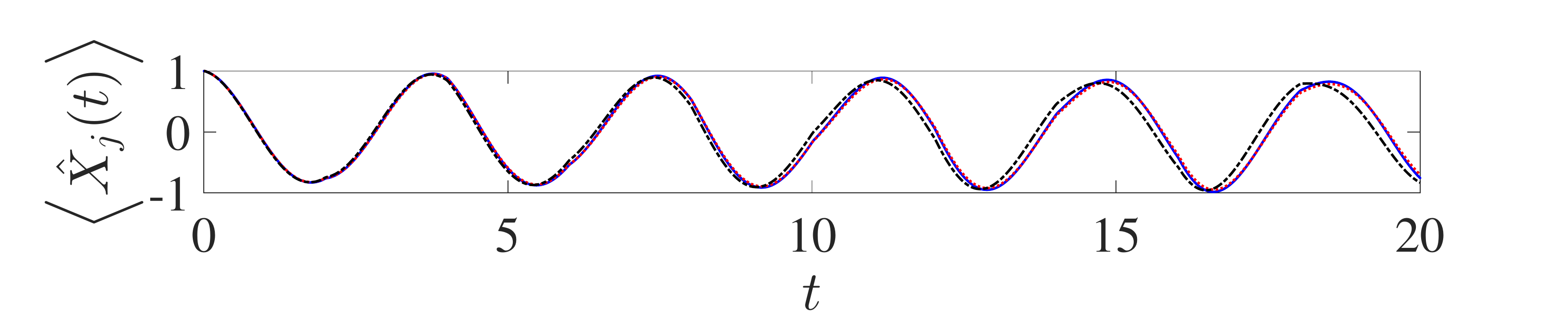}%
}
\hfill
\subfloat[\label{subfig:XtQuDuffingEps003Xg02XrXl0005}]{%
\includegraphics[scale=0.28]{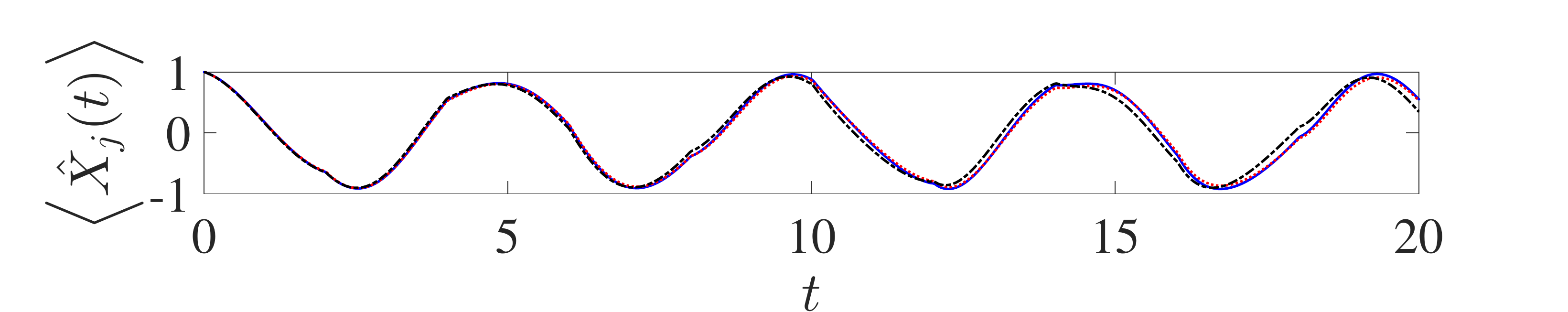}%
}
\caption{(Color online) Comparison of short-time dynamics between the results from linear theory (black dash-dot), MSPT (red dotted) and numerical (blue solid) of $\braket{\hat{X}_j(t)}$ for the same parameters as in Fig.~(\ref{Fig:FXtXj005XrXl1Em3IC11}) and for a) $\chi_g=0$, b) $\chi_g=0.01$, c) $\chi_g=0.1$ and d) $\chi_g=0.2$. The oscillation frequency and decay rate of the most dominant pole (transmon-like) are controlled by the hybridization strength. For a) where $\chi_g=0$, there is no dissipation and the transmon is isolated. The decay rate increases with $\chi_g$ such that the Q-factor for the transmon-like resonance reaches $Q_j\equiv\beta_j/\alpha_j \approx 625.3$ in Fig. d).} 
\label{Fig:TimeDynAnalVsODE45}
\end{figure}

In Fig.~\ref{Fig:TimeDynAnalVsODE45}, we compared the numerical results to both linear and MSPT solutions up to 10 resonator round-trip times and for different values of $\chi_g$. For $\chi_g=0$, the transmon is decoupled and behaves as a free Duffing oscillator. This corresponds to the first row in Fig.~(\ref{subfig:FXtXj005XrXl1Em3IC11View1}) where there is only one frequency component and MSPT provides the correction given in Eq.~(\ref{eqn:QuDuffingNoMem-bar(W)}). As we observe in Fig.~\ref{subfig:XtQuDuffingEps003Xg0XrXl0005} the MSPT results lie on top of the numerics, while the linear solution shows a visible lag by the 10th round-trip. Increasing $\chi_g$ further, brings more frequency components into play. As we observe in Fig.~\ref{Fig:FXtXj005XrXl1Em3IC11}, for $\chi_g=0.01$ the most resonant mode of the resonator has a non-negligible $u_1$. Therefore, we expect to observe weak beating in the dynamics between this mode and the dominant transmon-like resonance, which is shown in Fig.~\ref{subfig:XtQuDuffingEps003Xg001XrXl0005}. Figures~\ref{subfig:XtQuDuffingEps003Xg01XrXl0005} and \ref{subfig:XtQuDuffingEps003Xg02XrXl0005} show stronger couplings where many resonator modes are active and a more complex beating is observed. In all these cases, the MSPT results follow the pure numerical results more closely than the linear solution confirming the improvement provided by perturbation theory. 
\subsection{System output}
\label{Sec:SysOutput}

\begin{figure}[h!]
\centering
\subfloat[\label{subfig:FXcavXj005XrXl1Em3IC11View1}]{%
\includegraphics[scale=0.26]{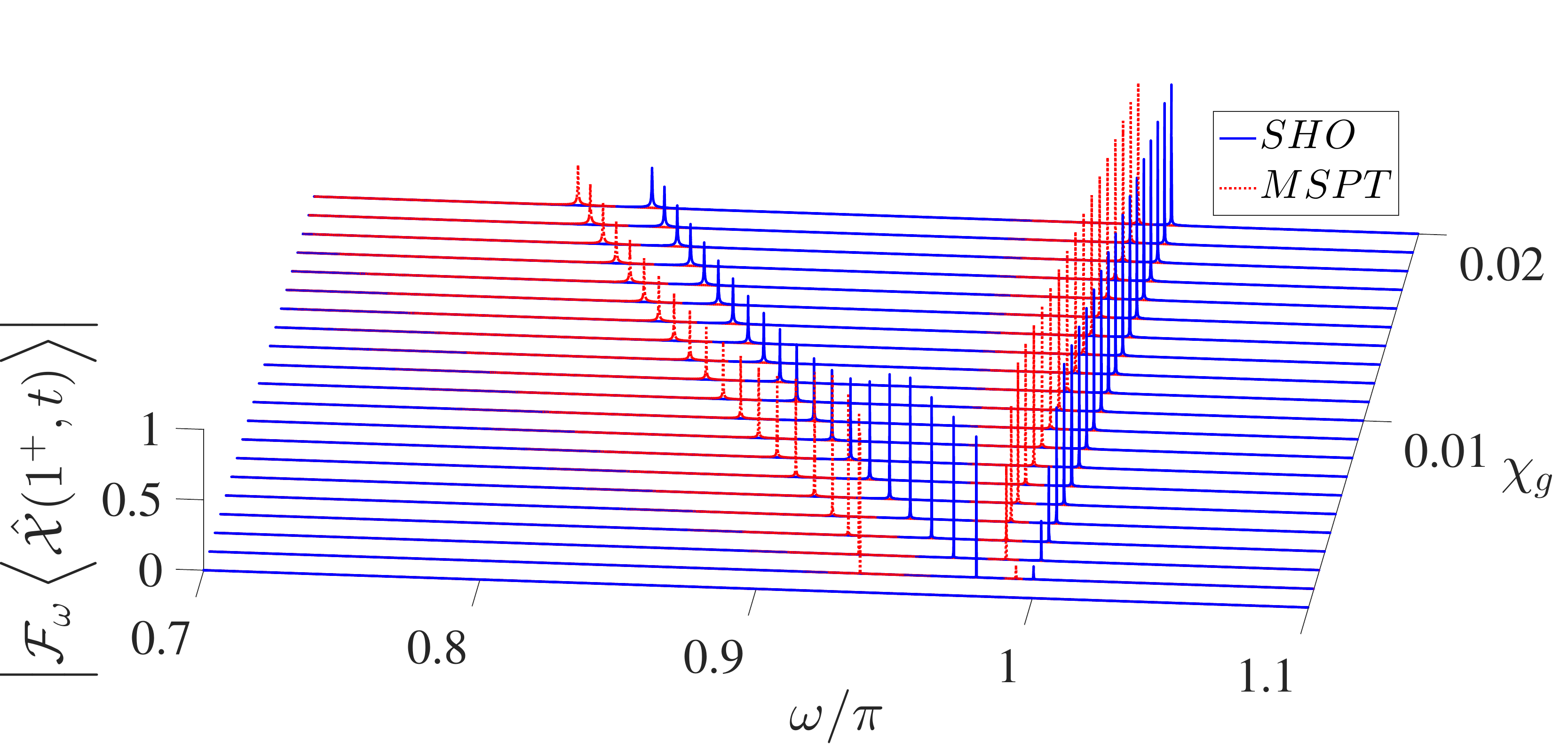}%
}\hfill 
\subfloat[\label{subfig:FXcavXj005XrXl1Em3IC11View2}]{%
\includegraphics[scale=0.26]{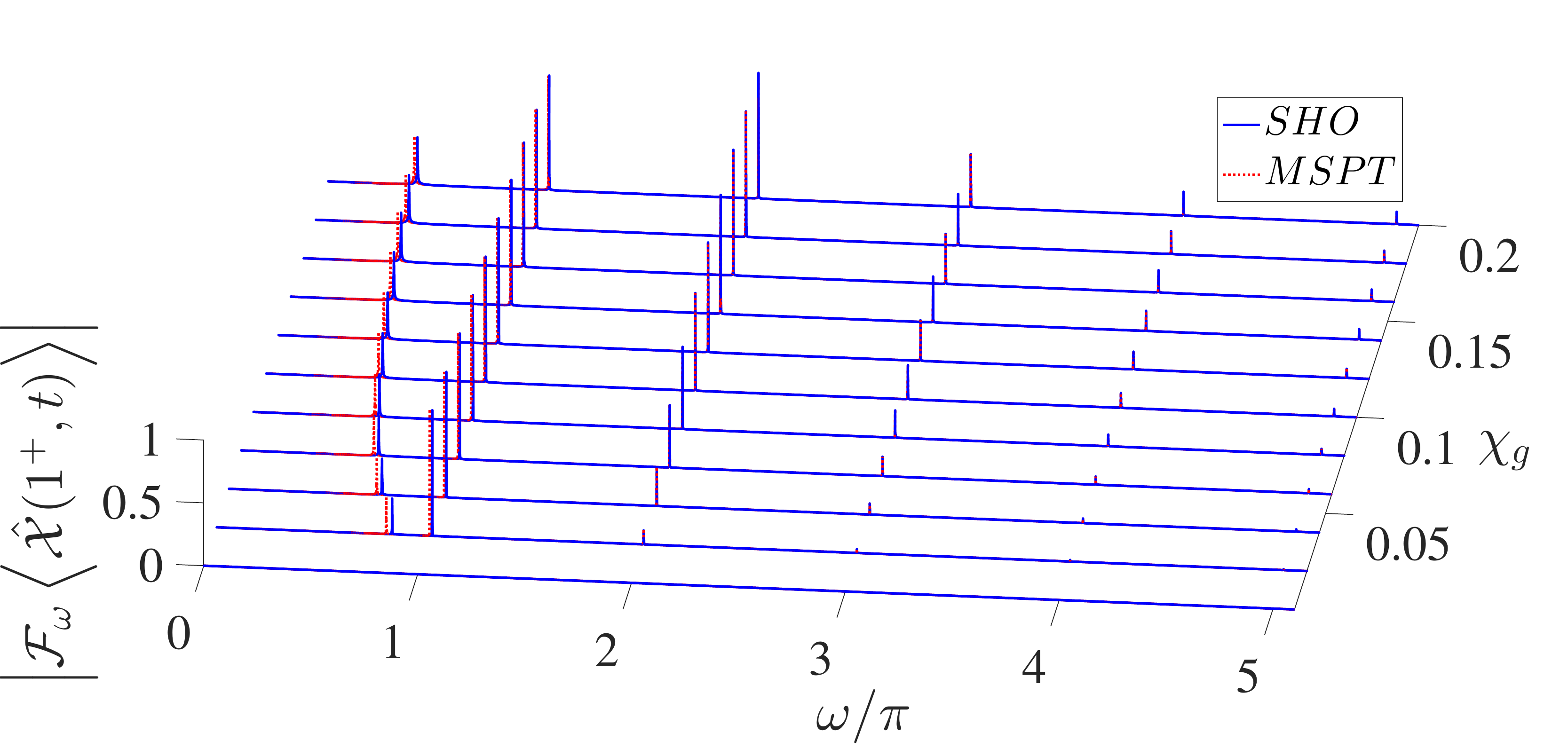}%
}  
\caption{(Color online) Fourier transform of $\braket{\hat{\mathcal{X}}(1^+,t)}$ for the linear solution (red dashed) and the MSPT (blue solid) for the same parameters as in Fig.~\ref{Fig:FXtXj005XrXl1Em3IC11}. The maximum value of $\left|\mathcal{F}_{\omega}\braket{\hat{X}_j(t)}\right|$ at each $\chi_g$ is set to $1$.}
\label{Fig:FXcavXj005XrXl1Em3IC11}
\end{figure}
Up to this point, we studied the dynamics of the spontaneous emission problem in terms of one of the quadratures of the transmon qubit, i.e. $\braket{\hat{\mathcal{X}}_j(t)}$. In a typical experimental setup however, the measuarable quantities are the quadratures of the field outside the resonator \cite{Clerk_Introduction_2010}. We devote this section to the computation of these quantities. 

The expression of the fields $\hat{\varphi}(x,t)$ can be directly inferred from the solution of the inhomogeneous wave Eq.~(\ref{eqn:Res Dyn}) using the impulse response (GF) defined in Eq.~\ref{Eq:Def of G(x,t|x0,t0)}. We note that this holds irrespective of whether one is solving for the classical or as is the case here, for the quantum fields. Taking the expectation value of this solution (App.~\ref{SubApp:Eff Dyn of transmon}) with respect to the initial density matrix~(\ref{eqn:SE-IC}) we find
\begin{align}
\braket{\hat{\varphi}(x,t)}=\chi_s\omega_j^2\int_{0}^{t}dt' G(x,t|x_0,t')\braket{\sin[\hat{\varphi}_j(t')]}.
\label{eqn:<varphi(x,t)>-SysOutput}
\end{align} 
Dividing both sides by $\phi_{\text{zpf}}$ and keeping the lowest order we obtain the resonator response as
\begin{align}
\braket{\hat{\mathcal{X}}^{(0)}(x,t)}=\chi_s\omega_j^2\int_{0}^{t}dt' G(x,t|x_0,t')\braket{\hat{\mathcal{X}}_j^{(0)}(t')},
\label{eqn:<varphi(x,t)>-SysOutput}
\end{align} 
where $\hat{\mathcal{X}}_j^{(0)}(t)$ is the lowest order MSPT solution~(\ref{eqn:PertCorr-X^(0)(t) MSPT Sol}), which takes into account the frequency correction to $\mathcal{O}(\varepsilon)$. Taking the Laplace transform decouples the convolution 
\begin{align}
\braket{\hat{\tilde{\mathcal{X}}}^{(0)}(x,s)}=\chi_s\omega_j^2\tilde{G}(x,x_0,s)\braket{\hat{\tilde{\mathcal{X}}}_j^{(0)}(s)},
\label{eqn:<varphi(x,t)>-SysOutput}
\end{align} 
which indicates that the resonator response is filtered by the GF. 

Figure~\ref{Fig:FXcavXj005XrXl1Em3IC11} shows the field outside the right end of the resonator, $\braket{\hat{\tilde{\mathcal{X}}}(x=1^+,s=i\omega)}$, in both linear and lowest order MSPT approximations. This quadrature can be measured via heterodyne detection \cite{Bishop_Nonlinear_2009}. Note that the hybridized resonances are the same as those of $\braket{\hat{\mathcal{X}}_j(t)}$ shown in Fig.~\ref{Fig:FXtXj005XrXl1Em3IC11}. What changes is the relative strength of the residues. The GF has poles at the bare cavity resonances and therefore the more hybridized a pole is, the smaller its residue becomes. 
\section{Conclusion}
\label{Sec:Conclusion}
In this paper, we introduced a new approach for studying the effective non-Markovian Heisenberg equation of motion of a transmon qubit coupled to an open multimode resonator beyond rotating wave and two level approximations. The main motivation to go beyond a two level representation lies in the fact that a transmon is a weakly nonlinear oscillator. Furthermore, the information regarding the electromagnetic environment is encoded in a single function, i.e. the electromagnetic GF. As a result, the opening of the resonator is taken into account analytically, in contrast to the Lindblad formalism where the decay rates enter only phenomenologically.

We applied this theory to the problem of spontaneous emission as the simplest possible example. The weak nonlinearity of the transmon allowed us to solve for the dynamics perturbatively in terms of $(\mathcal{E}_c/\mathcal{E}_j)^{1/2}$ which appears as a measure of nonlinearity. Neglecting the nonlinearity, the transmon acts as a simple harmonic oscillator and the resulting linear theory is exactly solvable via Laplace transform. By employing Laplace transform, we avoided Markov approximation and therefore accounted for the exact hybridization of transmon and resonator resonances. Up to leading nonzero order, the transmon acts as a quantum Duffing oscillator. Due to the hybridization, the nonlinearity of the transmon introduces both self-Kerr and cross-Kerr corrections to all hybdridized modes of the linear theory. Using MSPT, we were able to obtain closed form solutions in Heisenberg picture that do not suffer from secular behavior. A direct numerical solution confirmed the improvement provided by the perturbation theory over the harmonic theory. Surprisingly, we also learned that the linear theory becomes more accurate for stronger coupling since the nonlinearity is suppressed in the qubit-like resonance due to being shared between many hybdridized modes. 

The theory developed here illustrates how far one can go without the concept of photons. Many phenomena in the domain of quantum electrodynamics, such as spontaneous or stimulated emission and resonance fluorescence, have accurate semiclassical explanations in which the electric field is treated classically while the atoms obey the laws of quantum mechanics. For instance, the rate of spontaneous emission can be related to the local density of electromagnetic modes in the weak coupling limit. While it is now well understood that the electromagnetic fluctuations are necessary to start the spontaneous emission process \cite{Haake_Delay-time_1981}, it is important to ask to what extent a quantized electromagnetic field effects the qubit dynamics \cite{scully_concept_1972}. We find here that although the electromagnetic degrees of freedom are integrated out and the dynamics can systematically be reduced to the Hilbert space of the transmon, the quantum state of the electromagnetic environment reappears in the initial and boundary conditions when computing observables.  

Although we studied only the spontaneous emission problem in terms of quadratures, our theory can be applied to a driven-dissipative problem as well and all the mathematical machinery developed in this work can be used in more generic situations. In order to maintain a reasonable amount of material in this paper, we postpone the results of the driven-dissipative problem, as well as the study of  correlation functions to future work.
\section{Acknowledgements}
We appreciate helpful discussions with O. Malik on implementing the numerical results of Sec.~\ref{Sec:NumSimul}. This research was supported by the US Department of Energy, Office of Basic Energy Sciences, Division of Materials Sciences and Engineering under Award No. DE-SC0016011.
\appendix
\section{Quantum equations of motion}
\label{App:Quantum EOM}

The classical Lagrangian for the system shown in Fig.~\ref{Fig:cQED-open} can be found as sum of the Lagrangians for each circuit element. In the following, we use the convention of working with  flux variables \cite{Bishop_Circuit_2010, Devoret_Quantum_2014} as the generalized coordinate for our system. For an arbitrary node $n$ in the circuit, the flux variable $\Phi_n(t)$ is defined as
\begin{align}
\Phi_n(t)\equiv \int_{0}^{t}dt' V_n(t'),
\label{Eq:Def of Phi_n(t)}
\end{align}
while $V_n(t')$ stands for the voltage at node $n$. 

The classical Euler-Lagrange equations of motion can then be found by setting the variation of Lagrangian with respect to each flux variable to zero. For the transmon and the resonator we find
\begin{align}
&\ddot{\Phi}_j+\frac{1}{C_g+C_j}\frac{\partial U_j(\Phi_j)}{\partial \Phi_j}=\gamma \partial_{t}^2\Phi(x_0,t),
\label{Eq:Transmon Dyn in terms of Phi_j}
\end{align}
\begin{align} 
&\partial_{x}^2\Phi(x,t)-lc(x,x_0)\partial_{t}^2\Phi(x,t) =l\gamma \delta(x-x_0)\frac{\partial U_j(\Phi_j)}{\partial \Phi_j}.
\label{Eq:Resonatr-dynamics in terms of Phi}
\end{align}
where $U_j(\Phi_j)$ stands for the Josephson potential as
\begin{align}
U_j(\Phi_j)=-E_j\cos{\left(\frac{2\pi}{\Phi_0}\Phi_j\right)},
\label{Eq:Josephson Potential}
\end{align} 
and $\Phi_0\equiv \frac{h}{2e}$ is the superconducting flux quantum. Furthermore, $C_s \equiv C_gC_j/(C_g+C_j)$ is the series capacitance of $C_j$ and $C_g$ and  $\gamma\equiv C_g/(C_g+C_j)$. Moreover, $l$ and $c$ are the inductance and capacitance per length of the resonator and waveguides while $c(x,x_0)\equiv c+C_s\delta(x-x_0)$ represents the modified capacitance per length due to coupling to transmon. 

In addition, we find two wave equations for the flux field of the left and right waveguides as
\begin{align}
\partial_{x}^2\Phi_{R,L}(x,t)-lc\partial_{t}^2\Phi_{R,L}(x,t) =0,
\label{Eq:Side Bath dynamics}
\end{align}
The boundary conditions (BC) are derived from continuity of current at each end as
\begin{subequations}
\begin{align}
\begin{split}
-\frac{1}{l} \left.\partial_{x}\Phi\right|_{x=L^-}&=-\frac{1}{l} \left.\partial_{x}\Phi_R\right|_{x=L^+}\\
&=C_R\partial_t^2\left[\Phi(L^-,t)-\hat{\Phi}_R(L^+,t)\right],
\end{split}
\label{Eq:BC-Conservation of current at 0 in terms of Phi}
\end{align}
\begin{align}
-\frac{1}{l} \left.\partial_{x}\Phi\right|_{x=0^+}&=-\frac{1}{l} \left.\partial_{x}\Phi_L\right|_{x=0^-}\\
&=C_L\partial_t^2\left[\Phi_L(0^-,t)-\Phi(0^+,t)\right],
\label{Eq:BC-Conservation of current at 1 in terms of Phi}
\end{align}
\end{subequations}
continuity of flux at $x=x_0$
\begin{align}
\Phi(x=x_0^-,t)=\Phi(x=x_0^+,t), 
\end{align}
and conservation of current at $x=x_0$ as
\begin{align}
\left. \partial_x \Phi\right|_{x=x_0^+} - \left.\partial_x \Phi \right|_{x=x_0^-} -lC_s\partial_t^2 \Phi(x_0,t)=l\gamma \frac{\partial U_j(\Phi_j)}{\partial \Phi_j}.
\end{align}

In order to find the quantum equations of motion, we follow the common procedure of canonical quantization \cite{Devoret_Quantum_2014}: 
\begin{itemize}
\item[1)] Find the conjugate momenta $Q_n\equiv \frac{\delta \mathcal{L}}{\delta \dot{\Phi}_n}$ 
\item[2)] Find the classical Hamiltonian via a Legendre transformation as $\mathcal{H}=\sum\limits_{n}Q_n\dot{\Phi_n}-\mathcal{L}$
\item[3)] Find the Hamiltonian operator by promoting the classical conjugate variables to quantum operators such that $\{\hat{\Phi}_m,\hat{Q}_n\}=\delta_{mn}\to [\hat{\Phi}_m,\hat{Q}_n]=i\hbar\delta_{mn}$. We use a hat-notation to distinguish operators from classical variables.
\end{itemize} 

The derivation for the quantum Hamiltonian of the the closed version of this system where $C_{R,L}\to 0$ can be found in \cite{Malekakhlagh_Origin_2016} (see App. A, B, and C). Note that nonzero end capacitors $C_{R,L}$ leave the equations of motion for the resonator and waveguides unchanged, but modify the BC of the problem at $x=0,L$. The resulting equations of motion for the quantum flux operators $\hat{\Phi}_j$, $\hat{\Phi}(x,t)$ and $\hat{\Phi}_{R,L}(x,t)$ have the exact same form as the classical Euler-Lagrange equations of motion.

Next, we define unitless parameters and variables as
\begin{align}
\begin{split}
&\bar{x}\equiv\frac{x}{L},\quad
\bar{t}\equiv\frac{t}{\frac{L}{v_{p}}},\quad
\bar{\omega}\equiv\frac{\omega}{v_p}L,\\
&\hat{\varphi}\equiv 2\pi \frac{\hat{\Phi}}{\Phi_0}, \quad \hat{n}\equiv\frac{\hat{Q}}{2e}
\end{split}
\label{Eq:unitless vars}
\end{align}
where $v_p\equiv 1/\sqrt{lc}$ is the phase velocity of the resonator and waveguides. Furthermore, we define unitless capacitances as
\begin{align}
\chi_i\equiv\frac{C_i}{cL}, \quad i=R,L,j,g,s 
\end{align}
as well as a unitless modified capacitance per length as
\begin{align}
\chi(\bar{x},\bar{x}_0)\equiv 1+\chi_s \delta(\bar{x}-\bar{x}_0).
\label{Eq:Def of chi(x,x0)}
\end{align}
Then, the unitless equations of motion for our system are found as 
\begin{subequations}
\begin{align}
\hat{\ddot{\varphi}}_j(\bar{t})+(1-\gamma)\bar{\omega}_j^2\sin{[\hat{\varphi}_j(\bar{t})]}=\gamma \partial_{\bar{t}}^2\hat{\varphi}(\bar{x}_0,\bar{t}),
\label{Eq:Transmon Dyn}
\end{align}
\begin{align}
\begin{split}
\left[\partial_{\bar{x}}^2-\chi(\bar{x},\bar{x}_0)\partial_{\bar{t}}^2\right]\hat{\varphi}(\bar{x},\bar{t})=\chi_s\bar{\omega}_j^2 \sin{[\varphi_j(\bar{t})]}\delta(\bar{x}-\bar{x}_0),
\end{split}
\label{Eq:Res Dyn}
\end{align}
\begin{align}
\partial_{\bar{x}}^2\hat{\varphi}_{R,L}(\bar{x},\bar{t})-\partial_{\bar{t}}^2\hat{\varphi}_{R,L}(\bar{x},\bar{t}) =0,
\label{Eq:Side Res Dyn}
\end{align}
\end{subequations}
with the unitless BCs given as
\begin{subequations}
\begin{align}
\begin{split}
-\left.\partial_{\bar{x}}\hat{\varphi}\right|_{\bar{x}=1^-}&=-\left.\partial_{\bar{x}}\hat{\varphi}_R\right|_{\bar{x}=1^+}\\
&=\chi_R\partial_{\bar{t}}^2\left[\hat{\varphi}(1^-,\bar{t})-\hat{\varphi}_R(1^+,\bar{t})\right],
\end{split}
\label{Eq:BC-Conservation of current at 1}
\end{align}
\begin{align}
\begin{split}
-\left.\partial_{\bar{x}}\hat{\varphi}\right|_{\bar{x}=0^+}&=-\left.\partial_{\bar{x}}\hat{\varphi}_L\right|_{\bar{x}=0^-}\\
&=\chi_L\partial_{\bar{t}}^2\left[\hat{\varphi}_L(0^-,\bar{t})-\hat{\varphi}(0^+,\bar{t})\right],
\label{Eq:BC-Conservation of current at 0}
\end{split}
\end{align}
\begin{align}
\hat{\varphi}(\bar{x}=\bar{x}_0^-,\bar{t})=\hat{\varphi}(\bar{x}=\bar{x}_0^+,\bar{t}),
\end{align}
\begin{align}
\begin{split}
\left. \partial_{\bar{x}} \hat{\varphi}\right|_{\bar{x}=\bar{x}_0^+} &- \left.\partial_{\bar{x}} \hat{\varphi} \right|_{\bar{x}=\bar{x}_0^-}
-\chi_s\partial_{\bar{t}}^2 \hat{\varphi}(\bar{x}_0,\bar{t})\\
&=\chi_s \bar{\omega}_j^2 \sin{[\varphi_j(\bar{t})]}.
\end{split}
\label{Eq:BC-conservation of current at x0}
\end{align}
\end{subequations}

In Eqs.~(\ref{Eq:Transmon Dyn}) and (\ref{Eq:Res Dyn}), we have defined the unitless oscillation frequency $\bar{\omega}_j$ as
\begin{align}
\bar{\omega}_j^2  \equiv lc L^2\frac{E_j}{C_j} \left(\frac{2\pi}{\Phi_0}\right)^2=8\mathcal{E}_c\mathcal{E}_j,
\label{Eq:Def of bar(W)_j}
\end{align}
where $\mathcal{E}_c$ and $\mathcal{E}_j$ stand for the unitless charging and Josephson energy given as
\begin{align}
\mathcal{E}_{j,c}\equiv \sqrt{lc} L\frac{E_{j,c}}{\hbar},\
\end{align}
with $E_c\equiv \frac{e^2}{2C_j}$. 

In what follows, we work with the unitless Eqs.~(\ref{Eq:Transmon Dyn}-\ref{Eq:Side Res Dyn}) and BCs~(\ref{Eq:BC-Conservation of current at 1}-\ref{Eq:BC-conservation of current at x0}) and drop the bars.
\section{Effective dynamics of the transmon via a Heisenberg picture Green's function method}
\label{App:Eff Dyn of transmon}
In order to find the effective dynamics of the transmon qubit, one has to solve for the flux field $\hat{\varphi}(x,t)$ and substitute the result back into the RHS of time evolution of the qubit given by Eq.~(\ref{Eq:Transmon Dyn}). It is possible to perform this procedure in terms of the resonator GF. In Sec.~\ref{SubApp:Def of G} we define the resonator GF. In Sec.~\ref{SubApp:Spec Rep of G-open} we study the spectral representation of the GF in terms of a suitable set of non-Hermitian modes. In Sec.~\ref{SubApp:Eff Dyn of transmon}, we discuss the derivation of the effective dynamics of transmon in terms of the resonator GF. Finally, in Secs.~\ref{SubApp:SE Eff Dyn} and \ref{SubApp:Spec Rep of K} we discuss how the generic dynamics is reduced for the problem of spontaneous emission.
\subsection{Definition of $G(x,t|x',t')$}
\label{SubApp:Def of G}
The resonator GF is defined as the response of the linear system of Eqs.~(\ref{Eq:Res Dyn}-\ref{Eq:Side Res Dyn}) to a $\delta$-function source in space-time as
\begin{align}
\begin{split}
\left[\partial_x^2 -\chi(x,x_0)\partial_t^2\right]&G(x,t|x_0,t_0)=\delta(x-x_0)\delta(t-t_0),
\end{split}
\label{Eq:Def of G(x,t|x0,t0)}
\end{align}
with the same BCs as Eqs.~(\ref{Eq:BC-Conservation of current at 1}-\ref{Eq:BC-conservation of current at x0}). 
Using the Fourier transform conventions
\begin{subequations}
\begin{align}
&\tilde{G}(x,x_0,\omega)=\int_{-\infty}^{\infty}dt G(x,t|x_0,t_0) e^{+i\omega(t-t_0)}, \\
&G(x,t|x_0,t_0)=\int_{-\infty}^{\infty}\frac{d\omega}{2\pi} \tilde{G}(x,x_0,\omega)e^{-i\omega(t-t_0)},
\end{align}
\end{subequations}
Eq.~(\ref{Eq:Def of G(x,t|x0,t0)}) transforms into a Helmholtz equation 
\begin{align}
\begin{split}
\left[\partial_x^2 +\omega^2\chi(x,x_0) \right]\tilde{G}(x,x_0,\omega)=\delta(x-x_0).
\end{split}
\label{Eq:Helmholtz Eq for G(x,x0,W)}
\end{align}
Moreover, the BCs are transformed by replacing $\partial_x\to \partial_x$ and $\partial_t \to -i\omega $ as
\begin{subequations}
\begin{align}
&\left.\tilde{G}\right|_{x=x_0^+}=\left.\tilde{G}\right|_{x=x_0^-}, 
\label{Eq:Cont of G(x,x0,W) at x0}\\
&\left.\partial_x \tilde{G} \right|_{x=x_0^+}-\left.\partial_x \tilde{G}\right|_{x=x_0^-}+\chi_s\omega^2\left.\tilde{G}\right|_{x=x_0}=1 ,
\label{Eq:Cont of dxG(x,x0,W) at x0}
\end{align}
\begin{align}
\begin{split}
\left.\partial_x\tilde{G}\right|_{x=1^-}&=\left.\partial_x \tilde{G}\right|_{x=1^+}\\
&=\chi_R \omega^2 \left(\left.\tilde{G}\right|_{x=1^-}-\left.\tilde{G}\right|_{x=1^+} \right), 
\end{split}
\label{Eq:Cont of dxG(x,x0,W) at 1}
\end{align}
\begin{align}
\begin{split}
\left.\partial_x \tilde{G}\right|_{x=0^-}&= \left.\partial_x \tilde{G}\right|_{x=0^+}\\
&=\chi_L \omega^2 \left(\left.\tilde{G}\right|_{x=0^-}-\left.\tilde{G}\right|_{x=0^+} \right). 
\end{split}
\label{Eq:Cont of dxG(x,x0,W) at 0}
\end{align}
\end{subequations}

Note that BCs (\ref{Eq:Cont of G(x,x0,W) at x0}-\ref{Eq:Cont of dxG(x,x0,W) at 0}) do not specify what happens to $\tilde{G}(x,x_0,\omega)$ at $x\to \pm \infty$. We model the baths by imposing outgoing BCs at infinity as
\begin{align}
\left. \partial_x \tilde{G}(x,x_0,\omega)\right|_{x\to \pm\infty}=\pm i\omega \tilde{G}(x\to\pm\infty,x_0,\omega),
\label{Eq:Outgoing BC for G(x,x0,W))}
\end{align}
which precludes any reflections from the waveguides to the resonator. 

\subsection{Spectral representation of GF for a closed resonator}
\label{SubApp:Spec Rep of G-closed}
It is helpful to revisit spectral representation of GF for the closed version of our system by setting $\chi_R=\chi_L=0$. This imposes Neumann BC $\partial_x \tilde{G}|_{x=0,1}=0$ and the resulting differential operator becomes Hermitian. The idea of spectral representation is to expand $\tilde{G}$ in terms of a discrete set of normal modes that obey the homogeneous wave equation
\begin{subequations}
\begin{align}
&\partial_x^2\tilde{\Phi}_n(x)+\chi(x,x_0)\omega_n^2\tilde{\Phi}_n(x)=0,\\
\label{Eq:Helmholtz Eq for Phi_n(x)}
&\left.\partial_x \tilde{\Phi}_n(x)\right|_{x=0,1}=0.
\end{align}
\end{subequations}
Then, the real valued eigenfrequencies obey the transcendental equation
\begin{align}
\sin{(\omega_n)}+\chi_s\omega_n\cos{(\omega_n x_0)}\cos{\left[\omega_n (1-x_0)\right]}=0.
\label{Eq:Hermitian Eigenfrequencies}
\end{align}
The eigenfunctions read
\begin{align}
\tilde{\Phi}_n(x)\propto
\begin{cases}
\cos{\left[\omega_n (1-x_0)\right]}\cos{(\omega_n x)},&0<x<x_0\\
\cos{(\omega_n x_0)}\cos{\left[\omega_n (1-x)\right]},&x_0<x<1
\end{cases}
\label{Eq:Sol of Phi_n(x)-closed}
\end{align}
where the normalization is fixed by the orthogonality condition
\begin{align}
\int_{0}^{1}dx\chi(x,x_0)\tilde{\Phi}_m(x)\tilde{\Phi}_n(x)=\delta_{mn}.
\label{Eq:Closed Orthogonality Condition}
\end{align}

Note that eigenfunctions of a Hermitian differential operator form a complete orthonormal basis. This allows us to deduce the spectral representation of $\tilde{G}(x,x',\omega)$ \cite{Morse_Methods_1953, Economou_Green_1984, Hassani_Mathematical_2013} as
\begin{align}
\tilde{G}(x,x',\omega)=\sum\limits_{n\in \mathbb{N}}\frac{\tilde{\Phi}_n(x)\tilde{\Phi}_n(x')}{\omega^2-\omega_n^2}=\sum\limits_{n\in \mathbb{Z}\atop n\neq 0}\frac{1}{2\omega}\frac{\tilde{\Phi}_n(x)\tilde{\Phi}_n(x')}{\omega-\omega_n},
\label{Eq:Spectral rep of G-closed}
\end{align}
where the second representation is written due to relations $\omega_{-n}=-\omega_{n}$ and $\tilde{\Phi}_{-n}(x)=\tilde{\Phi}_{n}(x)$.
\subsection{Spectral representation of GF for an open resonator}
\label{SubApp:Spec Rep of G-open} 
A spectral representation can also be found for the GF of an open resonator in terms of a discrete set of non-Hermitian modes that carry a constant flux away from the resonator. The Constant Flux (CF) modes \cite{Tureci_SelfConsistent_2006} have allowed a consistent formulation of the semiclassical laser theory for complex media such as random lasers \cite{Tureci_Strong_2008}. The non-Hermiticity originates from the fact that the waveguides are assumed to be infinitely long, hence no radiation that is emitted from the resonator to the waveguides can be reflected back. This results in discrete and complex-valued poles of the GF.  The CF modes satisfy the same homogeneous wave equation
\begin{align}
\partial_x^2\tilde{\Phi}_n(x,\omega)+\chi(x,x_0)\omega_n^2(\omega)\tilde{\Phi}_n(x,\omega)=0,
\label{Eq:Helmholtz Eq for Phi_n(x)}
\end{align}
but with open BCs the same as Eqs.~(\ref{Eq:Cont of G(x,x0,W) at x0}-\ref{Eq:Outgoing BC for G(x,x0,W))}). Note that the resulting CF modes $\tilde{\Phi}_n(x,\omega)$ and eigenfrequencies $\omega_n(\omega)$ parametrically depend on the source frequency $\omega$. 

Considering only an outgoing plane wave solution for the left and right waveguides based on (\ref{Eq:Outgoing BC for G(x,x0,W))}), the general solution for $\tilde{\Phi}_n(x,\omega)$ reads
\begin{align}
\tilde{\Phi}_n(x,\omega)=
\begin{cases}
A_{n}^<e^{i\omega_n(\omega) x}+B_{n}^< e^{-i\omega_n(\omega) x},&0<x<x_0\\
A_{n}^>e^{i\omega_n(\omega) x}+B_{n}^> e^{-i\omega_n(\omega) x},&x_0<x<1\\
C_ne^{i\omega x},&x>1 \\
D_ne^{-i\omega x},&x<0 \\
\end{cases}
\label{Eq:General Ansantz for Phi_n(x)}
\end{align}
Applying BCs (\ref{Eq:Cont of G(x,x0,W) at x0}-\ref{Eq:Cont of dxG(x,x0,W) at 0}) leads to a characteristic equation 
\begin{widetext}
\begin{align}
\begin{split}
&\sin\left[\omega_n(\omega)\right]+(\chi_R+\chi_L)\omega_n(\omega)\left\{\cos[\omega_n(\omega)]-\frac{\omega_n(\omega)}{\omega}\sin[\omega_n(\omega)]\right\}\\
&-\chi_R\chi_L\omega_n^2(\omega)\left\{2i\frac{\omega_n(\omega)}{\omega}\cos[\omega_n(\omega)]+\left[1+\frac{\omega_n^2(\omega)}{\omega^2}\right]\sin[\omega_n(\omega)]\right\}\\
&+\chi_s\omega_n(\omega)\left\{\cos[\omega_n(\omega)x_0]-\chi_L\frac{\omega_n(\omega)}{\omega}\left\{i\omega_n(\omega)\cos[\omega_n(\omega)x_0]+\omega\sin[\omega_n(\omega)x_0]\right\}\right\}\\
&\times\left\{\cos[\omega_n(\omega)(1-x_0)]-\chi_R\frac{\omega_n(\omega)}{\omega}\left\{i\omega_n(\omega)\cos[\omega_n(\omega)(1-x_0)]+\omega\sin[\omega_n(\omega)(1-x_0)]\right\}\right\}=0,
\end{split}
\label{Eq:CF NHEigfreq}
\end{align}
which gives the parametric dependence of CF frequencies on $\omega$. Then, the CF modes $\tilde{\Phi}_n(x,\omega)$ are calculated as
\begin{align}
\tilde{\Phi}_n(x,\omega)\propto
\begin{cases}
e^{-i \omega_n(\omega) (x-x_0+1)} \left[e^{2 i \omega_n(\omega) x}+(1-2 i \omega_n(\omega) \chi _L)\right] \left[e^{2 i \omega_n(\omega)(1-x_0)}+\left(1-2 i \omega_n(\omega) \chi_R\right)\right],&0<x<x_0\\
e^{-i \omega_n(\omega) (x_0-x+1)} \left[e^{2 i \omega_n(\omega) x_0}+(1-2 i \omega_n(\omega) \chi _L)\right] \left[e^{2 i \omega_n(\omega)(1-x)}+\left(1-2 i \omega_n(\omega) \chi_R\right)\right],&x_0<x<1\\
-2i\chi_R\omega_n(\omega)e^{-i\omega_n(\omega) (1+x_0)} \left[e^{+2i\omega_n(\omega) x_0}+(1-2i \chi_L \omega_n(\omega))\right]e^{+i\omega x},&x>1 \\
-2i\chi_L\omega_n(\omega)e^{-i\omega_n(\omega)(1-x_0)} \left[e^{2i\omega_n(\omega) (1-x_0)}+(1-2i \chi_R\omega_n(\omega))\right]e^{-i\omega x}. &x<0 \\
\end{cases}
\label{Eq:General NHEigFun}
\end{align}
\end{widetext}
These modes satisfy the biorthonormality condition
\begin{align}
\begin{split}
\int_{0}^{1}dx\chi(x,x_0)\bar{\tilde{\Phi}}_m^*(x,\omega)\tilde{\Phi}_n(x,\omega)=\delta_{mn},
\label{Eq:Open Ortho Cond-unsimp}
\end{split}
\end{align}
where $\{\bar{\tilde{\Phi}}_m(x,\omega)\}$ satisfy the Hermitian adjoint of eigenvalue problem~(\ref{Eq:Helmholtz Eq for Phi_n(x)}). In other words, $\tilde{\Phi}_n(x,\omega)$ and $\bar{\tilde{\Phi}}_n(x,\omega)$ are the right and left eigenfunctions and obey $\bar{\tilde{\Phi}}_n(x,\omega)=\tilde{\Phi}_n^*(x,\omega)$. The normalization of Eq.~(\ref{Eq:General NHEigFun}) is then fixed by setting $m=n$.

In terms of the CF modes, the spectral representation of the GF can then be constructed
\begin{align}
\tilde{G}(x,x',\omega)=\sum\limits_{n}\frac{\tilde{\Phi}_n(x,\omega)\bar{\tilde{\Phi}}_n^*(x',\omega)}{\omega^2-\omega_n^2(\omega)}.
\label{Eq:Spectral rep of G-Open}
\end{align}
Examining Eq.~(\ref{Eq:Spectral rep of G-Open}), we realize that there are two sets of poles of $\tilde{G}(x,x',\omega)$ in the complex $\omega$ plane. First, from setting the denominator of Eq.~(\ref{Eq:Spectral rep of G-Open}) to zero which gives $\omega=\omega_n(\omega)$. These are the quasi-bound eigenfrequencies that satisfy the transcendental characteristic equation  
\begin{align}
\begin{split}
&\left[e^{2i\omega_n}-(1-2i\chi_L\omega_n)(1-2i\chi_R\omega_n)\right]\\
&+\frac{i}{2}\chi_s\omega_n[e^{2i\omega_n x_0}+(1-2i\chi_L\omega_n)]\\
&\times[e^{2i\omega_n (1-x_0)}+(1-2i\chi_R\omega_n)]=0.
\end{split}
\label{Eq:Generic NHEigfreq}
\end{align}
The quasi bound solutions $\omega_n$ to Eq.~(\ref{Eq:Generic NHEigfreq}) reside in the lower half of complex $\omega$-plane and  come in symmetric pairs with respect to the $\Im\{\omega\}$ axis, i.e. both $\omega_{n}$  and $-\omega_{n}^*$ satisfy the transcendental Eq.~(\ref{Eq:Generic NHEigfreq}). Therefore, we can label the eigenfrequencies as
\begin{align}
\omega_n=\begin{cases}
-i\kappa_0, \quad & n=0\\
+\nu_n-i\kappa_n, \quad &n\in+\mathbb{N}\\
-\nu_n-i\kappa_n, \quad &n\in-\mathbb{N}
\end{cases}
\end{align}
where $\nu_n$ and $\kappa_n$ are positive quantities representing the oscillation frequency and decay rate of each quasi-bound mode. Second, there is an extra pole at $\omega=0$ which comes from the $\omega$-dependence of CF states $\tilde{\Phi}_n(x,\omega)$. We confirmed these poles by solving for the explicit solution $\tilde{G}(x,x',\omega)$ that obeys Eq.~(\ref{Eq:Helmholtz Eq for G(x,x0,W)}) with BCs~(\ref{Eq:Cont of G(x,x0,W) at x0}-\ref{Eq:Outgoing BC for G(x,x0,W))}) with Mathematica.
\subsection{Effective dynamics of transmon qubit}
\label{SubApp:Eff Dyn of transmon}

Note that Eqs.~(\ref{Eq:Res Dyn}-\ref{Eq:Side Res Dyn}) are linear in terms of $\hat{\varphi}(x,t)$ and $\hat{\varphi}_{R,L}(x,t)$ . Therefore, it is possible to eliminate these linear degrees of freedom and express the formal solution for $\hat{\varphi}(x,t)$ in terms of $\hat{\varphi}_j(t)$ and $G(x,t|x',t')$. At last, by plugging the result into the RHS of Eq.~(\ref{Eq:Res Dyn}) we find a closed equation for $\hat{\varphi}_j(t)$.

Let us denote the source term that appears on the RHS of Eq.~(\ref{Eq:Res Dyn}) as 
\begin{align}
S\left[\hat{\varphi}_j(t)\right]\equiv \chi_s\omega_j^2\sin{[\hat{\varphi}_j(t)]}.
\end{align}
Then, we write two equations for $\hat{\varphi}(x,t)$ and $G(x,t|x',t')$ \cite{Morse_Methods_1953} (See Sec. $7.3$) as
\begin{subequations}
\begin{align}
&\left[\partial_{x'}^2-\chi(x',x_0)\partial_{t'}^2\right]\hat{\varphi}(x',t')=S\left[\hat{\varphi}_j(t')\right]\delta(x'-x_0), 
\label{Eq:Eff Dyn-wave Eq for varphi_j}\\
&\left[\partial_{x'}^2 -\chi(x,x')\partial_{t'}^2\right]G(x,t|x',t')=\delta(x-x') \delta(t-t').
\label{Eq:Eff Dyn-wave Eq for GF}
\end{align}
\end{subequations}
In Eq.~(\ref{Eq:Eff Dyn-wave Eq for GF}) we have employed the reciprocity property of the GF 
\begin{align}
G(x,t|x',t')=G(x',-t'|x,-t),
\label{Eq:Reciprocity property of Green's function}
\end{align}
which holds since Eq.~(\ref{Eq:Eff Dyn-wave Eq for GF}) is invariant under
\begin{align}
x\leftrightarrow x', \quad t\leftrightarrow-t'.
\label{Eq:Sym of Wave Eq}
\end{align}

Multiplying Eq.~(\ref{Eq:Eff Dyn-wave Eq for varphi_j}) by $G(x,t|x',t')$ and Eq.~(\ref{Eq:Eff Dyn-wave Eq for GF}) by $\hat{\varphi}(x',t')$ and integrating over the dummy variable $x'$ in the interval $[0^-,1^+]$ and over $t'$ in the interval $[0,t^+]$ and finally taking the difference gives
\begin{align}
\begin{split}
&\int_{0}^{t^+}dt' \int_{0^-}^{1^+}dx'\left\{\underbrace{\left(G\partial_{x'}^2\hat{\varphi}-\hat{\varphi}\partial_{x'}^2G\right)}_{\bf(a)}\right.\\
&+\underbrace{\left[\chi(x,x')\hat{\varphi}\partial_{t'}^2G-\chi(x',x_0)G\partial_{t'}^2\hat{\varphi}\right]}_{\bf (b)}\\
&-\left.\underbrace{G S(\hat{\varphi}_j)\delta(x'-x_0)}_{\bf (c)}+\underbrace{\hat{\varphi}\delta(t-t')\delta(x-x')}_{\bf (d)}\right\}=0,
\end{split}
\label{Eq:Intermediate Step Derivation of Green's Foralism}
\end{align}
where we have used the shorthand notation $G\equiv G(x,t|x',t')$ and $\hat{\varphi}\equiv \hat{\varphi}(x',t')$. 

The term labeled as $(a)$ can be simplified further through integration by parts in $x'$ as
\begin{align}
\int_{0}^{t^+}dt' \left.\left(G\partial_{x'}\hat{\varphi}-\hat{\varphi}\partial_{x'}G\right)\right|_{x'=0^-}^{x'=1^+}
\end{align}
There are two contributions from term $(b)$. One comes from the constant capacitance per length in $\chi(x,x')$ and $\chi(x,x_0)$ that simplifies to
\begin{align}
\int_{0^-}^{1^+}\,dx' \left.\left(\hat{\varphi}\partial_{t'}G-G\partial_{t'}\hat{\varphi}\right)\right|_{t'=0},
\end{align}
where due to working with the retarded GF 
\begin{align}
G(x,t|x',t^+)=0,
\end{align}
hence the upper limit $t'=t^+$ vanishes. The second contribution comes from the Dirac $\delta$-functions in $\chi(x,x')$ and $\chi(x,x_0)$ which gives
\begin{align}
\begin{split}
\chi_s\int_{0}^{t^+}dt'&\left[\hat{\varphi}(x,t')\partial_{t'}^2G(x,t|x,t')\right.\\
&\left.-G(x,t|x_0,t')\partial_{t'}^2\hat{\varphi}(x_0,t')\right]
\end{split}
\end{align}
Terms $(c)$ and $(d)$ get simplified due to Dirac $\delta$-functions as
\begin{align}
\int_{0}^{t^+}\,dt'G(x,t|x_0,t')S[\hat{\varphi}_j(t')],
\end{align}
and $\hat{\varphi}(x,t)$, respectively. 

At the end, we find a generic solution for the flux field $\hat{\varphi}(x,t)$ in the domain $[0^-,1^+]$ as
\begin{widetext}
\begin{align}
\begin{split}
&\hat{\varphi}(x,t)=\underbrace{\int_{0}^{t^+}\,dt'G(x,t|x_0,t')S[\hat{\varphi}_j(t')]}_{Source \ Contribution}+\underbrace{\int_{0}^{t^+}\,dt' \left.\left[\hat{\varphi}(x',t')\partial_{x'}G(x,t|x',t')-G(x,t|x',t')\partial_{x'}\hat{\varphi}(x',t')\right]\right|_{x'=0^-}^{x'=1^+}}_{Boundary \ Contribution}\\
&+\underbrace{\int_{0^-}^{1^+}\,dx' \left.\left[\hat{\varphi}(x',t')\partial_{t'}G(x,t|x',t')-G(x,t|x',t')\partial_{t'}\hat{\varphi}(x',t')\right]\right|_{t'=0}}_{Initial \ Condition \ Contribution}\\
&+\underbrace{\chi_s\int_{0}^{t^+}dt'\left[\hat{\varphi}(x,t')\partial_{t'}^2G(x,t|x,t')-G(x,t|x_0,t')\partial_{t'}^2\hat{\varphi}(x_0,t')\right]}_{Feedback \ induced \ by \ transmon}.
\end{split}
\label{Eq:Generic Sol of varphi(x,t)}
\end{align}
\end{widetext}

According to Eq.~(\ref{Eq:Transmon Dyn}), the transmon is forced by the resonator flux field evaluated at $x=x_0$, i.e. $\hat{\varphi}(x_0,t)$. In the following, we rewrite the GF in terms of its Fourier representation for each term in Eq.~(\ref{Eq:Generic Sol of varphi(x,t)}) at $x=x_0$. The Fourier representation simplifies the boundary contribution further, while also allowing us to employ the spectral representation of GF discussed in Sec.~\ref{SubApp:Spec Rep of G-open}.

The source contribution can be written as
\begin{align}
\chi_s\int_{0}^{t}\,dt'\int_{-\infty}^{+\infty} \frac{d\omega}{2\pi} \tilde{G}(x_0,x_0,\omega)\omega_j^2 \sin{[\hat{\varphi}_j(t')]}e^{-i\omega(t-t')}.
\label{Eq:Freq rep of Source}
\end{align}

The boundary terms consist of two separate contributions at each end. Assuming that there is no radiation in the waveguides for $t<0$ we can write

\begin{subequations}
\begin{align}
&\hat{\varphi}_{R,L}(x,t)=\hat{\varphi}_{R,L}(x,t)\Theta(t),
\label{Eq:Causaltiy of Phi_R,L}\\
&\partial_x\hat{\varphi}_{R,L}(x,t)=\partial_x\hat{\varphi}_{R,L}(x,t)\Theta(t).
\label{Eq:Causaltiy of d_x Phi_R,L}
\end{align}
\end{subequations}

Using Eqs.~(\ref{Eq:Causaltiy of Phi_R,L}-\ref{Eq:Causaltiy of d_x Phi_R,L}) and causality of the GF, i.e. $G(x,t|x',t')\propto \Theta(t-t')$, we can extend the integration domain in $t'$ from $[0,t^+]$ to $[-\infty,\infty]$ without changing the value of integral since for an arbitrary integrable function $F(t,t')$, we have 
\begin{align}
\begin{split}
\int_{0}^{t^+}dt' F(t,t')&\theta(t')\theta(t-t')\\
&=\int_{-\infty}^{+\infty}dt' F(t,t')\theta(t')\theta(t-t').
\end{split}
\end{align} 
This extension of integration limits becomes handy when we write both $\hat{\varphi}_R(x',t')$ and $G(x_0,t|x',t')$ in terms of their Fourier transforms in time. Focusing on the right boundary contribution at $x'=1^+$ we get
\begin{align}
\begin{split}
&\int _{-\infty}^{+\infty}dt' \int_{-\infty}^{+\infty}\frac{d\omega_1}{2\pi} \int_{-\infty}^{+\infty}\frac{d\omega_2}{2\pi} \left[\hat{\tilde{\varphi}}_R(x',\omega_1)\partial_{x'}\tilde{G}(x_0,x',\omega_2)\right.
\\
&\left.\left.-\tilde{G}(x_0,x',\omega_2)\partial_{x'}\hat{\tilde{\varphi}}_R(x',\omega_1)\right]\right|_{x'=1^+}e^{-i\omega_1 t'}e^{-i\omega_2 (t-t')}.
\end{split}
\label{Eq:Halfway simplified Boundary contribution}
\end{align}

Next, we write $\hat{\tilde{\varphi}}_R(x',\omega)$ as the sum of ``incoming'' and ``outgoing'' parts
\begin{align}
\hat{\tilde{\varphi}}_R(1^+,\omega_1)=\hat{\tilde{\varphi}}_R^{inc}(1^+,\omega_1)+\hat{\tilde{\varphi}}_R^{out}(1^+,\omega_1),
\label{Eq:Splitting varphi_R(x,t) into out/inc}
\end{align}
defined as
\begin{subequations}
\begin{align}
&\partial_{x'}\hat{\tilde{\varphi}}_R^{out}(x'=1^+,\omega_1)=+i\omega_1\hat{\tilde{\varphi}}_R^{out}(x'=1^+,\omega_1),
\label{Eq:outgoing part of varphi_R(x,t)}\\
&\partial_{x'}\hat{\tilde{\varphi}}_R^{inc}(x'=1^+,\omega_1)=-i\omega_1\hat{\tilde{\varphi}}_R^{inc}(x'=1^+,\omega_1).
\label{Eq:incoming part of varphi_R(x,t)}
\end{align}
\end{subequations}
On the other hand, since we are using a retarded GF with outgoing BC we have
\begin{align}
\partial_{x'}\tilde{G}(x_0,x'=1^+,\omega_2)=+i\omega_2\tilde{G}(x_0,x'=1^+,\omega_2).
\label{Eq:Outgoing BC for G of varphi_R(x,t)}
\end{align}
By substituting Eqs.~(\ref{Eq:outgoing part of varphi_R(x,t)}, (\ref{Eq:incoming part of varphi_R(x,t)}) and (\ref{Eq:Outgoing BC for G of varphi_R(x,t)}) into Eq.~(\ref{Eq:Halfway simplified Boundary contribution}), the integrand becomes
\begin{align}
\begin{split}
&i(\omega_1+\omega_2)\tilde{G}(x_0,1^+,\omega_2)\hat{\tilde{\varphi}}_R^{inc}(1^+,\omega_1)\\&+i(\omega_2-\omega_1)\tilde{G}(x_0,1^+,\omega_2)\hat{\tilde{\varphi}}_R^{out}(1^+,\omega_1) 
\end{split}
\end{align}
By taking the integral in $t'$ as $\int_{-\infty}^{\infty} dt' e^{i(\omega_2-\omega_1)t'}=2\pi\delta(\omega_1-\omega_2)$, Eq.~(\ref{Eq:Halfway simplified Boundary contribution}) can be simplified as
\begin{align}
\int_{-\infty}^{+\infty}\frac{d\omega}{2\pi} \left[2i\omega\tilde{G}(x_0,x'=1^+,\omega)\hat{\tilde{\varphi}}_R^{inc}(0^-,\omega) \right]e^{-i\omega t},
\label{Eq:Freq rep of right drive}
\end{align}
which indicates that only the incoming part of the field leads to a non-zero contribution to the field inside the resonator. A similiar expression holds for the left boundary with the difference that the incoming wave at the left waveguide is ``right-going" in contrast to the right waveguide
\begin{align}
\int_{-\infty}^{+\infty}\frac{d\omega}{2\pi} \left[2i\omega\tilde{G}(x_0,x'=0^-,\omega)\hat{\tilde{\varphi}}_L^{inc}(0^-,\omega) \right]e^{-i\omega t}.
\label{Eq:Freq rep of left drive} 
\end{align}
The initial condition (IC) terms can be expressed in a compact form as
\begin{align}
\begin{split}
\int_{x_1}^{x_2} dx'\int _{-\infty}^{\infty}\frac{d\omega}{2\pi} \left\{\chi(x',x_0)\tilde{G}(x_0,x',\omega)\right.\\
\left.\left[\hat{\dot{\varphi}}(x',0)-i\omega\hat{\varphi}(x',0)\right]\right\}e^{-i\omega t}.
\end{split}
\end{align}
Gathering all the contributions, plugging it in the RHS of Eq.~(\ref{Eq:Transmon Dyn}) and defining a family of memory kernels
\begin{subequations}
\begin{align}
\mathcal{K}_n(\tau)\equiv\gamma\chi_s\int_{-\infty}^{+\infty} \frac{d\omega}{2\pi} \omega^n\tilde{G}(x_0,x_0,\omega)e^{-i\omega\tau},
\label{Eq:Def of K_n(tau)}
\end{align}
and transfer functions
\begin{align}
&\mathcal{D}_R(\omega)\equiv -2i\gamma\omega^3\tilde{G}(x_0,1^+,\omega),
\label{Eq:Def of D_R(om)}\\
&\mathcal{D}_L(\omega)\equiv -2i\gamma\omega^3\tilde{G}(x_0,0^-,\omega),
\label{Eq:Def of D_L(om)}\\
&\mathcal{I}(x',\omega)\equiv \gamma\omega^2\chi(x',x_0)\tilde{G}(x_0,x',\omega),
\label{Eq:Def of D_L(om)}
\end{align}
\end{subequations}
the effective dynamics of the transmon is found to be
\begin{align}
\begin{split}
&\hat{\ddot{\varphi}}_j(t)+(1-\gamma)\omega_j^2\sin{\left[\hat{\varphi}_j(t)\right]}=\\
+&\frac{d^2}{dt^2}\int_{0}^{t}dt'\mathcal{K}_0(t-t')\omega_j^2\sin{\left[\hat{\varphi}_j(t')\right]}\\
+&\int_{-\infty}^{+\infty}\frac{d\omega}{2\pi}\mathcal{D}_R(\omega)\hat{\tilde{\varphi}}_R^{inc}(1^+,\omega)e^{-i\omega t}\\
+&\int_{-\infty}^{+\infty}\frac{d\omega}{2\pi}\mathcal{D}_L(\omega)\hat{\tilde{\varphi}}_L^{inc}(0^-,\omega)e^{-i\omega t}\\
+&\int_{0^-}^{1^+}dx'\int_{-\infty}^{+\infty}\frac{d\omega}{2\pi}\mathcal{I}(x',\omega)\left[i\omega\hat{\varphi}(x',0)-\hat{\dot{\varphi}}(x',0)\right]e^{-i\omega t}.
\end{split}
\label{Eq:Red Dyn before trace}
\end{align}
This is Eq.~(\ref{eqn:Eff Dyn before trace}) in Sec.~\ref{Sec:Eff Dyn Of Transmon}.
\subsection{Effective dynamics for spontaneous emission}
\label{SubApp:SE Eff Dyn}
Equation~(\ref{Eq:Red Dyn before trace}) is the most generic effective dynamics of a transmon coupled to an open multimode resonator. In this section, we find the effective dynamics for the problem of spontaneous emission where the system starts from the IC
\begin{align}
\hat{\rho}(0)=\hat{\rho}_j(0)\otimes \ket{0}_{ph}\bra{0}_{ph}.
\label{Eq:SE-IC}
\end{align}
In the absence of external drive and due to the interaction with the leaky modes of the resonator, the system reaches its ground state $\hat{\rho}_g\equiv\ket{0}_{j}\bra{0}_{j}\otimes \ket{0}_{ph}\bra{0}_{ph}$ in steady state.

Note that due the specific IC~(\ref{Eq:SE-IC}), there is no contribution from IC of the resonator in Eq.~(\ref{Eq:Red Dyn before trace}). To show this explicitly, recall that at $t=0$ the interaction has not turned on and we can represent $\hat{\varphi}(x,0)$ and $\hat{\dot{\varphi}}(x,0)$ in terms of a set of Hermitian modes of the resonator as \cite{Malekakhlagh_Origin_2016}    
\begin{subequations}
\begin{align}
&\hat{\varphi}(x,0)=\hat{\mathbf{1}}_j\otimes\sum\limits_n\left(\frac{\hbar}{2\omega_n^{(H)} cL}\right)^{1/2}\left[\hat{a}_n(0)+\hat{a}_n^{\dag}(0)\right]\tilde{\Phi}_n^{(H)}(x),\\
&\hat{\dot{\varphi}}(x,0)=\hat{\mathbf{1}}_j\otimes\sum\limits_n -i\left(\frac{\hbar \omega_n^{(H)}}{2 cL}\right)^{1/2}\left[\hat{a}_n(0)-\hat{a}_n^{\dag}(0)\right]\tilde{\Phi}_n^{(H)}(x),
\end{align}
\end{subequations}
where we have used superscript notation $(H)$ to distinguish Hermitian from non-Hermitian modes. By taking the partial trace over the photonic sector we find 
\begin{align}
\begin{split}
&\Tr_{ph}\left\{\hat{\rho}_{ph}\left[\hat{a}_n(0)\pm\hat{a}_n^{\dag}(0)\right]\right\}\\
&=\bra{0}_{ph}\left[\hat{a}_n(0)\pm\hat{a}_n^{\dag}(0)\right]\ket{0}_{ph}=0.
\end{split}
\label{Eq:Why IC terms vanish}
\end{align}
With no external drive, $\hat{\tilde{\varphi}}_{R,L}^{inc}$ do not have a coherent part and their expectation value vanish due to the same reasoning as Eq.~(\ref{Eq:Why IC terms vanish}). Therefore, the effective dynamics for the spontaneous emission problem reduces to
\begin{align}
\begin{split}
&\hat{\ddot{\phi}}_j(t)+(1-\gamma)\omega_j^2 \Tr_{ph} \left\{\hat{\rho}_{ph}(0)\sin{\left[\hat{\varphi}_j(t)\right]}\right\}\\
&=\frac{d^2}{dt^2}\int_0^{t}dt' \mathcal{K}_0(t-t') \omega_j^2 \Tr_{ph}\left\{\hat{\rho}_{ph}(0)\sin{\left[\hat{\varphi}_j(t')\right]}\right\}.
\end{split}
\end{align}

Taking the second derivative of the RHS using Leibniz integral rule, and bringing the terms evaluated at the integral limits to the LHS gives
\begin{align}
\begin{split}
&\hat{\ddot{\phi}}_j(t)-\omega_j^2\mathcal{K}_0(0)\Tr_{ph}\left\{\hat{\rho}_{ph}(0)\cos{\left[\hat{\varphi}_j(t)\right]}\hat{\dot{\varphi}}_j(t)\right\}\\
&+\omega_j^2\left[1-\gamma+i\mathcal{K}_1(0)\right]\Tr_{ph}\left\{\hat{\rho}_{ph}(0)\sin{\left[\hat{\varphi}_j(t)\right]}\right\}\\
&=-\int_0^{t}dt'\mathcal{K}_2(t-t')\omega_j^2\Tr_{ph}\left\{\hat{\rho}_{ph}(0)\sin{\left[\hat{\varphi}_j(t')\right]}\right\},
\end{split}
\label{Eq:SE Nonlinear Problem}
\end{align}
where we have used Eq.~(\ref{Eq:Def of K_n(tau)}) to rewrite time-derivatives of $\mathcal{K}_0(\tau)$ in terms of $\mathcal{K}_n(\tau)$.
\subsection{Spectral representation of $\mathcal{K}_0$, $\mathcal{K}_1$ and $\mathcal{K}_2$}
\label{SubApp:Spec Rep of K}
In this section, we express the contributions from the kernels $\mathcal{K}_0(0)$, $\mathcal{K}_1(0)$ and $\mathcal{K}_2(\tau)$ appearing in Eq.~(\ref{Eq:SE Nonlinear Problem}) in terms of the spectral representation of the GF. For this purpose, we use the partial fraction expansion of the GF in agreement with \cite{Leung_Time-independent_1994, Leung_Two-component_1997, 
Servini_Second_2004, Muljarov_Brillouin_2010, 
Doost_Resonant_2013, Kristensen_Normalization_2015} in terms of its simple poles discussed in Sec.~\ref{SubApp:Spec Rep of G-open} as
\begin{align}
\tilde{G}(x,x',\omega)=\sum\limits_{n\in\mathbb{Z}}\frac{1}{2\omega}\frac{\tilde{\Phi}_n(x)\tilde{\Phi}_n(x')}{\omega-\omega_n},
\label{Eq:QB rep of G-Open}
\end{align}
where $\tilde{\Phi}_n(x)\propto \tilde{\Phi}_n(x,\omega=\omega_n)$ is the quasi-bound eigenfunction.

\begin{figure}
\subfloat[\label{subfig:IntegContour1}]{%
\includegraphics[scale=0.40]{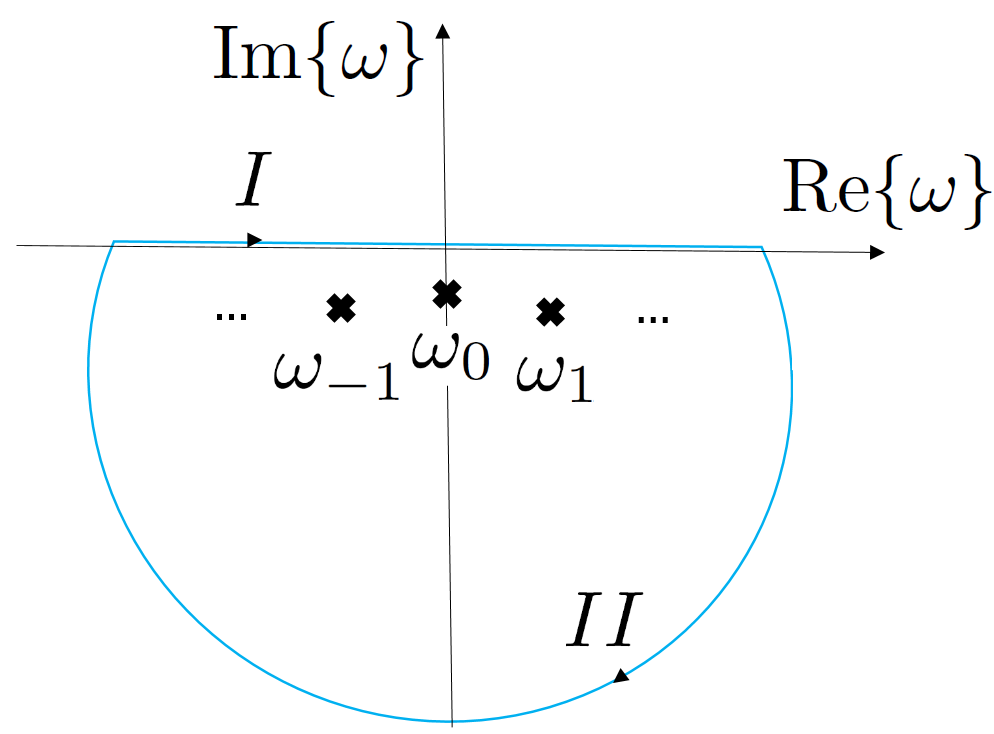}%
}\hfill
\subfloat[\label{subfig:IntegContour2}]{%
\includegraphics[scale=0.40]{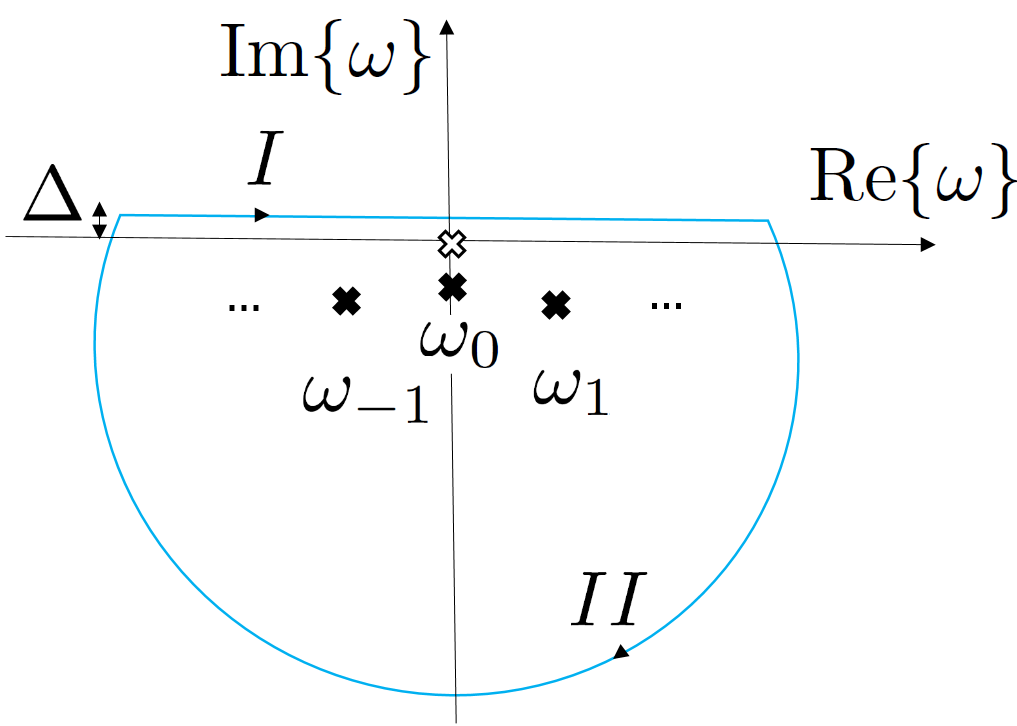}%
}
\caption{Integration contours: a) Integration contour that encloses the poles of $\omega^2\tilde{G}(x_0,x_0,\omega)$ and $\omega\tilde{G}(x_0,x_0,\omega)$; b) integration contour for $\tilde{G}(x_0,x_0,\omega)$, which has an extra pole at $\omega=0$.} 
\label{Fig:Inegration Contour for the Kernel}
\end{figure}

Let us first calculate $\mathcal{K}_2(\tau)$. By choosing an integration contour in the complex $\omega$-plane shown in Fig. \ref{subfig:IntegContour1} and applying Cauchy's residue theorem \cite{Mitrinovic_Cauchy_1984, Hassani_Mathematical_2013} we find
\begin{align}
\begin{split}
&\oint_C d\omega \omega^2 \tilde{G}(x_0,x_0,\omega)e^{-i\omega \tau}\\
&=\int_{I} d\omega \omega^2 \tilde{G}(x_0,x_0,\omega)e^{-i\omega \tau}+\int_{II} d\omega \omega^2 \tilde{G}(x_0,x_0,\omega)e^{-i\omega \tau}\\
&=-2\pi i\sum\limits_{n=0}^{\infty}\frac{1}{2}\left[\omega_n [\tilde{\Phi}_n(x_0)]^2 e^{-i\omega_n\tau}-\omega_n^*[\tilde{\Phi}_n^*(x_0)]^2e^{+i\omega_n^*\tau}\right]\\
&=-2\pi\sum\limits_{n=0}^{\infty}|\omega_n||\tilde{\Phi}_n(x_0)|^2\sin{[\nu_n\tau+\theta_n-2\delta_n(x_0)]}e^{-\kappa_n\tau},
\end{split}
\end{align} 
where due to nonzero opening of the resonator, both $\omega_n$ and $\tilde{\Phi}_n(x)$ are in general complex valued. Therefore, we have defined
\begin{align}
&\theta_n\equiv \arctan{\left(\frac{\kappa_n}{\nu_n}\right)},
\label{Eq:Def. of theta_n}\\
&\delta_n(x)\equiv \arctan{\left(\frac{\Im[{\tilde{\Phi}_n(x)}]}{\Re[{\tilde{\Phi}_n(x)}]}\right)}.
\label{Eq:Def. of delta_n(x)}
\end{align}
As the radius of the half-circle in Fig.~\ref{subfig:IntegContour1} is taken to infinity, $\int_{II}d\omega \omega^2 G(x_0,x_0,\omega)$ approaches zero. This can be checked by a change of variables  
\begin{align}
\omega=R_{II}e^{-i\psi}, \quad \psi \in [0,\pi]\rightarrow d\omega=-iR_{II}e^{-i\psi}d\psi
\end{align}
Substituting this into $\int_{II}$ and taking the limit $R_{II}\to \infty$ gives
\begin{align}
\begin{split}
&\lim\limits_{R_{II}\to\infty}\int_{II}d\omega\omega^2\tilde{G}(x_0,x_0,\omega)e^{-i\omega\tau}\\
&=\sum\limits_{n=0}^{\infty}\lim\limits_{R_{II}\to\infty} \int_{II} d\omega \frac{\omega(\omega+i\kappa_n)[\tilde{\Phi}_n(x_0)]^2}{(\omega-\omega_n)(\omega+\omega_n^*)}e^{-i\omega\tau}\\
&\propto\int_{0}^{\pi}d\psi\lim\limits_{R_{II}\to\infty}e^{-iR_{II}\tau\cos{(\psi)}}R_{II}e^{-R_{II}\tau\sin{(\psi)}}=0, \ \tau>0.
\end{split}
\end{align}
On the other hand, $\int_{I}$ in this limit reads 
\begin{align}
\begin{split}
&\lim\limits_{R_{II}\to\infty}\int_{I}d\omega\omega^2\tilde{G}(x_0,x_0,\omega)e^{-i\omega\tau}\\
&=\int_{-\infty}^{\infty}d\omega\omega^2\tilde{G}(x_0,x_0,\omega)e^{-i\omega\tau},
\end{split}
\end{align}
which is the quantity of interest. Therefore, we find
\begin{align}
\begin{split}
&\int_{-\infty}^{\infty}d\omega\omega^2\tilde{G}(x_0,x_0,\omega)e^{-i\omega\tau}\\
&=-2\pi\sum\limits_{n=0}^{\infty}|\omega_n||\tilde{\Phi}_n(x_0)|^2\sin{\left[\nu_n\tau+\theta_n-2\delta_n(x_0)\right]}e^{-\kappa_n\tau}.
\end{split}
\end{align}
From this, we obtain the spectral representation of $\mathcal{K}_2(\tau)$ as
\begin{align}
&\mathcal{K}_2(\tau)=-\sum\limits_{n=0}^{\infty} A_n \sin{\left[\nu_n\tau+\theta_n-2\delta_n(x_0)\right]}e^{-\kappa_n \tau},
\label{Eq:K^(2)(tau)}
\end{align}
with $A_n\equiv \gamma\chi_s \sqrt{\nu_n^2+\kappa_n^2} \left|\tilde{\Phi}_n(x_0)\right|^2$.

$\mathcal{K}_1(0)$ can be found through similar complex integration  
\begin{align}
\begin{split}
&\oint_C d\omega \omega \tilde{G}(x_0,x_0,\omega)\\
&=\int_{I} d\omega \omega \tilde{G}(x_0,x_0,\omega)+\int_{II} d\omega \omega \tilde{G}(x_0,x_0,\omega)\\
&=-2\pi i\sum\limits_{n=0}^{\infty}\left[\frac{[\tilde{\Phi}_n(x_0)]^2}{2}+\frac{[\tilde{\Phi}_{n}^*(x_0)]^2}{2}\right]\\
&=-2\pi i \sum\limits_{n=0}^{\infty}|\tilde{\Phi}_n(x_0)|^2\cos{[2\delta_{n}(x_0)]}
\end{split}
\end{align}
It can be shown again that $\int_{II}\to 0$ as $R_{II}\to\infty$ from which we find that
\begin{align}
\begin{split}
i\mathcal{K}_1(0)&=\gamma\chi_s\sum\limits_{n=0}^{\infty}|\tilde{\Phi}_n(x_0)|^2\cos{[2\delta_{n}(x_0)]}\\
&=\sum\limits_{n=0}^{\infty}\frac{A_n}{\sqrt{\nu_n^2+\kappa_n^2}}\cos{[2\delta_{n}(x_0)]}.
\label{Eq:iK^(1)(0)}
\end{split}
\end{align}

$\mathcal{K}_0(0)$ has an extra pole at $\omega=0$, so the previous contour is not well defined. Therefore, we shift the integration contour as shown in Fig. \ref{Fig:Inegration Contour for the Kernel}. Then, we have
\begin{align}
\begin{split}
&\oint_C d\omega \tilde{G}(x_0,x_0,\omega)\\
&=\int_{I}d\omega\tilde{G}(x_0,x_0,\omega)+\int_{II}d\omega\tilde{G}(x_0,x_0,\omega)\\
&=-2\pi i\sum\limits_{n=0}^{\infty}\frac{1}{2}\left[\frac{[\tilde{\Phi}_n(x_0)]^2}{\omega_n}-\frac{[\tilde{\Phi}_n^*(x_0)]^2}{\omega_n^*}\right]\\
&-2\pi i\sum\limits_{n=0}^{\infty}\frac{1}{2}\left[\frac{[\tilde{\Phi}_n(x_0)]^2}{-\omega_n}+\frac{[\tilde{\Phi}_n^*(x_0)]^2}{\omega_n^*}\right]=0,\\
\end{split}
\end{align}
where the first sum comes from the residues at $\omega=\omega_n$ and $\omega=-\omega_n^*$, while the last sum is the residue at $\omega=0$ and they completely cancel each other and we get
\begin{align}
\mathcal{K}_0(0)=0.
\label{Eq:K^(0)(0)}
\end{align}

From Eq.~(\ref{Eq:K^(0)(0)}) we find that the effective dynamics for the spontaneous emission problem simplifies to
\begin{align}
\begin{split}
&\hat{\ddot{\phi}}_j(t)+\omega_j^2\left[1-\gamma+i\mathcal{K}_1(0)\right]\Tr_{ph}\left\{\hat{\rho}_{ph}(0)\sin{\left[\hat{\varphi}_j(t)\right]}\right\}\\
&=-\int_0^{t}dt'\mathcal{K}_2(t-t')\omega_j^2\Tr_{ph}\left\{\hat{\rho}_{ph}(0)\sin{\left[\hat{\varphi}_j(t')\right]}\right\}.
\label{Eq:NL simplified SE Problem}
\end{split}
\end{align}
\section{Characteristic function $D_j(s)$ for the linear equations of motion}
\label{App:Char func D(s)}

Up to linear order, transmon acts as a simple harmonic oscillator and we find
we find 
\begin{align}
\begin{split}
\hat{\ddot{X}}_j(t)&+\omega_j^2\left[1-\gamma+i\mathcal{K}_1(0)\right]\hat{X}_j(t)\\
&=-\int_0^{t}dt'\mathcal{K}_2(t-t')\omega_j^2\hat{X}_j(t').
\label{Eq:Lin SE Problem}
\end{split}
\end{align}
Equation~(\ref{Eq:Lin SE Problem}) is a linear integro-differential equation with a memory integral on the RHS, appearing as the convolution of the memory kernel $\mathcal{K}_2$ with earlier values of $\hat{X}_j$. It can be solved by means of unilateral Laplace transform \cite{Abramowitz_Handbook_1964, Korn_Mathematical_2000, Hassani_Mathematical_2013} defined as
\begin{align}
\tilde{f}(s)\equiv \int_{0}^{\infty}dt e^{-st}f(t).
\end{align} 
Employing the following properties of Laplace transform:
\begin{itemize}
\item[1)] Convolution
\begin{align}
\begin{split}
&\mathfrak{L}\left\{\int_0^t dt' f(t')g(t-t')\right\}=\mathfrak{L}\left\{\int_0^t dt' f(t-t')g(t')\right\}\\
&=\mathfrak{L}\left\{f(t)\right\}\cdot \mathfrak{L}\left\{g(t)\right\}=\tilde{f}(s)\tilde{g}(s),
\end{split}
\end{align}
\item[2)] General derivative
\begin{align}
\mathfrak{L}\left\{\frac{d^N}{dt^N}f(t)\right\}=s^N\tilde{f}(s)-\sum\limits_{n=1}^{N}s^{N-n}\left.\frac{d^{n-1}}{dt^{n-1}}f(t)\right|_{t=0},
\end{align}
\end{itemize} 
we can transform the integro-differential Eq.~(\ref{Eq:Lin SE Problem}) into a closed algebraic form in terms of $\hat{\tilde{X}}_j(s)$ as
\begin{align}
\hat{\tilde{X}}_j(s)=\frac{s\hat{X}_j(0)+\hat{\dot{X}}_j(0)}{D_j(s)}=\frac{s\hat{X}_j(0)+\omega_j\hat{Y}_j(0)}{D_j(s)},
\label{Eq:Sol of X_j(s)}
\end{align}
where we have defined
\begin{subequations}
\begin{align}
&D_j(s)\equiv s^2+\Omega^2(s),
\label{Eq:Def of D(s)}\\
&\Omega^2(s)\equiv\omega_j^2\left[1-\gamma+i\mathcal{K}_1(0)+\tilde{\mathcal{K}}_2(s)\right].
\label{Eq:Def of Omega2(s)}
\end{align}
\end{subequations}
and $\hat{Y}_j$ is the normalized charge variable and is canonically conjugate to $\hat{X}_j$ such that $[\hat{X}_j(0),\hat{Y}_j(0)]=2i$.

Note that in order to solve for $\hat{X}_j(t)$ from Eq.~(\ref{Eq:Sol of X_j(s)}), one has to take the inverse Laplace transform of the resulting algebraic form in $s$. This requires studying the denominator first which determines the poles of the entire system up to linear order. Using the expressions for $\mathcal{K}_2(\tau_1)$ and $i\mathcal{K}_1(0)$ given in Eqs.~(\ref{Eq:K^(2)(tau)}) and (\ref{Eq:iK^(1)(0)}) we find
\begin{align}
\begin{split}
&i\mathcal{K}_1(0)+\tilde{\mathcal{K}}_2(s)=\sum\limits_{n\in\mathbb{N}}\frac{A_n}{\sqrt{\nu_n^2+\kappa_n^2}}\cos{[2\delta_{n}(x_0)]}\\
&-\sum\limits_{n \in \mathbb{N}} A_n\frac{\cos{[\theta_n-2\delta_{n}(x_0)]}\nu_n+\sin{[\theta_n-2\delta_{n}(x_0)]}(s+\kappa_n)}{(s+\kappa_n)^2+\nu_n^2}.
\end{split}
\label{Eq:iK1(0)+K2(s)}
\end{align}
Expanding the sine and cosine in the numerator of the second term in Eq.~(\ref{Eq:iK1(0)+K2(s)}) as
\begin{align}
\begin{split}
&\cos{[\theta_n-2\delta_{n}(x_0)]}\nu_n+\sin{[\theta_n-2\delta_{n}(x_0)]}(s+\kappa_n)\\
&=\left\{\cos{(\theta_n)}\cos{[2\delta_{n}(x_0)]}+\sin{(\theta_n)}\sin{[2\delta_{n}(x_0)]}\right\}\nu_n\\
&+\left\{\sin{(\theta_n)}\cos{[2\delta_{n}(x_0)]}-\cos{(\theta_n)}\sin{[2\delta_n(x_0)]}\right\}(s+\kappa_n)\\
&=\frac{\left\{\kappa_n\cos{[2\delta_{n}(x_0)]}-\nu_n\sin{[2\delta_{n}(x_0)]}\right\}s}{\sqrt{\nu_n^2+\kappa_n^2}}\\
&+\frac{(\nu_n^2+\kappa_n^2)\cos{[2\delta_{n}(x_0)]}}{\sqrt{\nu_n^2+\kappa_n^2}},
\end{split}
\end{align} 
Eq.~(\ref{Eq:iK1(0)+K2(s)}) simplifies to
\begin{align}
\begin{split}
&\sum\limits_{n=0}^{\infty}\frac{A_n}{\sqrt{\nu_n^2+\kappa_n^2}}\left\{\cos{[2\delta_{n}(x_0)]}-\frac{(\nu_n^2+\kappa_n^2)\cos{[2\delta_{n}(x_0)]}}{(s+\kappa_n)^2+\nu_n^2}\right.\\
&-\left. \frac{\left\{\kappa_n\cos{[2\delta_{n}(x_0)]}-\nu_n\sin{[2\delta_{n}(x_0)]}\right\}s}{(s+\kappa_n)^2+\nu_n^2}\right\}\\
&=\sum\limits_{n=0}^{\infty}M_n\frac{s\{\cos{[2\delta_{n}(x_0)]}s+\sin{[2\delta_{n}(x_0)]}\nu_n\}}{(s+\kappa_n)^2+\nu_n^2},
\end{split}
\end{align}
where we have defined
\begin{align}
M_n\equiv\frac{A_n}{\sqrt{\nu_n^2+\kappa_n^2}}=\gamma\chi_s|\tilde{\Phi}_n(x_0)|^2.
\label{Eq:Def of Mn}
\end{align}
Therefore, $D_j(s)$ simplifies to
\begin{align}
\begin{split}
&D_j(s)=s^2+\omega_j^2+\\
&\underbrace{\omega_j^2\left\{-\gamma+\sum\limits_{n=0}^{\infty}M_n\frac{s\{\cos{[2\delta_{n}(x_0)]}s+\sin{[2\delta_{n}(x_0)]}\nu_n\}}{(s+\kappa_n)^2+\nu_n^2}\right\}}_{Modification \ due \ to \ memory}.
\label{Eq:simplified D(s)}
\end{split}
\end{align}
\section{Multi-Scale Analysis}
\label{App:MSPT}
In order to understand the application of MSPT on the problem of spontaneous emission, we have broken down its complexity into simpler toy problems, discussing each in a separate subsection. In Sec.~\ref{SubApp:ClDuffingDiss}, we revisit the classical Duffing oscillator problem \cite{Bender_Advanced_1999} in the presence of dissipation, to study the interplay of nonlinearity and dissipation. In Sec.~\ref{SubApp:QuDuffingNoMem}, we discuss the free quantum Duffing oscillator to show how the non-commuting algebra of quantum mechanics alters the classical solution. Finally, in Sec.~\ref{SubApp:QuDuffQuHarm}, we study the full problem and provide the derivation for the MSPT solution~(\ref{eqn:PertCorr-X^(0)(t) MSPT Sol}).
\subsection{Classical Duffing oscillator with dissipation}
\label{SubApp:ClDuffingDiss}
Consider a classical Duffing oscillator 
\begin{align}
\ddot{X}(t)+\delta\,\omega\dot{X}(t)+\omega^2\left[X(t)-\varepsilon X^3(t)\right]=0,
\label{Eq:ClDuffing Osc}
\end{align}
with initial condition $X(0)=X_0$, $\dot{X}(0)=\omega Y_0$. In order to have a bound solution, it is sufficient that the initial energy of the system be less than the potential energy evaluated at its local maxima, $X_{max}\equiv\pm\sqrt{1/3\varepsilon}$ , i.e. $E_0 < U(X_{max})$ which in terms of the initial conditions $X_0$ and $Y_0$ reads
\begin{align}
\frac{1}{2}Y_0^2+\frac{1}{2}\left(X_0^2-\varepsilon X_0^4\right)<\frac{5}{36\varepsilon}.
\end{align}

Note that a naive use of conventional perturbation theory decomposes the solution into a series $X(t)=X^{(0)}(t)+\varepsilon X^{(1)}(t)+\ldots$, which leads to unbounded (secular) solutions in time. In order to illustrate this, consider the simple case where $\delta=0$, $X_0=1$ and $Y_0=0$. Then, we find
\begin{subequations}
\begin{align}
&\mathcal{O}(1):\ddot{X}^{(0)}(t)+\omega^2 X^{(0)}(t)=0,\\
&\mathcal{O}(\varepsilon):\ddot{X}^{(1)}(t)+\omega^2 X^{(1)}(t)=\omega^2[X^{(0)}(t)]^3,
\end{align} 
\end{subequations}
which leads to $X^{(0)}(t)=\cos(\omega t)$ and $X^{(1)}(t)=\frac{1}{32}\cos(\omega t)-\frac{1}{32}\cos(3\omega t)+\frac{3}{8}\omega t \sin(\omega t)$. The latter has a secular contribution that grows unbounded in time.

The secular terms can be canceled order by order by introducing multiple time scales, which amounts to a resummation of the conventional perturbation series \cite{Bender_Advanced_1999}. We assume small dissipation and nonlinearity, i.e. $\delta,\varepsilon \ll 1$. This allows us to define additional slow time scales $\tau\equiv\varepsilon t$ and $\eta\equiv\delta t$ in terms of which we can perform a multi-scale expansion for $X(t)$ as
\begin{subequations}
\begin{align}
\begin{split}
X(t)&=x^{(0)}(t,\tau,\eta)+\varepsilon x^{(1)}(t,\tau,\eta)\\
&+\delta y^{(1)}(t,\tau,\eta)+\mathcal{O}(\varepsilon^2,\delta^2,\varepsilon\delta).
\label{Eq:ClDuffing Expansion of X}
\end{split}
\end{align}
Using the chain rule, the total derivative $d/dt$ is also expanded as
\begin{align}
d_t=\partial_t+\varepsilon \partial_{\tau}+\delta\partial_{\eta}+\mathcal{O}(\varepsilon^2,\delta^2,\varepsilon\delta).
\label{Eq:ClDuffing Expansion of d/dt}
\end{align}
\end{subequations}
Plugging Eqs.~(\ref{Eq:ClDuffing Expansion of X}-\ref{Eq:ClDuffing Expansion of d/dt}) into Eq.~(\ref{Eq:ClDuffing Osc}) and collecting equal powers of $\delta$ and $\epsilon$ we find
\begin{subequations}
\begin{align}
&\mathcal{O}(1):\partial_t^2 x^{(0)}+\omega^2 x^{(0)}=0,
\label{Eq:ClDuffing-O(1)}\\
&\mathcal{O}(\delta):\partial_t^2 y^{(1)}+\omega^2 y^{(1)}=-\omega\partial_tx^{(0)}-2\partial_t\partial_{\eta}x^{(0)},
\label{Eq:ClDuffing-O(del)}\\
&\mathcal{O}(\varepsilon):\partial_t^2 x^{(1)}+\omega^2 x^{(1)}=\omega^2 \left[x^{(0)}\right]^3-2\partial_t\partial_{\tau}x^{(0)}.
\label{Eq:ClDuffing-O(eps)}
\end{align}
\end{subequations}

The general solution to $O(1)$ Eq.~(\ref{Eq:ClDuffing-O(1)}) reads
\begin{align}
x^{(0)}(t,\tau,\eta)=a(\tau,\eta)e^{-i\omega t}+a^*(\tau,\eta)e^{+i\omega t}.
\label{Eq:ClDuffing-O(1) Sol}
\end{align}
Plugging Eq.~(\ref{Eq:ClDuffing-O(1) Sol}) into Eq.~(\ref{Eq:ClDuffing-O(del)}) we find that in order to remove secular terms $a(\tau,\eta)$ satisfies 
\begin{align}
(2\partial_{\eta}+\omega)a(\tau,\eta)=0,
\label{ClDuffing-eta Sec Cond}
\end{align} 
which gives the $\eta$-dependence of $a(\tau,\eta)$ as
\begin{align}
a(\tau,\eta)=\alpha(\tau)e^{-\frac{\omega}{2}\eta}.
\label{Eq:ClDuffing-eta dep of a}
\end{align}

The condition that removes the secular term on the RHS of $\mathcal{O}(\varepsilon)$ Eq.~(\ref{Eq:ClDuffing-O(eps)}) reads
\begin{align}
2i\omega\partial_{\tau}a(\tau,\eta)+3\omega^2|a(\tau,\eta)|^2a(\tau,\eta)=0.
\label{ClDuffing-tau Sec Cond}
\end{align} 
Multipliying Eq.~(\ref{ClDuffing-tau Sec Cond}) by $a^*(\tau,\eta)$ and its complex conjugate by $a(\tau,\eta)$ and taking the difference gives
\begin{align}
\partial_{\tau}|a(\tau,\eta)|^2=0,
\end{align}
which together with Eq.~(\ref{Eq:ClDuffing-eta dep of a}) implies that
\begin{align}
|a(\tau,\eta)|^2=|\alpha(0)|^2e^{-\omega \eta}.
\end{align}
Then, $a(\tau,\eta)$ is found as 
\begin{align}
a(\tau,\eta)=\alpha(0)e^{-\frac{\omega}{2}\eta}e^{i\frac{3}{2}\omega|\alpha(0)|^2e^{-\omega \eta}\tau}.
\label{Eq:ClDuffing-sol of a(tau,eta)}
\end{align}

Replacing $\tau=\varepsilon t$ and $\eta=\delta t$, and, the general solution up to $\mathcal{O}(\varepsilon^2,\delta^2,\varepsilon\delta)$ reads
\begin{align}
\begin{split}
X^{(0)}(t)=x^{(0)}(t,\varepsilon t,\delta t)=e^{-\frac{\kappa}{2}t}\left[\alpha(0)e^{-i\bar{\omega}(t)t}+c.c.\right],
\end{split}
\label{Eq:ClDuffing-Sol of X^(0)(t)}
\end{align}
where we have defined the decay rate $\kappa\equiv\delta. \omega$ and a normalized frequency $\bar{\omega}(t)$ as
\begin{align}
\bar{\omega}(t)\equiv\left[1-\frac{3\varepsilon}{2}|\alpha(0)|^2e^{-\kappa t}\right]\omega.
\label{Eq:ClDuffing-Def of bar(om)(t)}
\end{align}
Furthermore, $\alpha(0)$ is determined based on initial conditions as $\alpha(0)=(X_0+iY_0)/2$. 
\begin{figure}
\subfloat[\label{subfig:XtClDuffingK001eps01}]{%
\includegraphics[scale=0.37]{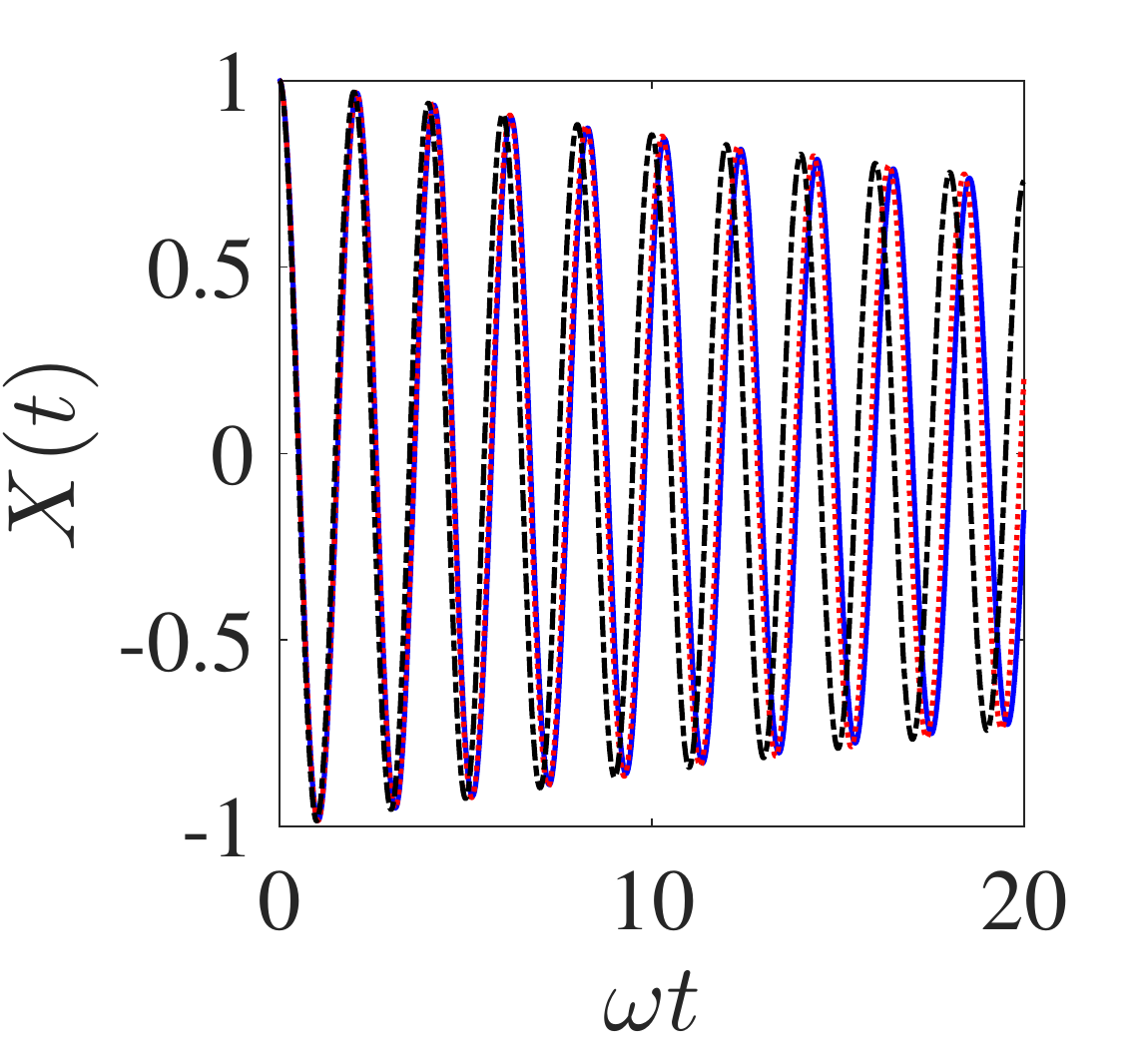}%
}\hfill 
\subfloat[\label{subfig:XtClDuffingK001eps02}]{%
\includegraphics[scale=0.37]{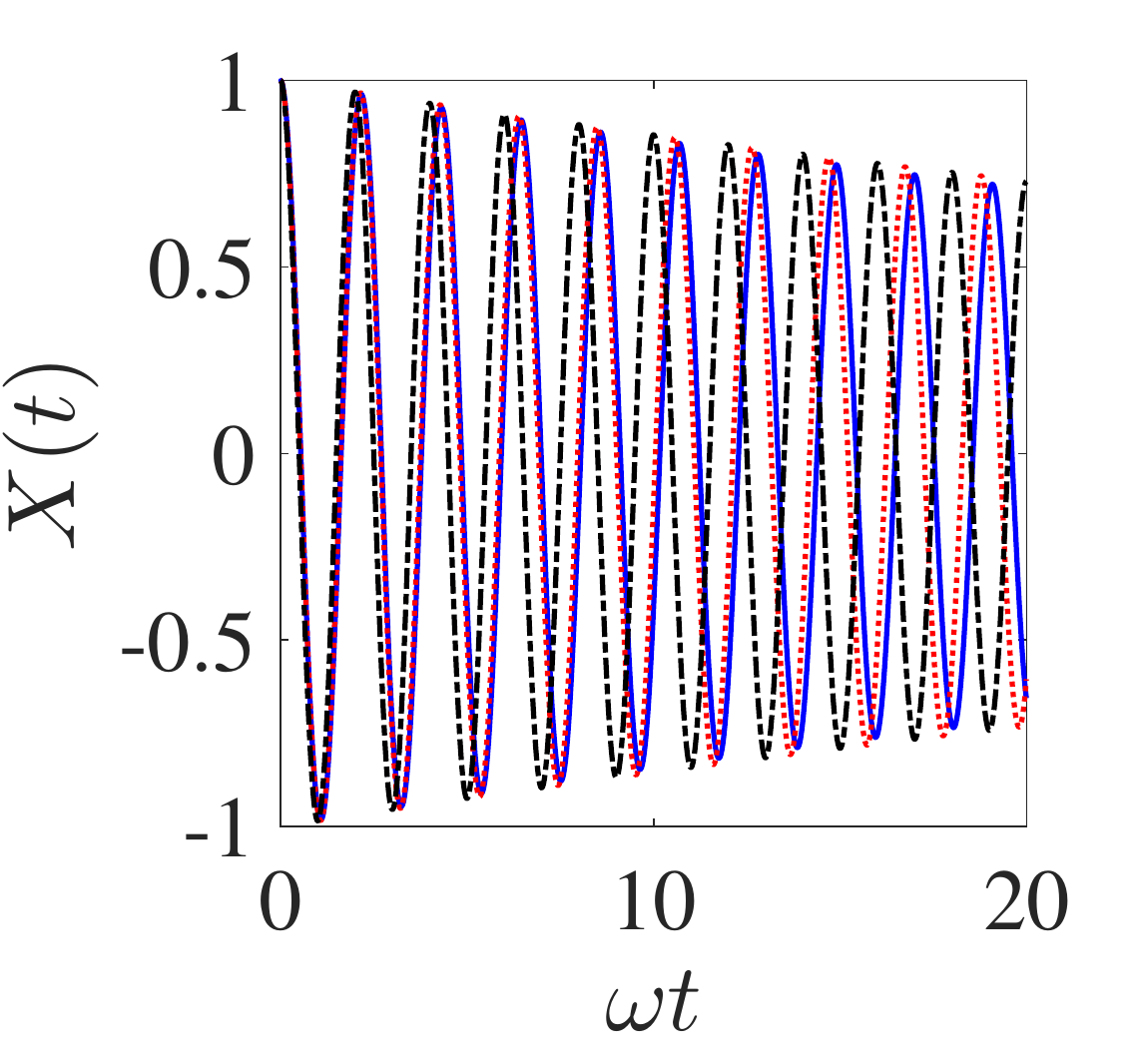}%
}
\caption{(Color online) Comparison of numerical solution (blue solid) with MSPT solution ~(\ref{Eq:ClDuffing-Sol of X^(0)(t)}) (red dotted) and linear solution, i.e. $\varepsilon=0$, (black dash-dot) of Eq.~(\ref{Eq:ClDuffing Osc}) for $\delta=0.01$ and ICs $X_0=1$, $Y_0=0$. a) $\varepsilon=0.1$ , b) $\varepsilon=0.2$.} 
\label{Fig:ClDuffing}
\end{figure}

A comparison between the numerical solution (blue), $\mathcal{O}(1)$ MSPT solution~(\ref{Eq:ClDuffing-Sol of X^(0)(t)}) (red) and linear solution (black) is made in Fig.~\ref{Fig:ClDuffing} for the first ten oscillation periods. The MSPT solution captures the true oscillation frequency better than the linear solution. However, it is only valid for $\omega t\ll \varepsilon^{-2}, \delta^{-2}, \varepsilon^{-1}\delta^{-1}$ up to this order in perturbation theory.   
\subsection{A free quantum Duffing oscillator}
\label{SubApp:QuDuffingNoMem}
Consider a free quantum Duffing oscillator that obeys 
\begin{align}
\hat{\ddot{X}}(t)+\omega^2\left[\hat{X}(t)-\varepsilon \hat{X}^3(t)\right]=0,
\label{Eq:QuDuffingNoMem Osc}
\end{align}
with operator initial conditions
\begin{align}
\hat{X}(0),\quad 
\hat{\dot{X}}(0)=\omega \hat{Y}(0)
\label{Eq:Eq:QuDuffingNoMem ICs}
\end{align}
such that $\hat{X}(0)$ and $\hat{Y}(0)$ are canonically conjugate variables and obey $[\hat{X}(0),\hat{Y}(0)]=2i\hat{\mathbf{1}}$. 

Next, we expand $\hat{X}(t)$ and $d/dt$ up to $\mathcal{O}(\varepsilon^2)$ as
\begin{subequations}
\begin{align}
&\hat{X}(t)=\hat{x}^{(0)}(t,\tau)+\varepsilon \hat{x}^{(1)}(t,\tau)+\mathcal{O}(\varepsilon^2),
\label{Eq:QuDuffingNoMem Expansion of X}\\
&d_t=\partial_t+\varepsilon \partial_{\tau}+\mathcal{O}(\varepsilon^2).
\label{Eq:QuDuffingNoMem Expansion of d/dt}
\end{align}
\end{subequations}
Plugging this into Eq.~(\ref{Eq:QuDuffingNoMem Osc}) and collecting equal powers of $\varepsilon$ gives
\begin{subequations}
\begin{align}
&\mathcal{O}(1):\partial_t^2 \hat{x}^{(0)}+\omega^2 \hat{x}^{(0)}=0,
\label{Eq:QuDuffingNoMem-O(1)}\\
&\mathcal{O}(\varepsilon):\partial_t^2 \hat{x}^{(1)}+\omega^2 \hat{x}^{(1)}=\omega^2 \left[\hat{x}^{(0)}\right]^3-2\partial_t\partial_{\tau}\hat{x}^{(0)}.
\label{Eq:QuDuffingNoMem-O(eps)}
\end{align}
\end{subequations}
Up to $\mathcal{O}(1)$, the general solution reads
\begin{align}
&\hat{x}^{(0)}(t,\tau)=\hat{a}(\tau)e^{-i\omega t}+\hat{a}^{\dag}(\tau)e^{+i\omega t}
\label{Eq:QuDuffingNoMem-O(1) Ansatz}
\end{align}
Furthermore, from the commutation relation $[\hat{x}(t,\tau),\hat{y}(t,\tau)]=2i\hat{\mathbf{1}}$ we find that
$[\hat{a}(\tau),\hat{a}^{\dag}(\tau)]=\hat{\mathbf{1}}$. Substituting Eq.~(\ref{Eq:QuDuffingNoMem-O(1) Ansatz}) into the RHS of Eq.~(\ref{Eq:QuDuffingNoMem-O(eps)}) and setting the secular term oscillating at $\omega$ to zero we obtain
\begin{align}
\begin{split}
&2i\omega\frac{d \hat{a}(\tau)}{d \tau}+\omega^2\left[\hat{a}(\tau)\hat{a}(\tau)\hat{a}^{\dag}(\tau)\right.\\
&\left. +\hat{a}(\tau)\hat{a}^{\dag}(\tau)\hat{a}(\tau)+\hat{a}^{\dag}(\tau)\hat{a}(\tau)\hat{a}(\tau)\right]=0,
\label{Eq:QuDuffingNoMem-Secular Cond 1}
\end{split}
\end{align}
The condition that removes secular term at $-\omega$, appears as Hermitian conjugate of Eq.~(\ref{Eq:QuDuffingNoMem-Secular Cond 1}).

Using $[\hat{a}(\tau),\hat{a}^{\dag}(\tau)]=1$, Eq.~(\ref{Eq:QuDuffingNoMem-Secular Cond 1}) can be rewritten in a compact form
\begin{align}
\frac{d \hat{a}(\tau)}{d \tau}-i\frac{3\omega}{4}\left[\hat{\mathcal{H}}(\tau)\hat{a}(\tau)+\hat{a}(\tau)\hat{\mathcal{H}}(\tau)\right]=0,
\label{Eq:QuDuffingNoMem-Sec Cond  Simpl}
\end{align}
where 
\begin{align}
\hat{\mathcal{H}}(\tau)\equiv \frac{1}{2}\left[\hat{a}^{\dag}(\tau)\hat{a}(\tau)+\hat{a}(\tau)\hat{a}^{\dag}(\tau)\right].
\label{Eq:QuDuffingNoMem-Def of H}
\end{align}

Next, we show that $\hat{\mathcal{H}}(\tau)$ is a conserved quantity. Pre- and post-multiplying Eq.~(\ref{Eq:QuDuffingNoMem-Secular Cond 1}) by $\hat{a}^{\dag}(\tau)$, pre- and post-multiplying Hermitian conjugate of Eq.~(\ref{Eq:QuDuffingNoMem-Secular Cond 1}) by $\hat{a}(\tau)$ and adding all the terms gives
\begin{align}
\frac{d\hat{\mathcal{H}}(\tau)}{d\tau}=0,
\end{align}
which implies that $\hat{\mathcal{H}}(\tau)=\hat{\mathcal{H}}(0)$. Therefore, we find the solution for $\hat{a}(\tau)$ as
\begin{align}
\hat{a}(\tau)=\mathcal{W}\left\{\hat{a}(0)\exp\left[+i\frac{3\omega}{2}\hat{\mathcal{H}}(0)\tau\right]\right\},
\label{Eq:QuDuffingNoMem-a_j(tau_1) sol}
\end{align}
where $\mathcal{W}\{\bullet\}$ represents Weyl-ordering of operators \cite{Schleich_Quantum_2011}. The operator ordering $\mathcal{W}\left\{\hat{a}(0)f\left(\hat{\mathcal{H}}(0)\tau\right)\right\}$ is defined as follows:
\begin{enumerate}
\item
Expand $f\left(\hat{\mathcal{H}}(0)\tau\right)$ as a Taylor series in powers of operator $\hat{\mathcal{H}}(0)\tau$, \\
\item
Weyl-order the series term-by-term as $\mathcal{W}\left\{\hat{a}(0)\left[\hat{\mathcal{H}}(0)\right]^n\right\}\equiv\frac{1}{2^n}\sum\limits_{m=0}^{n} {{n}\choose{m}} \left[\hat{\mathcal{H}}(0)\right]^m \hat{a}(0)\left[\hat{\mathcal{H}}(0)\right]^{n-m}$.
\end{enumerate}
The formal solution~(\ref{Eq:QuDuffingNoMem-a_j(tau_1) sol}) can be re-expressed in a closed form \cite{Bender_Resolution_1986, Bender_Continuous_1987, Bender_Polynomials_1988, Bender_Multiple_1996} using the properties of Euler polynomials \cite{Abramowitz_Handbook_1964} as 
\begin{align}
\hat{a}(\tau)=\frac{\hat{a}(0)e^{i\frac{3\omega}{2}\hat{\mathcal{H}}(0)\tau}+e^{i\frac{3\omega}{2}\hat{\mathcal{H}}(0)\tau}\hat{a}(0)}{2\cos\left(\frac{3\omega\tau}{4}\right)}.
\label{Eq:QuDuffingNoMem-Weyl Ord Simplified}
\end{align}
Plugging Eq.~(\ref{Eq:QuDuffingNoMem-Weyl Ord Simplified}) into Eq.~(\ref{Eq:QuDuffingNoMem-O(1) Ansatz}) and substituting $\tau=\varepsilon t$, we find the solution for $\hat{X}(t)$ up to $\mathcal{O}(\varepsilon)$ as
\begin{align}
\begin{split}
\hat{X}^{(0)}(t)=\hat{x}^{(0)}(t,\varepsilon t)&=\frac{\hat{a}(0)e^{-i\hat{\bar{\omega}} t}+e^{-i\hat{\bar{\omega}} t}\hat{a}(0)}{2\cos\left(\frac{3\omega}{4}\varepsilon t\right)}\\
&+\frac{\hat{a}^{\dag}(0)e^{+i\hat{\bar{\omega}} t}+e^{+i\hat{\bar{\omega}} t}\hat{a}^{\dag}(0)}{2\cos\left(\frac{3\omega}{4}\varepsilon t\right)},
\label{Eq:QuDuffingNoMem-X^(0)(t) sol}
\end{split}
\end{align}
where $\hat{\bar{\omega}}\equiv \omega[1-\frac{3\varepsilon}{2}\hat{\mathcal{H}}(0)]$ appears as a renormalized frequency operator. 

The physical quantity of interest is the expectation value of $\hat{X}^{(0)}(t)$ with respect to the initial density matrix $\hat{\rho}(0)$. The number basis of the simple harmonic oscillator is a complete basis for the Hilbert space of the Duffing oscillator such that
\begin{align}
\hat{\rho}(0)=\sum\limits_{mn}c_{mn}\ket{m}\bra{n}.
\end{align}
Therefore, calculation of $\braket{\hat{X}^{(0)}(t)}$ reduces to calculating the matrix element $\bra{m}\hat{a}(\varepsilon t)\ket{n}$. From Eq.~(\ref{Eq:QuDuffingNoMem-Weyl Ord Simplified}) we find that the only nonzero matrix element read
\begin{align}
\begin{split}
\bra{n-1}\hat{a}(\varepsilon t)\ket{n}&=\frac{\bra{n-1}\hat{a}(0)\ket{n}e^{i\frac{3\varepsilon\omega}{2}\bra{n}\hat{\mathcal{H}}(0)\ket{n}}}{2\cos\left(\frac{3\varepsilon\omega}{4}t\right)}\\
&+\frac{e^{i\frac{3\varepsilon\omega}{2}\bra{n-1}\hat{\mathcal{H}}(0)\ket{n-1}}\bra{n-1}\hat{a}(0)\ket{n}}{2\cos\left(\frac{3\varepsilon\omega}{4}t\right)}\\
&=\bra{n-1}\hat{a}(0)\ket{n}e^{i\frac{3n\varepsilon\omega}{2}t},
\end{split}
\end{align}
where we used that $\bra{n}\hat{\mathcal{H}}(0)\ket{n}=n+1/2$ is diagonal in the number basis.
\subsection{Quantum Duffing oscillator coupled to a set of quantum harmonic oscillators}
\label{SubApp:QuDuffQuHarm}
Quantum MSPT can also be applied to the problem of a quantum Duffing oscillator coupled to multiple harmonic oscillators. For simplicity, consider the toy Hamiltonian
\begin{align}
\begin{split}
\hat{\mathcal{H}} &\equiv \frac{\omega_j}{4}\left(\hat{\mathcal{X}}_j^2+\hat{\mathcal{Y}}_j^2-\frac{\varepsilon}{2}\hat{\mathcal{X}}_j^4\right)\\
&+\frac{\omega_c}{4}\left(\hat{\mathcal{X}}_c^2+\hat{\mathcal{Y}}_c^2\right)+g\hat{\mathcal{Y}}_j\hat{\mathcal{Y}}_c,
\end{split}
\label{Eq:QuDuffQuHarm-ToyH}
\end{align}
where the nonlinearity only exists in the Duffing sector of the Hilbert space labeled as $j$. Due to linear coupling there will be a hybridization of modes up to linear order. Therefore, Hamiltonian (\ref{Eq:QuDuffQuHarm-ToyH}) can always be rewritten in terms of the normal modes of its quadratic part as
\begin{align}
\begin{split}
\hat{\mathcal{H}} &\equiv \frac{\beta_j}{4}\left(\hat{\bar{\mathcal{X}}}_j^2+\hat{\bar{\mathcal{Y}}}_j^2\right)+\frac{\beta_c}{4}\left(\hat{\bar{\mathcal{X}}}_c^2+\hat{\bar{\mathcal{Y}}}_c^2\right)\\
&-\frac{\varepsilon\omega_j}{8}\left(u_j\hat{\bar{\mathcal{X}}}_j+u_c\hat{\bar{\mathcal{X}}}_c\right)^4,
\end{split}
\label{Eq:QuDuffQuHarm-ToyH NormModes}
\end{align}
where $u_{j,c}$ are real hybridization coefficients and $\hat{\bar{\mathcal{X}}}_{j,c}$ and $\hat{\bar{\mathcal{Y}}}_{j,c}$ represent $j$-like and $c$-like canonical operators. For $g=0$, $u_j\rightarrow 1$, $u_c\rightarrow 0$, $\hat{\bar{\mathcal{X}}}_{j,c}\rightarrow \hat{\mathcal{X}}_{j,c}$ and $\hat{\bar{\mathcal{Y}}}_{j,c}\rightarrow \hat{\mathcal{Y}}_{j,c}$. To find $u_{j,c}$ consider the Heisenberg equations of motion
\begin{subequations}
\begin{align}
&\hat{\ddot{\mathcal{X}}}_j(t)+\omega_j^2\hat{\mathcal{X}}_j(t)=-2g\omega_c\hat{\mathcal{X}}_c(t),
\label{Eq:QuDuffQuHarm-ddot(X)_j}\\
&\hat{\ddot{\mathcal{X}}}_c(t)+\omega_c^2\hat{\mathcal{X}}_c(t)=-2g\omega_j\hat{\mathcal{X}}_j(t).
\label{Eq:QuDuffQuHarm-ddot(X)_c}
\end{align}
\end{subequations}
Expressing $\vec{\mathcal{X}}\equiv(\hat{\mathcal{X}}_j \ \hat{\mathcal{X}}_c)^T$, the system above can be written as $\vec{\ddot{\mathcal{X}}}+V\vec{\mathcal{X}}=0$, where $V$ is a $2\times 2$ matrix. Plugging an Ansatz $\vec{\mathcal{X}}=\vec{\mathcal{X}}_0e^{i\lambda t}$ leads to an eigensystem whose eigenvalues are $\beta_{j,c}$ and whose eigenvectors give the hybridization coefficients $u_{j,c}$. 

The Heisenberg equations of motion for the hybdridized modes $\hat{\bar{\mathcal{X}}}_{l}(t)$, $l\equiv j,c$, reads
\begin{align}
\begin{split}
\hat{\ddot{\bar{\mathcal{X}}}}_{l}(t)+\beta_{l}^2\left\{\hat{\bar{\mathcal{X}}}_{l}(t)-\varepsilon_{l} \left[u_j\hat{\bar{\mathcal{X}}}_j(t)+u_c\hat{\bar{\mathcal{X}}}_c(t)\right]^3\right\}=0,
\end{split}
\label{Eq:QuDuffQuHarm Osc}
\end{align}
where due to hybridization, each oscillator experiences a distinct effective nonlinearity as $\varepsilon_{l}\equiv\frac{\omega_j}{\beta_{l}}u_{l}\varepsilon$. Therefore, we define two new time scales $\tau_{l}\equiv\varepsilon_{l}t$ in terms of which we can expand
\begin{subequations}
\begin{align}
\begin{split}
\hat{\bar{\mathcal{X}}}_{l}(t)&=\hat{\bar{x}}_{l}^{(0)}(t,\tau_j,\tau_c)+\varepsilon_{l}\hat{\bar{x}}_{l}^{(1)}(t,\tau_j,\tau_c)\\
&+\varepsilon_{l'}\hat{\bar{y}}_{l}^{(1)}(t,\tau_j,\tau_c)+\mathcal{O}(\varepsilon_j^2,\varepsilon_c^2,\varepsilon_j\varepsilon_c),
\end{split}
\label{Eq:QuDuffQuHarm-Expansion of X}
\end{align}
\begin{align}
d_t=\partial_t+\varepsilon _j\partial_{\tau_j}+\varepsilon_c\partial_{\tau_c}+\mathcal{O}(\varepsilon_j^2,\varepsilon_c^2,\varepsilon_j\varepsilon_c).
\label{Eq:QuDuffQuHarm Expansion of d/dt}
\end{align}
\end{subequations}
where we have used the notation that if $l=j$, $l'=c$ and vice versa. Up to $\mathcal{O}(1)$ we find
\begin{align}
\mathcal{O}(1):\partial_t^2 \hat{\bar{x}}_{l}^{(0)}+\beta_{l}^2 \hat{\bar{x}}_{l}^{(0)}=0,
\label{Eq:QuDuffQuHarm-O(1)}
\end{align}
whose general solution reads
\begin{align}
\begin{split}
\hat{\bar{x}}_{l}^{(0)}(t,\tau_j,\tau_c)&=\hat{\bar{a}}_{l}(\tau_j,\tau_c)e^{-i\beta_{l}t}\\
&+\hat{\bar{a}}_{l}^{\dag}(\tau_j,\tau_c)e^{+i\beta_{l}t}.
\end{split}
\end{align}
where
\begin{align}
[\hat{\bar{a}}_{l_1},\hat{\bar{a}}_{l_2}^{\dag}]=\delta_{l_1l_2}\hat{\mathbf{1}}, \ [\hat{\bar{a}}_{l_1},\hat{\bar{a}}_{l_2}]=[\hat{\bar{a}}_{l_1}^{\dag},\hat{\bar{a}}_{l_2}^{\dag}]=0.
\label{Eq:QuDuffQuHarm-Commut Rels} 
\end{align}
There are $\mathcal{O}(\varepsilon_{l})$ and $\mathcal{O}(\varepsilon_{l'})$ equations of for each normal mode as
\begin{subequations}
\begin{align}
\begin{split}
\mathcal{O}(\varepsilon_{l}) \ of \ l: &\partial_t^2 \hat{\bar{x}}_{l}^{(1)}+\beta_{l}^2 \hat{\bar{x}}_{l}^{(1)}=-2\partial_t\partial_{\tau_{l}}\hat{\bar{x}}_{l}^{(0)}\\
&-\beta_{l}^2\left[u_j\hat{\bar{x}}_j^{(0)}+u_c\hat{\bar{x}}_{c}^{(0)}\right]^3=0,
\end{split}
\label{Eq:QuDuffQuHarm-O(eps_pm) of pm}
\end{align}
\begin{align}
\mathcal{O}(\varepsilon_{l'}) \ of \ l:\partial_t^2 \hat{\bar{y}}_{l}^{(1)}+\beta_{l}^2 \hat{\bar{y}}_{l}^{(1)}=-2\partial_t\partial_{\tau_{l'}}\hat{\bar{x}}_{l}^{(0)}.
\label{Eq:QuDuffQuHarm-O(eps_mp) of pm}
\end{align}
\end{subequations}
By setting the secular terms on the RHS of Eq.~(\ref{Eq:QuDuffQuHarm-O(eps_mp) of pm}) we find that $\partial_{\tau_{l'}}\hat{b}_{l}=0$ which means that $q$ and $c$ sectors are only modified with their own time scale, i.e. $\hat{\bar{a}}_{l}=\hat{\bar{a}}_{l}(\tau_{l})$. Applying the same procedure on Eq.~(\ref{Eq:QuDuffQuHarm-O(eps_pm) of pm}) and using commutation relations (\ref{Eq:QuDuffQuHarm-Commut Rels}) we find
\begin{align}
\begin{split}
\frac{d \hat{\bar{a}}_{l}}{d \tau_{l}}-i\frac{3\beta_{l}}{4}&\left\{u_{l}^3\left[\hat{\bar{\mathcal{H}}}_{l}\hat{\bar{a}}_{l}+\hat{\bar{a}}_{l}\hat{\bar{\mathcal{H}}}_{l}\right]\right. \\
&+\left. 2u_{l}u_{l'}^2\left[\hat{\bar{\mathcal{H}}}_{l'}\hat{\bar{a}}_{l}+\hat{\bar{a}}_{l}\hat{\bar{\mathcal{H}}}_{l'}\right]\right\}=0,
\end{split}
\label{Eq:QuDuffQuHarm-Sec Cond}
\end{align}
where
\begin{align}
\hat{\bar{\mathcal{H}}}_{l}(\tau_{l})\equiv \frac{1}{2}\left[\hat{\bar{a}}_{l}^{\dag}(\tau_{l})\hat{\bar{a}}_{l}(\tau_{l})+\hat{\bar{a}}_{l}(\tau_{l})\hat{\bar{a}}_{l}^{\dag}(\tau_{l})\right].
\label{Eq:QuDuffQuHarm-Def of H_pm}
\end{align}
By pre- and post-multiplying Eq.~(\ref{Eq:QuDuffQuHarm-Def of H_pm}) by $\hat{\bar{a}}_{l}^{\dag}(\tau_{l})$ and its Hermitian conjugate by $\hat{\bar{a}}_{l}(\tau_{l})$ and adding them we find that
\begin{align}
\frac{d\hat{\bar{\mathcal{H}}}_{l}(\tau_{l})}{d\tau_{l}}=0,
\end{align}
which means that the sub-Hamiltonians of each normal mode remain a constant of motion up to this order in perturbation. Therefore, in terms of effective Hamiltonians
\begin{align}
\hat{\bar{h}}_{l}(0)\equiv u_{l}^3\hat{\bar{\mathcal{H}}}_{l}(0)+2 u_{l}u_{l'}^2\hat{\bar{\mathcal{H}}}_{l'}(0),
\label{Eq:QuDuffQuHarm-Eff h}
\end{align}
Eq.~(\ref{Eq:QuDuffQuHarm-Sec Cond Simpl}) simplifies to
\begin{align}
\frac{d \hat{\bar{a}}_{l}}{d \tau_{l}}-i\frac{3\beta_{l}}{4}\left[\hat{\bar{h}}_{l}(0)\hat{\bar{a}}_{l}+\hat{\bar{a}}_{l}\hat{\bar{h}}_{l}(0)\right]=0.
\label{Eq:QuDuffQuHarm-Sec Cond Simpl}
\end{align}
Equation~(\ref{Eq:QuDuffQuHarm-Sec Cond Simpl}) has the same form as Eq.~(\ref{Eq:QuDuffingNoMem-Sec Cond  Simpl}) while the Hamiltonian $\mathcal{H}(0)$ is replaced by an effective Hamiltonian $\hat{\bar{h}}_{l}(0)$. Therefore, the formal solution is found as the Weyl ordering 
\begin{align}
\hat{\bar{a}}_{l}(\tau)=\mathcal{W}\left\{\hat{\bar{a}}_{l}(0)\exp\left[+i\frac{3\beta_{l}}{2}\hat{\bar{h}}_{l}(0)\tau_{l}\right]\right\}.
\label{Eq:QuDuffQuHarm-b_(eta) Sol}
\end{align}
Note that since $[\hat{\bar{a}}_{l},\hat{\bar{\mathcal{H}}}_{l'}(0)]=0$, the Weyl ordering only acts partially on the Hilbert space of interest which results in a closed form solution
\begin{align}
\hat{\bar{a}}_{l}(\tau_{l})=\frac{\hat{\bar{a}}_{l}(0)e^{i\frac{3\beta_{l}}{2}\hat{\bar{h}}_{l}(0)\tau_{l}}+e^{i\frac{3\beta_{l}}{2}\hat{\bar{h}}_{l}(0)\tau_{l}}\hat{\bar{a}}_{l}(0)}{2\cos\left(\frac{3 u_{l}^3\beta_{l}\tau_{l}}{4}\right)}.
\label{Eq:QuDuffQuHarm-Weyl Ord Simplified}
\end{align}
At last, $\hat{\bar{\mathcal{X}}}_{l}^{(0)}(t)$ is found by replacing $\tau_{l}=\varepsilon_{l} t$ as
\begin{align}
\begin{split}
\hat{\bar{\mathcal{X}}}_{l}^{(0)}(t)=\hat{\bar{x}}_{l}^{(0)}(t,\varepsilon_{l}t)&=\frac{\hat{\bar{a}}_{l}(0)e^{-i\hat{\bar{\beta}}_{l} t}+e^{-i\hat{\bar{\beta}}_{l} t}\hat{\bar{a}}_{l}(0)}{2\cos\left(\frac{3 u_{l}^3\beta_{l}\varepsilon_{l}}{4}t\right)}\\
&+\frac{\hat{\bar{a}}_{l}^{\dag}(0)e^{+i\hat{\bar{\beta}}_{l} t}+e^{+i\hat{\bar{\beta}}_{l} t}\hat{\bar{a}}^{\dag}(0)}{2\cos\left(\frac{3 u_{l}^3\beta_{l}\varepsilon_{l}}{4}t\right)},
\end{split}
\label{Eq:QuDuffQuHarm-Q_(eta)^(0)(t) sol}
\end{align}
where $\hat{\bar{\beta}}_{l}\equiv \beta_{l}\left[1-\frac{3\varepsilon_{l}}{2}\hat{\bar{h}}_{l}(0)\right]$. Plugging the expressions for $\varepsilon_{l}$ and $\hat{\bar{h}}_{l}(0)$, we find the explicit operator renormalization of each sector as
\begin{subequations}
\begin{align}
&\hat{\bar{\beta}}_j=\beta_j-\frac{3\varepsilon}{2}\omega_j\left[u_j^4\hat{\bar{\mathcal{H}}}_j(0)+2u_j^2u_c^2\hat{\bar{\mathcal{H}}}_c(0)\right],
\label{Eq:QuDuffQuHarm-bar(nu)_j}\\
&\hat{\bar{\beta}}_{c}=\beta_{c}-\frac{3\varepsilon}{2}\omega_j\left[u_c^4\hat{\bar{\mathcal{H}}}_c(0)+2u_c^2u_j^2\hat{\bar{\mathcal{H}}}_j(0)\right].
\label{Eq:QuDuffQuHarm-bar(nu)_c}
\end{align}
\end{subequations}
Equations.~(\ref{Eq:QuDuffQuHarm-bar(nu)_j}-\ref{Eq:QuDuffQuHarm-bar(nu)_c}) are symmetric under $j\leftrightarrow c$, implying that in the normal mode picture all modes are renormalized in the same manner. The terms proportional to $u_{j,c}^4$ and $u_{j,c}^2u_{c,j}^2$ are the self-Kerr and cross-Kerr contributions, respectively.

This analysis can be extended to the case of a Duffing oscillator coupled to multiple harmonic oscillators without further complexity, since the Hilbert spaces of the distinct normal modes do not mix to lowest order in MSPT. Consider the full Hamiltonian of our cQED system as
\begin{align}
\begin{split}
\hat{\mathcal{H}} &\equiv \frac{\omega_j}{4}\left(\hat{\mathcal{X}}_j^2+\hat{\mathcal{Y}}_j^2-\frac{\varepsilon}{2}\hat{\mathcal{X}}_j^4\right)\\
&+\sum\limits_{n}\frac{\omega_n}{4}\left(\hat{\mathcal{X}}_n^2+\hat{\mathcal{Y}}_n^2\right)+\sum\limits_{n}g_n\hat{\mathcal{Y}}_j\hat{\mathcal{Y}}_n,
\end{split}
\label{Eq:QuDuffQuHarm-H}
\end{align}
where here we label transmon operators with $j$ and all modes of the cavity by $n$. The coupling $g_n$ between transmon and the modes is given as \cite{Malekakhlagh_Origin_2016}
\begin{align}
g_n=\frac{1}{2}\gamma\sqrt{\chi_j}\sqrt{\omega_j\omega_n}\tilde{\Phi}_n(x_0).
\end{align}
Then, the Hamiltonian can be rewritten in a new basis that diagonalizes the quadratic part as
\begin{align}
\begin{split}
\hat{\mathcal{H}} &\equiv \frac{\beta_j}{4}\left(\hat{\bar{\mathcal{X}}}_j^2+\hat{\bar{\mathcal{Y}}}_j^2\right)+\sum\limits_n\frac{\beta_n}{4}\left(\hat{\bar{\mathcal{X}}}_n^2+\hat{\bar{\mathcal{Y}}}_n^2\right)\\
&-\frac{\varepsilon\omega_j}{8}\left(u_j \hat{\bar{\mathcal{X}}}_j+\sum\limits_nu_n \hat{\bar{\mathcal{X}}}_n\right)^4.
\end{split}
\label{Eq:QuDuffQuHarm-H NormModes}
\end{align}
The procedure to arrive at $u_{j,c}$ and $\beta_{j,c}$ is a generalization of the one presented under Eqs.~(\ref{Eq:QuDuffQuHarm-ddot(X)_j}-\ref{Eq:QuDuffQuHarm-ddot(X)_c}).

The Heisenberg dynamics of each normal mode is then obtained as
\begin{align}
\begin{split}
\hat{\ddot{\bar{\mathcal{X}}}}_{l}(t)+\beta_{l}^2\left\{\hat{\bar{\mathcal{X}}}_{l}(t)-\varepsilon_{l} \left[u_j\hat{\bar{\mathcal{X}}}_j(t)+\sum\limits_n u_n\hat{\bar{\mathcal{X}}}_n(t)\right]^3\right\}=0,
\end{split}
\label{Eq:QuDuffQuHarm Osc}
\end{align}
where $\varepsilon_{l}\equiv\frac{\omega_j}{\beta_{l}}u_{l}\varepsilon$ for $l\equiv j,n$. Up to lowest order in perturbation, the solution for $\hat{\bar{\mathcal{X}}}_{l}^{(0)}(t)$ has the exact same form as Eq.~(\ref{Eq:QuDuffQuHarm-Q_(eta)^(0)(t) sol}) with operator renormalization $\hat{\bar{\beta}}_{j}$
\begin{subequations}
\begin{align}
\hat{\bar{\beta}}_{j}=\beta_{j}-\frac{3\varepsilon}{2}\omega_j\left[u_j^4\hat{\bar{\mathcal{H}}}_j(0)+\sum\limits_{n}2u_j^2u_n^2\hat{\bar{\mathcal{H}}}_n(0)\right],
\label{Eq:QuDuffQuHarm-bar(beta)_j}
\end{align}
and $\hat{\bar{\beta}}_{n}$ as
\begin{align}
\begin{split}
\hat{\bar{\beta}}_{n}=\beta_{n}-\frac{3\varepsilon}{2}\omega_j&\left[u_n^4\hat{\bar{\mathcal{H}}}_n(0)+2u_n^2u_j^2\hat{\bar{\mathcal{H}}}_j(0)\right.\\
&+\left.\sum\limits_{m\neq n}2u_n^2u_m^2\hat{\bar{\mathcal{H}}}_m(0)\right].
\end{split}
\label{Eq:QuDuffQuHarm-bar(beta)_n}
\end{align}
\end{subequations}

In App.~\ref{SubApp:ClDuffingDiss}, we showed that adding another time scale for the decay rate and doing MSPT up to leading order resulted in the trivial solution ~(\ref{Eq:ClDuffing-Sol of X^(0)(t)}) where the dissipation only appears as a decaying envelope. Therefore, we can immediately generalize the MSPT solutions~(\ref{Eq:QuDuffQuHarm-bar(beta)_j}-\ref{Eq:QuDuffQuHarm-bar(beta)_n}) to the dissipative case where the complex pole $p_{j}=-\alpha_j-i\beta_j$ of the transmon-like mode is corrected as
\begin{subequations}
\begin{align}
\hat{\bar{p}}_{j}=p_{j}+i\frac{3\varepsilon}{2}\omega_j\left[u_j^4\hat{\bar{\mathcal{H}}}_j(0)e^{-2\alpha_j t}+\sum\limits_{n}2u_j^2u_n^2\hat{\bar{\mathcal{H}}}_n(0)e^{-2\alpha_n t}\right],
\label{Eq:QuDuffQuHarm-bar(p)_j}
\end{align}
and resonator-like mode $p_{n}=-\alpha_n-i\beta_n$ as
\begin{align}
\begin{split}
\hat{\bar{p}}_{n}=p_{n}+i\frac{3\varepsilon}{2}\omega_j&\left[u_n^4\hat{\bar{\mathcal{H}}}_n(0)e^{-2\alpha_n t}+2u_n^2u_j^2\hat{\bar{\mathcal{H}}}_j(0)e^{-2\alpha_j t}\right.\\
&+\left.\sum\limits_{m\neq n}2u_n^2u_m^2\hat{\bar{\mathcal{H}}}_m(0)e^{-2\alpha_m t}\right].
\end{split}
\label{Eq:QuDuffQuHarm-bar(p)_n}
\end{align}
\end{subequations}
Then, the MSPT solution for $\hat{\mathcal{X}}_j^{(0)}(t)$ is obtained as
\begin{align}
\begin{split}
\hat{\mathcal{X}}_j^{(0)}(t)&=u_j\frac{\hat{\bar{a}}_j(0)e^{\hat{\bar{p}}_j t}+e^{\hat{\bar{p}}_j t}\hat{\bar{a}}_j(0)}{2\cos\left(\frac{3\omega_j}{4}u_j^4\varepsilon t e^{-2\alpha_j t}\right)}+H.c.\\
&+\sum\limits_n\left[u_n\frac{\hat{\bar{a}}_n(0)e^{\hat{\bar{p}}_n t}+e^{\hat{\bar{p}}_n t}\hat{\bar{a}}_n(0)}{2\cos\left(\frac{3\omega_j}{4}u_n^4\varepsilon t e^{-2\alpha_n t}\right)}+H.c.\right].
\label{Eq:QuDuffQuHarm-X_j^(0)(t) MSPT Sol}
\end{split}
\end{align}
Note that if there is no coupling, $u_j=1$ and $u_n=0$ and we retrieve the MSPT solution of a free Duffing oscillator given in Eq.~(\ref{Eq:QuDuffingNoMem-X^(0)(t) sol}).
\section{Reduced equation for the numerical solution}
\label{App:RedNumEq}
In this appendix, we provide the derivation for Eq.~(\ref{eqn:NumSim-QuDuffingOscMemReduced}) based on which we did the numerical solution for the spontaneous emission problem. Substituting Eq.~(\ref{eqn:Expansion of Sine}) into Eq.~(\ref{eqn:NL SE Problem}) we obtain the effective dynamics up to $\mathcal{O}(\varepsilon^2)$ as
\begin{align}
\begin{split}
&\hat{\ddot{X}}_j(t)+\omega_j^2\left[1-\gamma+i\mathcal{K}_1(0)\right]\left[\hat{X}_j(t)-\varepsilon \Tr_{ph}{\{\hat{\rho}_{ph}(0)\hat{\mathcal{X}}_j^3(t)\}}\right]\\
&=-\int_{0}^{t}dt'\omega_j^2 \mathcal{K}_2(t-t')[\hat{X}_j(t')-\varepsilon \Tr_{ph}{\{\hat{\rho}_{ph}(0)\hat{\mathcal{X}}_j^3(t')\}}].
\label{Eq:RedNumEq-QuDuffingOscMem}
\end{split}
\end{align}
If we are only interested in the numerical results up to linear order in $\varepsilon$ then we can write
\begin{align}
\hat{\mathcal{X}}_j(t)=\hat{\mathcal{X}}_j^{(0)}(t)+\varepsilon\hat{\mathcal{X}}_j^{(1)}(t)+\mathcal{O}(\varepsilon^2),
\end{align}
and we find that
\begin{align}
\begin{split}
\varepsilon\Tr_{ph}\left\{\hat{\rho}_{ph}(0)\hat{\mathcal{X}}_j^3(t)\right\}&=\varepsilon\Tr_{ph}\left\{\hat{\rho}_{ph}(0)\left[\hat{\mathcal{X}}_j^{(0)}(t)\right]^3\right\}\\
&+\mathcal{O}\left(\varepsilon^2\right).
\end{split}
\end{align}
Note that in this appendix $\hat{\mathcal{X}}_j^{(0)}(t)$ differs the MSPT notation in the main body and represents the linear solution. We know the exact solution for $\hat{\mathcal{X}}_j^{(0)}(t)$ via Laplace transform as
\begin{align}
\begin{split}
\hat{\mathcal{X}}_j^{(0)}(t)&=\mathfrak{L}^{-1}\left\{\frac{s\hat{\mathcal{X}}_j^{(0)}(0)+\omega_j\hat{\mathcal{Y}}_j^{(0)}(0)}{D_j(s)}\right\}\\
&+\mathfrak{L}^{-1}\left\{\frac{\sum\limits_n \left[a_n(s)\hat{\mathcal{X}}_n^{(0)}(0)+b_n(s)\hat{\mathcal{Y}}_n^{(0)}(0)\right]}{D_j(s)}\right\}\\
&=\mathfrak{L}^{-1}\left\{ \frac{s\hat{X}_j^{(0)}(0)+\omega_j\hat{Y}_j^{(0)}(0)}{D_j(s)}\right\}\otimes\hat{\mathbf{1}}_{ph}\\
&+\hat{\mathbf{1}}_j\otimes\mathfrak{L}^{-1}\left\{\frac{\sum\limits_n \left[a_n(s)\hat{X}_n^{(0)}(0)+b_n(s)\hat{Y}_n^{(0)}(0)\right]}{D_j(s)}\right\},
\end{split}
\label{Eq:RedNumEq-FormalSol Of Mathcal(X)_j}
\end{align}
where we have employed the fact that at $t=0$, the Heisenberg and Schr\"odinger operators are the same and have the following product form
\begin{subequations}
\begin{align}
&\hat{\mathcal{X}}_j^{(0)}(0)=\hat{X}_j^{(0)}(0)\otimes \hat{\mathbf{1}}_{ph},\\
&\hat{\mathcal{Y}}_j^{(0)}(0)=\hat{Y}_j^{(0)}(0)\otimes \hat{\mathbf{1}}_{ph},\\
&\hat{\mathcal{Y}}_n^{(0)}(0)=\hat{\mathbf{1}}_{j}\otimes\hat{Y}_n^{(0)}(0),\\
&\hat{\mathcal{X}}_n^{(0)}(0)=\hat{\mathbf{1}}_{j}\otimes\hat{X}_n^{(0)}(0).
\end{align}
\end{subequations}
The coefficients $a_n(s)$ and $b_n(s)$ can be found from the circuit elements and are proportional to light-matter coupling $g_n$. However, for the argument that we are are trying to make, it is sufficient to keep them in general form. 

Note that equation~(\ref{Eq:RedNumEq-FormalSol Of Mathcal(X)_j}) can be written formally as
\begin{align}
\hat{\mathcal{X}}_j^{(0)}(t)=\hat{X}_j^{(0)}(t)\otimes \hat{\mathbf{1}}_{ph}+\hat{\mathbf{1}}_{j} \otimes \hat{X}_{j,ph}(t).
\end{align}
Therefore, $\left[\hat{\mathcal{X}}_j^{(0)}(t)\right]^3$ is found as
\begin{align}
\begin{split}
&\left[\hat{\mathcal{X}}_j^{(0)}(t)\right]^3=\left[\hat{X}_j^{(0)}(t)\right]^3\otimes \hat{\mathbf{1}}_{ph}+\hat{\mathbf{1}}_{j} \otimes \hat{X}_{j,ph}^3(t)\\
&+3\left\{\left[\hat{X}_j^{(0)}(t)\right]^2\otimes \hat{X}_{j,ph}(t)+\hat{X}_j^{(0)}(t) \otimes \hat{X}_{j,ph}^2(t)\right\}.
\end{split}
\end{align}
Finally, we have to take the partial trace with respect to the photonic sector. For the initial density matrix $\hat{\rho}_{ph}(0)=\ket{0}_{ph}\bra{0}_{ph}$
\begin{align}
\braket{\hat{X}_{j,ph}(t)}_{ph}=\braket{\hat{X}_{j,ph}^3(t)}_{ph}=0.
\end{align}
The only nonzero expectation values in $\braket{\hat{\mathcal{X}}_{j,ph}^2(t)}_{ph}$ are $\braket{\hat{X}_n^2(0)}_{ph}=\braket{\hat{Y}_n^2(0)}_{ph}=1$. Therefore, the partial trace over the cubic nonlinearity takes the form
\begin{align}
\begin{split}
&\Tr_{ph}\left\{\hat{\rho}_{ph}(0)\left[\hat{\mathcal{X}}_j^{(0)}(t)\right]^3\right\}=\\
&\left[\hat{X}_j^{(0)}(t)\right]^3
+3\mathfrak{L}^{-1}\left\{\frac{\sum\limits_n\left[a_n^2(s)+b_n^2(s)\right]}{D_j(s)}\right\}\hat{X}_j^{(0)}(t).
\end{split}
\end{align}
The first term is the reduced transmon operator cubed. The second term is the sum over vacuum fluctuations of the resonator modes. Neglecting these vacuum expectation values we can write
\begin{align}
\begin{split}
\Tr_{ph}\left\{\hat{\rho}_{ph}(0)\left[\hat{\mathcal{X}}_j^{(0)}(t)\right]^3\right\}&\approx\left[\hat{X}_j^{(0)}(t)\right]^3\\
&=\hat{X}_j^3(t)+\mathcal{O}(\varepsilon^2),
\end{split}
\label{Eq:RedNumEq-trace simplification}
\end{align}
Substituting Eq.~(\ref{Eq:RedNumEq-trace simplification}) into Eq.~(\ref{Eq:RedNumEq-QuDuffingOscMem}) gives
\begin{align}
\begin{split}
&\hat{\ddot{X}}_j(t)+\omega_j^2\left[1-\gamma+i\mathcal{K}_1(0)\right]\left[\hat{X}_j(t)-\varepsilon\left[\hat{X}_j(t)\right]^3\right]\\
&=-\int_{0}^{t}dt'\omega_j^2 \mathcal{K}_2(t-t')\left[\hat{X}_j(t')-\varepsilon\left[\hat{X}_j(t')\right]^3\right]+\mathcal{O}(\varepsilon^2).
\label{Eq:RedNumEq-QuDuffingOscMemReduced}
\end{split}
\end{align}
\bibliography{SpEmAll}
\end{document}